\documentclass[aps,prb,superscriptaddress,floats,reprint,nofootinbib]{revtex4-2}
\usepackage[colorlinks=true,
    linkcolor=blue,
    urlcolor=blue,
    citecolor=green]{hyperref}

\usepackage[english]{babel}
\usepackage[utf8]{inputenc}
\usepackage{amsfonts,graphicx,soul,mathrsfs}
\usepackage{url}
\usepackage[dvipsnames]{xcolor}
\usepackage[normalem]{ulem}
\usepackage{epstopdf}
\usepackage{textcomp}
\usepackage{xcolor}
\usepackage{soul}
\usepackage{physics}
\usepackage{graphics}
\usepackage{comment}
\usepackage{amssymb}
\usepackage{amsmath}
\usepackage{tensor}

\usepackage{subcaption}

\usepackage{animate}
\newcommand{\e}{\mathrm{e}}

\newcommand{\be}{\begin{equation}}
\newcommand{\ee}{\end{equation}}
\newcommand{\bear}{\be\begin{array}}
\newcommand{\bw}{\begin{widetext}}
\newcommand{\ew}{\end{widetext}}
\newcommand{\bea}{\begin{eqnarray}}
\newcommand{\eea}{\end{eqnarray}}

\newcommand{\bp}{{\bf p}}
\newcommand{\br}{{\bf r}}

\newcommand{\bk}{{\bf k}}

\newcommand{\la}{\langle}
\newcommand{\ra}{\rangle}

\usepackage{xcolor}

\newcommand{\dst}{\displaystyle}
\newcommand{\fr}[2]{\frac{{\dst #1}}{{\dst #2}}}

\usepackage{dcolumn}
\usepackage{bm}

\usepackage{latexsym}
\usepackage{indentfirst}

\usepackage{bibunits}
\defaultbibliographystyle{apsrev4-2}
\defaultbibliography{refs}

\begin{document}
\begin{bibunit}  

\title{Phase-space description of photon emission}

\author{D.V. Karlovets}
\email{dmitry.karlovets@metalab.ifmo.ru}
\affiliation{School of Physics and Engineering,
ITMO University, 197101, St. Petersburg, Russia}
\affiliation{Petersburg Nuclear Physics Institute named by B.P. Konstantinov of National Research Centre ``Kurchatov Institute'', 188300, Gatchina, Russia}

\author{A.A. Shchepkin}
\email{a.shchepkin@metalab.ifmo.ru}
\affiliation{School of Physics and Engineering,
ITMO University, 197101, St. Petersburg, Russia}
\affiliation{Petersburg Nuclear Physics Institute named by B.P. Konstantinov of National Research Centre ``Kurchatov Institute'', 188300, Gatchina, Russia}

\author{A.D. Chaikovskaia}
\affiliation{School of Physics and Engineering,
ITMO University, 197101, St. Petersburg, Russia}

\author{ \\ D.V. Grosman}
\affiliation{School of Physics and Engineering,
ITMO University, 197101, St. Petersburg, Russia}

\author{D.A. Kargina}
\affiliation{School of Physics and Engineering,
ITMO University, 197101, St. Petersburg, Russia}
\affiliation{Petersburg Nuclear Physics Institute named by B.P. Konstantinov of National Research Centre ``Kurchatov Institute'', 188300, Gatchina, Russia}

\author{U.G. Rybak}
\affiliation{Peter the Great St. Petersburg Polytechnic University, 194021, St. Petersburg, Russia}

\author{G.K. Sizykh}
\affiliation{School of Physics and Engineering,
ITMO University, 197101, St. Petersburg, Russia}
\affiliation{Petersburg Nuclear Physics Institute named by B.P. Konstantinov of National Research Centre ``Kurchatov Institute'', 188300, Gatchina, Russia}

\begin{abstract}
Interactions between charged particles and light occur in real space and time, yet quantum field theory usually describes them in momentum space. While this approach is well suited for calculating emission probabilities and cross sections, it is insensitive to spatial and temporal phenomena such as radiation formation, quantum coherence, and wavepacket spreading. These effects are becoming increasingly important for experiments involving electrons, photons, atoms, and ions, particularly with the advent of attosecond spectroscopy and metrology. Here, we propose a general method for describing the emission of photons in phase space via a Wigner function. Several effects for Cherenkov radiation are predicted, absent in classical realm or in quantum theory in momentum space, such as a finite spreading time of the photon, finite duration of the flash and a quantum shift of the photon arrival time. The photon spreading time turns out to be negative near the Cherenkov angle, the flash duration is defined by the electron packet size, and the temporal shift can be both positive and negative. The characteristic time scales of these effects lie in the atto- and femtosecond ranges, thereby illustrating atomic origins of these macroscopic phenomena. The near-field distribution of the photon field resembles the electron packet shape, thus making ``snapshots'' of the emitter wave function. Our approach can easily be generalized to the other types of radiation and extended to scattering, decay, and annihilation processes, bringing tomographic methods of quantum optics to particle physics.

\end{abstract}

\maketitle

\section{Introduction}

Radiation emitted by charged particles is an inherently space- and time-dependent phenomenon, and yet quantum field theory describes such processes in momentum space, where scattering amplitudes and cross sections are most commonly formulated. Although this approach is widespread, it masks spatial and temporal features of photon emission, such as a finite formation length, spatial and temporal coherence, wavepacket spreading and so forth, essential in modern experiments involving attosecond metrology with ultrashort electron beams and structured wave packets.

The phase-space description of a photon field is more general, and can be based on Glauber-Sudarshan $P$-distribution, Husimi $Q$-distribution, and Wigner function (see, e.g., the review \cite{Lee1995}). These quasiprobability functions, being the different representations of the photon density matrix, are widely used in quantum optics and quantum information for a visualization of quantum states (see, e.g., \cite{Linowski2024}). Phase-space methods have found a lot of applications across quantum optics, quantum communication and molecular dynamics. Moreover, the measurements of a Wigner function, including its negative regions, are accessible experimentally \cite{Nogues2000_Wig_meas, Mallick2025}.

Despite their versatility, these techniques have rarely been applied to radiation processes in QED, especially when the spatial structure of a source or a detector plays a role. This gap is especially pronounced for Cherenkov radiation, where the classical Tamm-Frank theory \cite{Konkov, Mono, Pafomov, NIM} describes spectral and angular features, however provides no insight into how spatial coherence of the electron wave packet influences the emission in real space and time.

In this work, we develop a general phase-space approach for single-photon emission based on the Wigner function of the emitted field. The method explicitly incorporates the structured electron packet and the measurement procedure, thereby treating both projective and generalized measurements. Applying this formalism to Cherenkov radiation, we recover the classical results in the far field while predicting several effects with no direct analogue in classical electrodynamics or customary momentum-space QED. They are negative spreading time -- and, therefore, negative formation length -- of the photon near the Cherenkov angle, finite flash duration and a quantum shift in photon arrival time. These effects originate from the spatial coherence encoded in the Wigner function and are apparent near the Cherenkov angle, where constructive interference of partial waves leads to slow spreading of the photon field. 

Remarkably, in the near field, the photon spatial profile resembles the shape of the emitting electron packet, effectively generating ``snapshots'' of the electron wave function, an effect absent in the far-field theory. The characteristic time scales -- from attoseconds to femtoseconds -- remain imprinted by the atomic-scale origins of the medium response \cite{Karlovets2021,Ehberger,Cho,Latychevskaia,Cho_Oshima,Dienstbier}. Importantly, this phase-space approach can be extended to scattering, decay, and annihilation processes \cite{PRA2019}. 

This work is a continuation of the paper \cite{Karlovets2025}. Here, we provide more detailed derivations of the Cherenkov photon Wigner function, taking the magnetic counterpart into account. We also study the features of Cherenkov emission in a dispersive medium and visualize the Wigner function with respect to time and emission angle, thereby investigating its evolution and negative regions.

{The paper is organized as follows. In Section \ref{sec:Gen_meas} we introduce the concept of the generalized measurements and, as a special case, positive operator valued measurements (POVM) of the emitted photon state. Section \ref{Contrad} specifies the model assumptions for the correct description of the photon state. In Section \ref{sec:Phot_em} the phase-space formalism for the emission is introduced via the Wigner function of the emitted Cherenkov photon, while sections \ref{sec:Near-field} and \ref{sec:parax} study the near-field and paraxial approximations of the Wigner function, respectively, and the Cherenkov radiation in strongly dispersive media. In Section \ref{sec:Images} we demonstrate the formation and evolution of the photon Wigner function via its visual representations and discuss its key properties including the regions of negativity. Section \ref{sec:conc} summarizes our conclusions and outline possible directions for future work.}

Throughout the paper we use the relativistic system of units $\hbar = c = m_e = 1$, and the Coulomb gauge of photon vector potential is implied.

\section{Generalized measurements \\ of the evolved quantum states}
\label{sec:Gen_meas}
Let us consider the photon emission in QED with a final state consisting of an electron and a Cherenkov photon in a transparent medium:
\bea
e \to e' + \gamma.
\label{FOprocess}
\eea
This process can be extended to the general emission, scattering, or annihilation process with two final particles. A bipartite final state in the interaction picture is obtained by acting on the initial state by an evolution operator $\hat{S}^{(1)}$ within the first order of the perturbation theory \cite{BLP, PS}. This state is 
\bea
|e',\gamma\ra = \left(\hat{1} + \hat{S}^{(1)}\right)|\text{in}\ra, \label{final_state}
\eea
where $|\text{in}\ra = |e_{\text{in}}\ra\otimes|0_{\gamma}\ra$. There is no interaction between the final particles in the asymptotic limit, so the interaction picture coincides with the Heisenberg one, in which the operators depend on time whereas the state vectors do not.

One can rewrite the second term in Eq.(\ref{final_state}) into the two-particle Hilbert space,
\bea
& \dst \hat{S}^{(1)}|\text{in}\ra = \sum\limits_{f_e, f_\gamma} |f_e\ra\otimes|f_\gamma\ra\, S_{fi}^{(1)},
\label{2partWF} 
\eea
where $S_{fi}^{(1)} = \la f_e,f_\gamma|\hat{S}^{(1)}|\text{in}\ra$. The complete sets $\{|f_e\ra\}$, $\{|f_\gamma\ra\}$ can represent the plane-wave states with the momenta $\bp', \bk$ and the helicities $\lambda'=\pm 1/2,\lambda_\gamma = \pm 1$, so that 
\bea
|e',\gamma\ra = |\text{in}\ra + \sum_{\lambda', \lambda_\gamma} \int\bar{d}^3\bk\,\bar{d}^3\bp'\, |\bp',\lambda'\ra\otimes|\bk,\lambda_\gamma\ra\, S_{fi}^{(1)}. \quad
\label{2partWFPW}
\eea
where $\bar{d}^3 \equiv d^3/(2\pi)^3$. If the final electron state is projected onto a bra $\la f_{e}^{(\text{det})}|$, which \textit{belongs to the orthonormal set of states $\{|f_e\ra\}$} from Eq.\eqref{2partWF}, we get $\la f_{e}^{(\text{det})}|f_e\ra = \delta_{f_{e}^{(\text{det})}f_e}$ and the state of the photon only is
\bea
& \dst |\gamma\ra = \la f_{e}^{(\text{det})}|e_{\text{in}}\ra|0_{\gamma}\ra + \sum\limits_{f_\gamma} |f_\gamma\ra \la f_{e}^{(\text{det})},f_\gamma|\hat{S}^{(1)}|\text{in}\ra \cr
& \dst \equiv \la f_{e}^{(\text{det})}|e_{\text{in}}\ra|0_{\gamma}\ra + \sum\limits_{f_\gamma} |f_\gamma\ra\, S_{fi}^{(1)},
\label{1partPh}
\eea
where $\la f_{e}^{(\text{det})}|e_{\text{in}}\ra$ is \textit{not} necessarily Kronecker-symbol $\delta_{\text{in},\text{det}}$ because the initial electron state may not belong to the set from Eq.(\ref{2partWF}). This detection scheme is called \textit{a projective} -- von Neumann or standard -- measurement of the final electron state. However, the detector state of the electron $\la f_{e}^{\text{(det)}}|$ actually may \textit{not} belong to any orthonormal set, which implies that $\la f_{e}^{\text{(det)}}|f_e\ra \ne \delta_{f_{e}^{(\text{det})}f_e}$. This scheme represents the so-called \textit{generalized measurements}, which also encompasses projective ones as a special case. 

In this framework, the detector state can represent a coherent superposition of states from an orthonormal set with the simplest example being a coherent state, which implies that the packet is measured (localized) in two reciprocal spaces: for instance, in the momentum space and in the coordinate one. Thereby, one can obtain complementary information about the electron state in both the representations, whereas the projective measurement takes place only in one space -- say, when measuring a three-momentum $\bp$ of a packet -- its spatial distribution remains unobserved (see \cite{Barnett}). Now we have the following more general expression for the photon state:
\bea
|\gamma\ra = \la f_{e}^{(\text{det})}|e_{\text{in}}\ra|0_{\gamma}\ra + \sum\limits_{f_e,f_\gamma} |f_\gamma\ra\,\la f_{e}^{\text{(det)}}|f_e\ra\, S_{fi}^{(1)}.
\label{1partPhGen}
\eea
In terms of plane waves, when $|f_\gamma\ra = |\bk,\lambda_\gamma\ra$ and $|f_e\ra = |\bp',\lambda'\ra$, the evolved photon state reads:
\bw
\bea
& \dst |\gamma\ra = \la f_{e}^{(\text{det})}|e_{\text{in}}\ra|0_{\gamma}\ra + \sum\limits_{\lambda',\lambda_\gamma}\int\bar{d}^3\bk\,\bar{d}^3\bp'\, |\bk,\lambda_\gamma\ra\,\left(f_e^{(\text{det})}(\bp',\lambda')\right)^*\, S_{fi}^{(1)},\cr
& \dst S_{fi}^{(1)} \equiv  S_{fi}^{(1)}(\bp',\lambda',\bk,\lambda_\gamma) = \la \bk,\lambda_\gamma;\bp',\lambda'|\hat{S}^{(1)}|\text{in}\ra,\quad \left(f_e^{(\text{det})}(\bp',\lambda')\right)^* = \la f_{e}^{\text{(det)}}|\bp',\lambda'\ra.
\label{1partPhGenPW}
\eea
\ew
\textit{The electron detector function} $f_e^{(\text{det})}(\bp',\lambda')$, which is generally complex, can be described as a wave packet with finite uncertainties of all the momenta components, each being on-shell with the energy $\varepsilon_e = \sqrt{m_e^2 + (\bp')^2}$. Eq.(\ref{1partPhGenPW}) with the generalized measurement of the electron state can be written in a way similar to Eq.(\ref{1partPh}) with the projective measurement as follows:
\bea
& \dst |\gamma\ra = \la f_{e}^{(\text{det})}|e_{\text{in}}\ra|0_{\gamma}\ra + \sum\limits_{f_\gamma} |f_\gamma\ra\, S_{fi}^{\text{GM}} = \cr
& \dst \la f_{e}^{(\text{det})}|e_{\text{in}}\ra|0_{\gamma}\ra + \sum\limits_{\lambda_\gamma}\int\bar{d}^3\bk\, |\bk,\lambda_\gamma\ra\, S_{fi}^{\text{GM}}(\bk,\lambda_\gamma), \label{gammaGM} \\
& \dst 
S_{fi}^{\text{GM}}(\bk,\lambda_\gamma) = \sum\limits_{\lambda'}\int\bar{d}^3\bp' \left(f_e^{(\text{det})}(\bp',\lambda')\right)^*S_{fi}^{(1)}.
\label{SGM}
\eea
Clearly, Eq.(\ref{1partPh}) is a special case of Eq.(\ref{gammaGM}). When the electron is detected in a pure state within the generalized measurement scheme, the photon state in the lowest order of the perturbation theory is defined by Eq.(\ref{gammaGM}) as
\bea
& \dst \la\gamma|\gamma\ra = |\la f_{e}^{(\text{det})}|e_{\text{in}}\ra|^2 + \sum\limits_{\lambda_\gamma}\int\bar{d}^3\bk\,\left|S_{fi}^{\text{GM}}(\bk,\lambda_\gamma)\right|^2,
\eea
We call $ |\gamma\ra$ \textit{the evolved state of the photon field}, even though it is not normalized to unity as the integrated emission probability is less than one.

Equations \eqref{1partPhGenPW} and \eqref{gammaGM} allow one to find a photon evolved state when the electron is jointly detected in the given state $|f_{e}^{\text{(det)}}\ra$ described as a packet with finite uncertainties of the quantum numbers from a set. Importantly, when measuring a quantum number with a finite error -- for example, a component of momentum, an angular momentum projection, etc. --  the information about the system is not lost only if the measurement of the conjugate quantum numbers is done \textit{jointly} so that the corresponding uncertainty relation is minimized. Otherwise the information about the system is partially lost and it can be described only with a density matrix. 

\begin{figure}[h!]
	\includegraphics[width=0.75\linewidth] {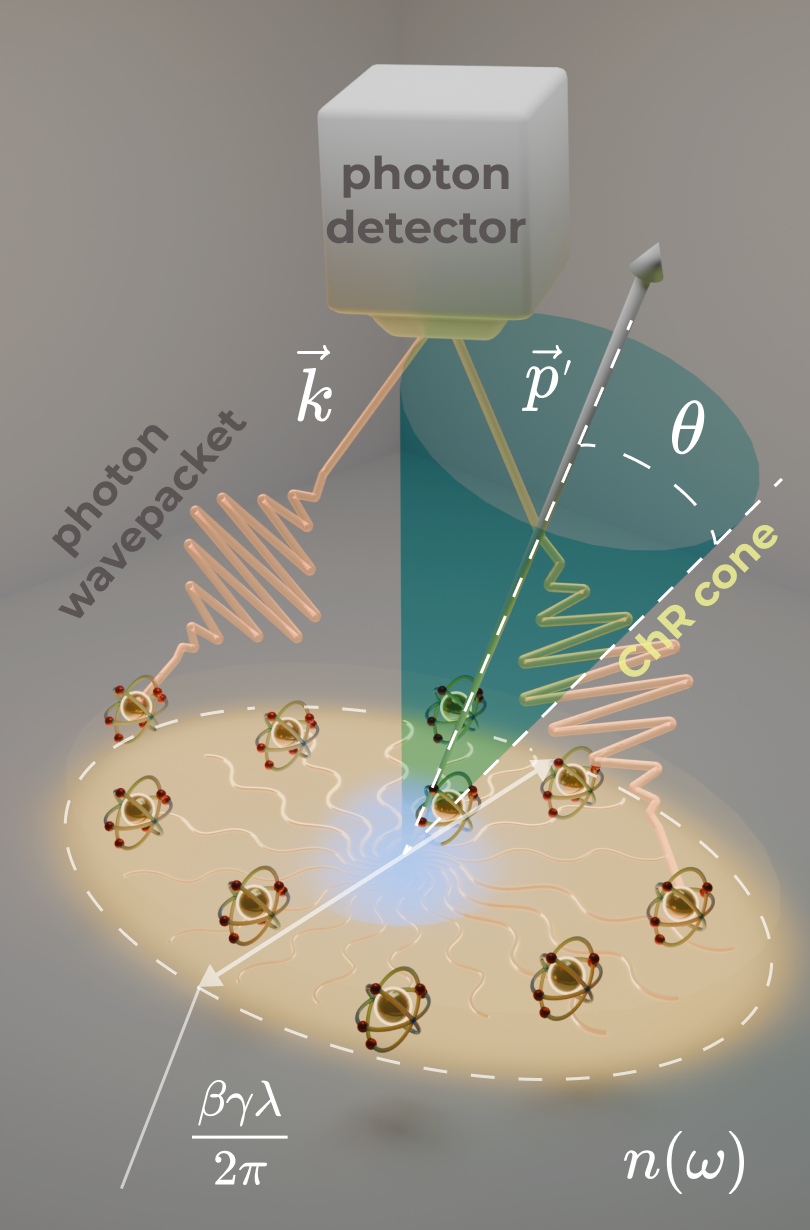}
	\caption{The Cherenkov radiation generated by the atoms, excited by an incident electron in the medium with the refractive index $n(\omega)$. The emitted photon with the momentum $\bk$ is registered by the photon detector. The final electron with the momentum $\bp'$ may or may not be detected in this scheme, which in turn impacts the photon state. The characteristic emission size is equal to $\beta\gamma\lambda/2\pi$ \cite{Jackson} (see Section \ref{sec:FDaSh}, Eq. \eqref{emit_size})}
	\label{IllustCher}
\end{figure}

The studied problem is illustrated in Fig. \ref{IllustCher}. The Cherenkov photon state is emitted by the atoms of the medium excited by an incident electron. The final photon is supposed to be registered by a photon detector, whereas the final electron (the electron after the emission) may or may not be registered. Whether or not the final electron is ultimately detected affects the photon state due to the entanglement between the two final particles.

In the scenario when the quantum numbers of the final electron are not jointly measured, the scheme of the positive-operator-valued measurements (POVM) can be used. This is a special case for the generalized measurement  scheme, where the information about the system after the measurement of the electron state is partially lost. Hence, the final photon can only be described with a density matrix defined as follows (see,e.g. \cite{Lvovsky}):

\bea
\hat{\rho}_{\gamma}^{(\text{POVM})} = \text{Tr}\{\hat{F}^{(\text{det})}_e\,\hat{\rho}_{e\gamma}\},
\label{rhoPOVM}
\eea
where $\hat{\rho}_{e\gamma}$ is a density matrix of the final bipartite system, which for the pure state becomes $\hat{\rho}_{e\gamma} \to |e',\gamma\ra \la e',\gamma|$, and the trace is taken over the electron component. The operator $\hat{F}^{(\text{det})}_e$ describes the detection of the final electron and it can be written as an \textit{incoherent} mixture of states $|f_e\ra$ from the corresponding orthogonal set. For instance, for the plane waves it is
\bea
\hat{F}^{(\text{det})}_e = \sum\limits_{\lambda'}\int\bar{d}^3\bp'\, F_e^{(\text{det})}(\bp',\lambda')\, |\bp',\lambda'\ra\la\bp',\lambda'|,
\label{POVMpw}
\eea
where $F_e^{(\text{det})}(\bp',\lambda')$ is a normalized probability to detect the electron with the momentum $\bp'$ and the helicity $\lambda'$.

Note that the choice of the detector state basis is crucial. For example, the orbital angular momentum (OAM) of an electron cannot be measured in the plane-wave basis. During a measurement, the particle state collapses into one of the detector eigenstates. Since states with well-defined linear momentum do not possess a  well-defined OAM, the latter cannot be meaningfully determined in this basis. In contrast, when the detector states are described with the angular momentum operator eigenfunctions
\bea
    \hat{J}_z|k_z,k_\perp,m,\Lambda\ra = m|k_z,k_\perp,m,\Lambda\ra,
\eea
the measurement projects the electron onto a state with an OAM, for example described with Bessel functions ($m,\, \Lambda$ are the total angular momentum and the helicity of a Bessel state, respectively, see Supplemental information for details \cite{SuppInf}, which include the references \cite{Lvovsky,JHEP, PRD,MIC,Yashin,EPJC}).  Also, POVM scheme, in contrast to projective measurements, allows one to operate with nonorthogonal states, preserving them after a measurement.


\section{Subtleties of the plane-wave formalism}
\label{Contrad}

When the initial electron is in a plane-wave state $|\text{in}\ra = |\bp,\lambda\ra$ with the 4-momentum $p = \{\varepsilon, 0, 0, |\bp|\}$ and the helicity $\lambda = \pm 1/2$, the matrix element is proportional to the delta function of the energy-momentum conservation \cite{BLP, EPJC},
\bea
& \dst S_{fi}^{(1)} = i(2\pi)^4 \delta^{(4)}(p-p'-k)\times\cr
& \dst \frac{\sqrt{4\pi}}{\sqrt{8\omega n^2 \varepsilon \varepsilon'}}\, M_{fi}(\bp, \bk,\lambda,\lambda_\gamma).
\label{Sfi1}
\eea
The triangle rule for transverse momenta (see Appendix \ref{ApTri}) allows us to rewrite the delta-function in Eq. \eqref{Sfi1} as follows
{\small\bea
& \dst \delta^{(4)} (p-p'-k) \propto \delta^{(2)}(\bp'_{\perp} + \bk_{\perp}) = \frac{\Theta(k_{\perp}, p'_{\perp})}{p'_{\perp}}\delta (p'_{\perp} - k_{\perp}) \times \cr 
& \dst \Bigg(\delta(\phi' - (\phi_k - \pi))\Big|_{\phi_k\in (\pi,2\pi)} + \delta(\phi' - (\phi_k + \pi))\Big|_{\phi_k\in (0,\pi)}\Bigg),
\label{delta_f}
\eea}
where $\Theta(k_{\perp}, p'_{\perp})$ equals unity when $k_{\perp}= p'_{\perp}$ and zero otherwise. The transverse momenta of the final electron and the photon are $\bp'_{\perp} = p'_{\perp}\{\cos\phi',\sin\phi',0\} $ and $\bk_{\perp} = k_{\perp}\{\cos\phi_k,\sin\phi_k, 0\}$, respectively.

The azimuthal delta function reveals correlation between the angles: when the electron is detected at a certain azimuthal angle $\phi'$, the photon angle $\phi_k$ is automatically put to a definite value, regardless of whether the photon is subsequently detected or not. Thereby the photon evolved state \textit{becomes a plane wave} with undefined OAM projection because the twisted state with a definite OAM implies an unknown azimuthal angle with its uncertainty equal to $2\pi$ \cite{Carruthers1968}. If the electron is detected in the cylindrical basis with no definite azimuthal angle, the matrix element is obtained by integrating Eq.\eqref{Sfi1} over $\phi'$ \cite{EPJC}:
\bea
& \dst S_{fi}^{\text{GM}} = \int\frac{d\phi'}{2\pi}\, S_{fi}^{(1)}.
\label{SNPW}
\eea
The azimuthal delta-function is erased and the evolved photon state turns out to be \textit{the Bessel beam} with a definite orbital angular momentum.

Let us assume now that the incoming electron plane-wave state has a finite -- even if vanishingly small -- transverse momentum $p_{\perp}\ne 0$, so that $\bp = \{p_{\perp} \cos\phi, p_{\perp}\sin\phi, p_z\}$, which can also take place if the electron is described as a wave packet. Then the energy-momentum delta-function can be written as follows (see the Appendix \ref{ApTri}):
{\small\bea
& \dst \delta^{(4)}(p-p'-k) = \cr
& \dst = \frac{\varepsilon' + \omega}{2\Delta} \delta(p_z - p_z' - k_z) \frac{\delta (p_{\perp} - \tilde{p}_{\perp})}{\tilde{p}_{\perp}} \Theta(p_{\perp}, k_{\perp}, p'_{\perp})\times\cr
& \dst \Big(\delta\left(\phi' - \phi + \alpha\right)\delta\left(\phi_k - \phi - \gamma\right) + \cr
& \dst \delta\left(\phi' - \phi - \alpha\right)\delta\left(\phi_k - \phi + \gamma\right)\Big)
\label{delta4}
\eea}
where 
\bea
& \dst \tilde{p}_{\perp} \equiv \tilde{p}_{\perp}(p'_{\perp}, p'_z, k_{\perp}, k_z) = \cr
& \dst = \sqrt{(p'_{\perp})^2 + k_{\perp}^2 + k^2 + 2 (\varepsilon'\omega - p'_z k_z)},
\label{pt}
\eea 
$\alpha$ is the angle between ${\bp}_{\perp}$ and ${\bp}'_{\perp}$, whereas $\gamma$ is the one between ${\bp}_{\perp}$ and ${\bk}_{\perp}$. The following identities also come in handy:
\bw
\bea
& \dst \delta\left(\phi' - \phi + \alpha\right)\delta\left(\phi_k - \phi - \gamma\right) + \delta\left(\phi' - \phi - \alpha\right)\delta\left(\phi_k - \phi + \gamma\right) = \cr & \dst = \delta\left(\phi' - \phi_k - (\vartheta-\pi)\right)\delta\left(\phi_k - \phi - \gamma\right) + \delta\left(\phi' - \phi_k + (\vartheta-\pi)\right)\delta\left(\phi_k - \phi + \gamma\right),
\label{delta22}
\eea
\ew
where $\vartheta = \pi - \alpha - \gamma$ is an angle between ${\bp}'_{\perp}$ and ${\bk}_{\perp}$. Here we encounter two azimuthal delta functions instead of one in Eq. \eqref{delta_f} and the integration over $\phi'$ in (\ref{SNPW}) does \textit{not} eliminate the delta function with $\phi_k$. The definite photon angle $\phi_k$ implies vanishing OAM, so the photon \textit{turns out to be in a plane-wave state}. 

Clearly, we have \textit{a puzzling contradiction}: the photon evolved state represents either the Bessel beam with an indefinite azimuthal angle and the definite OAM or the plane wave with a definite azimuthal angle and a vanishing OAM, depending on the transverse momentum of the incoming plane-wave electron. However, this transverse momentum \textit{can always be put to zero} by rotating the coordinate frame and the physical predictions cannot depend on a choice of coordinates. The resolution of the seeming paradox lies in fact that the real state of the incoming electron represents a well-normalized wave packet, defined, for instance, by Eq. \eqref{WPin}, and then the matrix element becomes as following
\bea
\dst S_{fi}^{\text{GM}} = \sum\limits_{\lambda}\int\bar{d}^3\bp\frac{d\phi'}{2\pi}\,f_e^{(\text{in})}(\bp,\lambda)\, S_{fi}^{(1)}.
\label{SWP}
\eea
Now the integration over $\phi$ and $\phi'$ eliminates both the delta functions in Eq.(\ref{delta4}) or Eq.(\ref{delta22}), so the photon evolved state \textit{becomes a twisted Bessel beam}.

One may well ask if the measurement of the electron state in the cylindrical basis is \textit{inevitable} condition to obtain an OAM of the photon or one can simply take the initial electron as a packet (\ref{WPin}) -- say, the twisted one -- and then detect the final electron as a plane wave $|\bp',\lambda'\ra$. The answer is clear: one integral over $\phi$ in (\ref{SWP}) is \textit{not} enough to eliminate both the delta functions in Eq.(\ref{delta22}), which is why the generalized measurement scheme must be applied to produce twisted photons even if the incoming electron is also a twisted packet. In other words, {vorticity of the radiating charged particle cannot be automatically transferred to the emitted photons if the final electron is projectively measured in the plane-wave state}. 

Indeed, when the incoming electron is a packet \eqref{WPin}, whereas the final one is a plane wave detected at the angle $\phi'$, we get the following photon wave function in momentum space 
\bw
\bea
& \dst \bm{\tilde{A}}(\bk, \omega)  =
i(2\pi)^4\frac{\pi}{\omega n^2}\sum\limits_{\lambda_\gamma,\lambda'}{\bm e}_{\bk, \lambda_\gamma}\int d^3\bp'\,\frac{1}{\sqrt{\varepsilon\varepsilon'}}\left(f_e^{(\text{det})}(\bp',\lambda')\right)^*\,\delta^{(4)}(p-p'-k) M_{fi} \propto \cr 
& \dst \propto \sum\limits_{\lambda_\gamma,\lambda'}{\bm e}_{\bk, \lambda_\gamma}\Big(\left(f_e^{(\text{det})}(\phi' = \phi + \alpha,\lambda')\right)^*\, M_{fi}(\phi = \phi_k - \gamma)\, \delta(\phi' - \phi_k - (\vartheta - \pi)) + \cr 
& \dst + \left(f_e^{(\text{det})}(\phi' = \phi - \alpha,\lambda')\right)^* M_{fi}(\phi = \phi_k + \gamma)\,\delta(\phi' - \phi_k + (\vartheta - \pi))\Big),
\label{Ak1}
\eea
\ew
where $p_z = p'_z + k_z, \, p_{\perp} = \tilde{p}_\perp$ (see Eq.(\ref{pt})), and $f_e^{(\text{det})}$ is a detector wave function (see, for instance, Eq.\eqref{1partPh}). Let us take the incoming packet as a Bessel beam 
\bea
f_e^{(\text{in})}(\bp,\lambda) \propto \delta(p_z-\tilde{p}_z) \frac{\delta(p_{\perp} - \varkappa)}{\varkappa}\, e^{im\phi}.
\eea
Then we employ the following identities:
\bea
& \dst \frac{\delta(p_{\perp} - \varkappa)}{\varkappa}\Big|_{p_{\perp} = \tilde{p}_{\perp}} = \frac{\delta(\omega + \varepsilon' - \varepsilon)}{\varepsilon} =\cr
& \dst = \frac{\varepsilon-\varepsilon'}{\varepsilon}\frac{n^2}{\sqrt{n^2(\varepsilon-\varepsilon')^2 - k_z^2}}\cr
& \dst \delta\left(k_{\perp} - \sqrt{n^2(\varepsilon-\varepsilon')^2 - k_z^2}\right). \quad
\eea
where $\varepsilon = \sqrt{m_e^2 + \varkappa^2 + (p'_z + k_z)^2}$. Thus we have three delta-functions in (\ref{Ak1}), 
\bea
& \dst {\tilde{\bm A}}(\bk, \omega) \propto \delta(p'_z + k_z -\tilde{p}_z)\, \delta(\phi'-\phi_k \pm (\vartheta - \pi)) \times \cr
& \dst \times \delta\left(k_{\perp} - \sqrt{n^2(\varepsilon-\varepsilon')^2 - (\tilde{p}_z - p'_z)^2}\right),
\label{Ak12}
\eea
so the photon momentum has three definite components, $\bk = (k_{\perp},\phi_k,k_z)$, which implies the vanishing OAM projection onto the $z$ axis. Should the initial electron state be a well-normalized Laguerre-Gaussian packet instead of the delocalized Bessel beam, it would simply smear the transverse delta function, but the result would hold. 

\section{Photon emission in phase space}
\label{sec:Phot_em}

The spatial energy density of the emitted photon is
\bea
& \dst W(\br, t) = \frac{1}{4\pi} \left(\left|\la 0|\hat{\bm E}(\br, t)|\gamma\ra\right|^2+\left|\la 0|\hat{\bm H}(\br, t)|\gamma\ra\right|^2\right),
\label{SED_finite}
\eea
where $\hat{\bm E}(\br,t)$ and $\hat{\bm H}(\br,t)$ are the electric and magnetic field operators, respectively, $|\gamma\ra$ is the quantum state with one photon and $|0\ra$ is a vacuum state (for details see Appendix \ref{SpatEnerDens}).

When the incoming electron is a general normalized wave packet:
\bea
& |\text{in}\ra = \dst \sum_{\lambda}\int \bar{d}^3\bp\, f_e^{(\text{in})}(\bp,\lambda)|\bp,\lambda\ra, \cr
& \la\text{in}|\text{in}\ra = 1,
\label{WPin}
\eea
and the outgoing one is detected as a plane wave $|\bp',\lambda'\ra$, which implies a \textit{projective measurement}, the evolved photon state cannot possess an OAM, as shown in Section \ref{Contrad}. However, it can still depend on the size and shape of the incoming electron packet, which reveal themselves in the finite part of the spatial energy density of the emitted photon field Eq. \eqref{E2H2}. For a first-order QED process \eqref{FOprocess} the spatial energy density Eq. \eqref{SED_finite} can be written as follows (see Eqs. \eqref{E_to_expand} and \eqref{H_to_expand}):
\bea
& \dst W(\br,t) = \int \bar{d}^3\bk\, \mathcal W(\br, \bk, t),
\eea
where $\mathcal W(\br,\bk,t)$ is
\bw
{\small\bea
& \dst \mathcal W(\br, \bk, t) = \frac{1}{4\pi} \sum\limits_{\lambda_\gamma,\tilde{\lambda}_{\gamma}}\int \bar{d}^3{\tilde \bk}\,\Big[\bm E^*_{\tilde{\lambda}_{\gamma}}(\bk - {\tilde \bk}/2)\cdot\bm E_{\lambda_\gamma}(\bk+{\tilde \bk}/2) + \bm{H}^*_{\tilde{\lambda}_{\gamma}}(\bk - {\tilde \bk}/2)\cdot\bm H_{\lambda_\gamma}(\bk+{\tilde \bk}/2)\Big]e^{-it \left(\omega(\bk + {\tilde \bk}/2)-\omega(\bk-{\tilde \bk}/2)\right) + i\br\cdot{\tilde \bk}}
\label{EdensityMod}
\eea}
\bea
& \dst \bm E_{\lambda_\gamma}(\bk) = \frac{i\omega \sqrt{4\pi}}{\sqrt{2\omega n^2}}\,{\bm e}_{\bk\lambda_\gamma}\sum\limits_{\lambda}\int\bar{d}^3\bp\,f_e^{(\text{in})}(\bp,\lambda) S_{fi}^{(1)} \cr
& \dst \bm H_{\lambda_\gamma}(\bk) = \frac{i\sqrt{4\pi}}{\sqrt{2\omega n^2}}\,\left[\bk\times{\bm e}_{\bk\lambda_\gamma}\right]\sum\limits_{\lambda}\int\bar{d}^3\bp\,f_e^{(\text{in})}(\bp,\lambda) S_{fi}^{(1)}
\label{EMod}
\eea
\ew
Here, $\mathcal{W}(\br,\bk,t)$ is a Wigner function of the photon evolved state, where, in addition to the previous work \cite{Karlovets2025}, the contribution of the magnetic field is also taken into account. The quantities $\bm E_{\lambda_\gamma}(\bk)$ and $\bm H_{\lambda_\gamma}(\bk)$ are positive-frequency components of the electric and magnetic fields, respectively, with the vector potential and its normalization from Eqs. \eqref{A_k_lamb} and \eqref{A_norm_cond}, and $f_e^{(\text{in})}(\bp,\lambda)$ is the wave function of the incoming electron packet in momentum representation from Eq.(\ref{WPin}).

Eq.\eqref{EdensityMod} shows that one marginal distribution of this Wigner function yields the energy density of the photon field in the coordinate space and time or -- in other words -- the probability density to detect the photon in a region of space centered at the point $\br$ in a moment of time $t$. The other marginal distribution
\bw
\bea
& \dst \int d^3\br\,\mathcal W(\br, \bk, t) = \frac{\omega}{2n^2} \left|\sum\limits_{\lambda}\int\bar{d}^3\bp\,f_e^{(\text{in})}(\bp,\lambda) S_{fi}^{(1)}\right|^2 = \cr 
& \dst = \frac{\omega}{2n^2} (2\pi)^2 \frac{T}{2\pi} \delta(\varepsilon(\bp)-\varepsilon'(\bp')-\omega(\bk))\, \frac{4\pi}{2\omega(\bk) n^2(\omega(\bk)) 2\varepsilon(\bp) 2\varepsilon'(\bp')}\,\left|\sum\limits_{\lambda}f_e^{(\text{in})}(\bp,\lambda)\,M_{fi}\right|^2_{\bp=\bp'+\bk}
\label{drW}
\eea
\ew
yields probability to detect the photon with the frequency $\omega$ and the wave vector $\bk$. Here, $M_{fi}$ - is an emission amplitude, defined in Eqs.\eqref{Mfi} and \eqref{Sfi1}, and $T$ is a normalization time arising from the squared delta-function representation (see Eq.\eqref{nrm_cnst_T}).

To understand what new information the marginal distribution (\ref{EdensityMod}) brings about, let us write the complex amplitude $M_{fi}$ in terms of its module and phase as follows
\bea
M_{fi} = \left|M_{fi}\right|\,\e^{i\zeta_{fi}},
\label{amp_phase}
\eea
where $\zeta_{fi} = \zeta_{fi}(\bp, \bk, \lambda_e,\lambda_\gamma)$ is a dynamic phase of the emission process depending on initial electron and photon momenta, $\bp,\,\bk$ and their helicities, $\lambda_e,\,\lambda_\gamma$, respectively (see, for instance, \cite{JHEP,TOTEM,PRD}). Importantly, this phase \textit{is non-vanishing} even at the tree level (see Appendix \ref{Amp}). The photon Wigner function is defined by the following \textit{master integral} of $\mathcal W_p(\br, \bp, \bk, t)$:
\bw
\bea
& \dst \mathcal W(\br, \bk, t) = (2\pi)^5\sum\limits_{\lambda_\gamma,\tilde{\lambda}_{\gamma}}\sum\limits_{\lambda\tilde{\lambda}} \int\bar{d}^3\bp\,\delta^{(3)}(\bp - \bp' - \bk) \int \frac{\bar{d}^3{\tilde \bk}}{4n^2\left(\omega(\bk+{\tilde \bk}/2)\right)n^2\left(\omega(\bk-{\tilde \bk}/2)\right)\varepsilon'\sqrt{\varepsilon(\bp + \tilde\bk/2)\varepsilon(\bp - \tilde\bk/2)}}\cr  
& \dst \left(\underbrace{{\bm e}^*_{\bk-{\tilde \bk}/2,\tilde{\lambda}_{\gamma}}\cdot{\bm e}_{\bk+{\tilde \bk}/2,\lambda_\gamma}}_{\text{electric counterpart}} + \underbrace{\frac{1}{\omega(\bk - {\tilde\bk}/2)\omega(\bk + {\tilde\bk}/2)}\left[\left(\bk - \frac{\tilde \bk}{2}\right)\times \bm{e}^*_{\bk - \tilde \bk/2,\, \tilde\lambda_\gamma}\right]\left[\left(\bk + \frac{\tilde \bk}{2}\right)\times \bm{e}_{\bk + \tilde \bk/2,\, \lambda_\gamma}\right]}_{\text{magnetic counterpart}}\right)\times\cr
& \dst \times \left(f_e^{(\text{in})}(\bp-{\tilde \bk}/2,\tilde{\lambda})\right)^* f_e^{(\text{in})}(\bp+{\tilde \bk}/2,\lambda)\,\delta\left(\varepsilon(\bp+{\tilde \bk}/2)-\varepsilon' - \omega(\bk + {\tilde \bk}/2)\right)\cr
& \dst \times \delta\left(\varepsilon(\bp-{\tilde \bk}/2)-\varepsilon' - \omega(\bk - {\tilde \bk}/2)\right)  M_{fi}\left(\bp+{\tilde \bk}/2,\lambda,\bk+{\tilde \bk}/2,\lambda_\gamma\right) M_{fi}^*\left(\bp-{\tilde \bk}/2,\tilde{\lambda},\bk-{\tilde \bk}/2,\tilde{\lambda}_{\gamma}\right)\cr
& \dst \times e^{-it \left(\omega(\bk+{\tilde \bk}/2)-\omega(\bk-{\tilde \bk}/2)\right) + i\br\cdot{\tilde \bk}} \equiv \cr
& \dst \equiv \int\bar{d}^3\bp\,(2\pi)^3 \delta^{(3)}(\bp - \bp' - \bk)\, \mathcal W_p(\br, \bp, \bk, t) = \mathcal W_p(\br, \bp = \bp' + \bk, \bk, t),
\label{WEE}
\eea
\ew
so that $\mathcal W(\br, \bk, t) = \mathcal W_p(\br, \bp = \bp' + \bk, \bk, t)$. Note that no approximations have been made so far. 

If the final electron is \textit{not measured}, which is typical for Cherenkov radiation (ChR), the spatio-temporal distribution of the emitted energy is obtained by tracing out the electron quantum numbers,
\bea
& \dst \sum\limits_{\lambda'}\int\bar{d}^3\bp'\,\frac{1}{4\pi} \left[\left|\la 0|\hat{\bm E}({\bm r}, t)|\gamma\ra\right|^2 + \left|\la 0|\hat{\bm H}({\bm r}, t)|\gamma\ra\right|^2\right] 
=  \cr
& \dst = \sum\limits_{\lambda'} \int\bar{d}^3\bp' \bar{d}^3\bk\, \mathcal W(\br, \bk, t) = \cr 
& \dst = \sum\limits_{\lambda'} \int\bar{d}^3\bp\, \bar{d}^3\bk\, \mathcal W_p(\br,\bp,\bk,t)\Big|_{\bp' = \bp - \bk}.
\label{EdEInt}
\eea
where $\mathcal{W}_p(\br,\bp,\bk,t)$ is the photon Wigner function dependent on the initial electron momentum.

To proceed further, we specify the model of the incoming electron packet, defined by the function $f_e^{(\text{in})}$, see Eq.(\ref{WPin}). Assume that it has a Gaussian envelope in momentum space with a phase $\varphi(\bp)$ and a definite helicity $\lambda_e$,
\bea
& \dst f_e^{(\text{in})} (\bp,\lambda) = f_e^{(\text{in})} (\bp)\, \delta_{\lambda,\lambda_e} = \delta_{\lambda,\lambda_e} \left(\frac{2\sqrt{\pi}}{\sigma}\right)^{3/2} \cr
& \dst \times \exp\left\{-\frac{(\bp - \la\bp\ra)^2}{2\sigma^2} + i\varphi(\bp)\right\},
\label{WPm}
\eea

Note that Eq. \eqref{drW} does not depend on the phases both of the complex amplitude $M_{fi}$ and of the incoming electron wave packet, $\varphi(\bp)$. When the latter is a plane wave with a definite momentum $\la\bp\ra$ and the helicity $\lambda_e$, we have $f_e^{(\text{in})}(\bp,\lambda) \propto \delta_{\lambda\lambda_e}\,\delta(\bp - \la\bp\ra)$ and Eq.(\ref{drW}) reproduces the standard probability of the plane-wave approximation, 
\bea
\int d^3\br\,\mathcal W(\br, \bk, t) \propto \left|S_{fi}^{(1)}(\la\bp\ra, \lambda_e, \bk,\lambda_\gamma)\right|^2.
\eea

\section{Near-field approximation: ``snapshots'' of the electron wave packet}
\label{sec:Near-field}
In the simplest approximation, we completely neglect ${\tilde \bk}$ everywhere in Eq.(\ref{WEE}) except for the space-time dependent phase and the electron function $f_e^{(\text{in})}$ itself. This physically means neglecting field dynamics, that is, spreading and interference of the partial waves (see Eqs. \eqref{NFA_E}). Then, as shown in Eq. \eqref{WF_NFA}, the following scalar combination appears in Eq. \eqref{WEE}
\bea
& \dst w_e^{(\text{in})}(\br, \bp, t) = \int\bar{d}^3 {\tilde \bk}\,f_e^{(\text{in})}(\bp + {\tilde \bk}/2)\left(f_e^{(\text{in})}(\bp - {\tilde \bk}/2)\right)^*\cr
& \dst \times \e^{-it(\varepsilon(\bp + {\tilde \bk}/2) - \varepsilon(\bp - {\tilde \bk}/2)) + i\br\cdot{{\tilde \bk}}},
\label{WWd3k}
\eea
which is the Wigner function of an incoming electron packet (see details in \cite{PRD}). As a result, we obtain the following compact expression:
\bea
& \dst \mathcal{W}(\br,\bk,t) \approx 4\pi T\,\sum\limits_{\lambda_\gamma}\int d\Gamma\,w_e^{(\text{in})}(\br, \bp, t)\cr 
& \dst \times\left[1 + n^2(\omega(\bk))\right] \, \frac{\left|M_{fi}(\bp, \bk,\lambda_e,\lambda_\gamma)\right|^2}{4n^4(\omega(\bk)) 2\varepsilon'2\varepsilon(\bp)},
\label{E2}
\eea
where $d\Gamma = (2\pi)^4\,\delta^{(4)}(p - p' - k)\,\bar{d}^3\bp\bar{d}^3\bk$.

Now, we recall that the Wigner function $w_e^{(\text{in})}$ is concentrated in the vicinity of $\bp \approx \la\bp\ra$, which allows us to take the rest of the functions at $\bp = \la\bp\ra$. Hence, the integral over the momentum of the Wigner functions yields spatial distribution of the electron probability density at the time instant $t$ (see \cite{PRD}),
\bea
\int \bar{d}^3\bp\,w_e^{(\text{in})}(\br, \bp, t) = \left|\psi_e^{(\text{in})}(\br, t)\right|^2,
\label{psixt}
\eea
and 
{\small\bea
& \dst \frac{\mathcal{W}(\br,\bk,t)}{T} \to 8\pi^2\,\delta(\varepsilon(\la\bp\ra) - \varepsilon' - \omega(\bk = \la\bp\ra - \bp')) \times \cr
& \dst \to \left|\psi_e^{(\text{in})}(\br, t)\right|^2\,\sum\limits_{\lambda_\gamma}\,\frac{\left|M_{fi}(\la\bp\ra, {\bk} = \la\bp\ra - \bp',\lambda_e,\lambda_\gamma)\right|^2}{\left(2 n^2\omega(\bk = \la\bp\ra - \bp')\right)^2 2\varepsilon'2\varepsilon(\la\bp\ra)}.
\label{E2Mono}
\eea}
or simply
\bea
& \dst \mathcal{W}(\br,\bk,t) \propto \left|\psi_e^{(\text{in})}(\br, t)\right|^2
\label{E2Monopr}
\eea
This way, the photon field represents \textit{a snapshot} of the electron wave function. However, the above simple and illustrative formula contains a singular function and it does not take dynamic effects into account. One can say that it has been obtained in the near-field approximation and when taking $\bp = \la\bp\ra$ we have loosely assumed that the electron Wigner function as a function of its momentum is sharper than the delta-function. The latter assumption can be lifted if either the final electron is {\it not} detected, which can be done by integrating over the momenta $\bp'$ (see Eq.(\ref{EdEInt})):
\bea
& \dst \sum\limits_{\lambda'}\int\bar{d}^3\bp'\mathcal{W}(\br,\bk,t) 
\propto \left|\psi_e^{(\text{in})}(\br, t)\right|^2,
\eea
or, alternatively, the final electron is detected as a wave packet in the generalized measurement scheme, which would imply an additional integration and replace the delta-function with a smooth function describing the electron detector. In any way, if the observation point $(\br, t)$ is very close to the radiating electron packet, the photon energy distribution is defined by the electron probability density $\left|\psi_e^{(\text{in})}({\br}, t)\right|^2$ in the same point.

\section{Paraxial Wigner function: WKB approach}
\label{sec:parax}

\subsection{Spreading time and the correlation radius}

Within the same model of the incoming electron packet (\ref{WPm}), let us now make a more accurate approach, the Wentzel-Kramers-Brillouin (WKB) approximation. This approach can also be called the paraxial approximation, which implies that the width of the imcoming electron packet is much less than the electron mass, $\sigma \ll m_e$. Here, the terms $\mathcal O\left(\bk\right)$ in the squared modulus of the amplitude are neglected, however those of $\mathcal O\left(\bk^2\right)$ remain kept in the phase of the exponent in Eq. \eqref{WEE}. Hence, the term in brackets and the amplitudes $M_{fi}$ from Eq. \eqref{WEE} can be simplified to Eqs. \eqref{eexpan} and \eqref{Mexpan}, respectively. The calculations are very similar to those of the quasi-classical approximation in relativistic quantum mechanics \cite{Bagrov} and there are regions in phase space -- analogous to the WKB turning points -- where this approximation fails to work. In reality, this region lies within the emission angles $\theta_k \ll \gamma^{-1}$, where $\gamma = \varepsilon/m_e$, which are of no practical interest for studies of ChR.

Let us suppose that the electron packet is a helicity eigenstate, $f_e^{(\text{in})}(\bp\pm\bk/2,\lambda) = f_e^{(\text{in})}(\bp\pm\bk/2)\, \delta_{\lambda\lambda_e}$, so that $\tilde{\lambda} = \lambda = \lambda_e$. Neglecting the linear correction in Eq. \eqref{eexpan} one ends up with the following master integral:
\bw
\bea
& \dst \mathcal W_p(\br, \bp, \bk, t) = (2\pi)^2\sqrt{4\pi}\left[1+n^2(\omega(\bk))\right] \sum\limits_{\lambda_\gamma}\frac{|M_{fi}(\bp,\lambda_e,\bk,\lambda_\gamma)|^2}{(2n^2(\omega(\bk)))^2 2\varepsilon' 2\varepsilon(\bp)}\, \int\bar{d}^3{\tilde \bk} \frac{dt'}{2\pi}\frac{d\tau}{2\pi}\, \left(f_e^{(\text{in})}(\bp-{\tilde \bk}/2)\right)^* f_e^{(\text{in})}(\bp+{\tilde \bk}/2) \cr
& \dst \times \exp\left\{it'(\varepsilon(\bp) - \varepsilon' - \omega(\bk)) + i{\tilde \bk}\cdot(\br - {\bm u}_p t + (\partial_{\bp} + \partial_{\bk})\zeta_{fi}) + i{\tilde \bk}\cdot \tau ({\bm u}_p - {\bm u}_k) + it' \frac{1}{2}\frac{{\tilde k}_i}{2}\frac{{\tilde k}_j}{2} \left(\partial^2_{ij}\varepsilon - \partial^2_{ij}\omega\right)\right\},
\label{MI}
\eea
\ew
where the approximate equality ${\tilde \bk}\cdot {\bm u}_p \approx {\tilde \bk}\cdot {\bm u}_k$ is taken into account, which takes place within the paraxial approximation. Here
\bea
& \dst 
{\bm u}_p = \frac{\partial \varepsilon(\bp)}{\partial\bp} = \fr{\bp}{\varepsilon(\bp)},\, {\varepsilon(\bp)} = \sqrt{m_e^2+\bp^2},\ {\bm u}_k = \frac{\partial \omega(\bk)}{\partial\bk},\cr
& \dst \partial^2_{ij}\varepsilon(\bp) \equiv \frac{\partial^2\varepsilon(\bp)}{\partial p_i\partial p_j} = \frac{1}{\varepsilon(\bp)} (\delta_{ij} - ({\bm u}_p)_i({\bm u}_p)_j).
\label{upuk}
\eea
Let us first treat a medium with {\it weak dispersion}, for which 
\bea
\frac{\omega}{n(\omega)}\frac{dn(\omega)}{d\omega} \ll 1,
\label{Disp_cond}
\eea
whereas we postpone the investigation of an arbitrary dispersion for Sec. \ref{ArbDisp}. The scenario of Eq. \eqref{Disp_cond} gives the following expressions for ${\bm u}_k$ and its derivative (recall that $\bk^2 = n^2(\omega) \omega^2$):
\bea
& \dst {\bm u}_k =\frac{\partial\omega}{\partial\bk} \approx \frac{\bk}{n^2 \omega} = \frac{\bk/|\bk|}{n}.\cr \label{uk_nodisp}
& \dst \partial^2_{ij}\omega \equiv \frac{\partial^2\omega}{\partial k_i\partial k_j} \approx \frac{1}{n^2 \omega} \left(\delta_{ij} - \frac{k_ik_j}{\bk^2}\right).
\eea
The photon group velocity in the medium, $|{\bm u}_k| = 1/n < 1$ as $n>1$. Taking the incoming packet from Eq. \eqref{WPm}, and expanding its complex phase in series over $\tilde\bk$ as follows 
\bea
& \dst f_e^{(\text{in})}(\bp + {\tilde \bk}/2,\lambda)\left(f_e^{(\text{in})}(\bp - {\tilde \bk}/2,\tilde{\lambda})\right)^* = \cr
& \dst = \delta_{\lambda,\lambda_e}\delta_{\tilde{\lambda},\lambda_e} \left(\frac{2\sqrt{\pi}}{\sigma}\right)^{3} \exp\left\{-\frac{(\bp - \la\bp\ra)^2}{\sigma^2} -\left(\frac{{\tilde \bk}}{2\sigma}\right)^2\right\} \cr
& \dst \times \exp\left\{ i{\tilde \bk}\cdot\frac{\partial \varphi(\bp)}{\partial \bp} + \mathcal O ({\tilde \bk}^3)\right\},
\eea
the Wigner function reads
\bw
\bea
& \dst \mathcal W_p(\br, \bp, \bk, t) = (2\pi)^2 \sqrt{4\pi}\left[1+n^2(\omega(\bk))\right]\sum\limits_{\lambda_\gamma} \left(\frac{2\sqrt{\pi}}{\sigma}\right)^3 \frac{|M_{fi}(\bp,\lambda_e,\bk,\lambda_\gamma)|^2}{(2n^2(\omega(\bk)))^2 2\varepsilon' 2\varepsilon(\bp)}\,\exp\left\{-\frac{(\bp-\la\bp\ra)^2}{\sigma^2}\right\} \cr
& \dst\times \int\bar{d}^3{\tilde \bk}\frac{dt'}{2\pi}\frac{d\tau}{2\pi}\, \exp\left\{it'(\varepsilon(\bp) - \varepsilon' - \omega(\bk)) -\mathcal {\bm A}\cdot {\tilde \bk} - \frac{1}{2}{\tilde k}_i{\tilde k}_j B_{ij}\right\},\cr
\label{MII}
\eea
\bea
& \dst \mathcal {\bm A}(t,\tau) = -i \Big(\br - {\bm u}_p t + (\partial_{\bp} + \partial_{\bk})\zeta_{fi} + \partial_{\bp}\varphi + \tau ({\bm u}_p - {\bm u}_k)\Big),\cr
& \dst B_{ij}(t') = \delta_{ij}\, \frac{1}{2\sigma^2} - \frac{it'}{4}\left(\partial^2_{ij}\varepsilon - \partial^2_{ij}\omega\right) \approx
\delta_{ij} \Big(\frac{1}{2\sigma^2} + \frac{it'}{4}\Big(\frac{1}{\omega n^2} - \frac{1}{\varepsilon}\Big)\Big) + \frac{it'}{4}\Big(\frac{1}{\varepsilon} - \frac{1}{\omega}\Big)({\bm u}_p)_i({\bm u}_p)_j.
\label{A_B}
\eea
\ew
Taking the integral over $\tau$ in Eq. \eqref{MII} we obtain a Gouy phase $g(t')$ of the photon field in its formation zone (for details see \ref{int_tau}). This phase arises when the emitting particle represents a wave packet (see, e.g., \cite{Karlovets2021}). Note that it depends neither on the electron phase $\varphi(\bp)$ nor on that of the amplitude, $\zeta_{fi}$. For Cherenkov emission it can be represented in terms of two partial Gouy phases $g_1(t')$ and $g_2(t')$
\bea
g(t') = g_1(t') + g_2(t') = \arctan\fr{t'}{t_d} + \arctan\fr{t'}{\tilde t_d},
\label{gouy_phase}
\eea
where
\bea
& \dst t_d = \frac{2}{\sigma^2}\frac{({\bm u}_p - {\bm u}_k)^2}{\left(\frac{1}{\omega n^2}- \frac{1}{\varepsilon}\right) ({\bm u}_p - {\bm u}_k)^2 + \left(\frac{1}{\varepsilon}- \frac{1}{\omega}\right) \left[{\bm u}_p\times {\bm u}_k\right]^2},\cr
& \dst \tilde{t}_d = t_d\Big|_{{\bm u}_p || {\bm u}_k} = \fr{2}{\sigma^2}\fr{\omega n^2}{1-n^2\omega/\varepsilon}.
\label{gtp}
\eea
Here, $t_d$ is the \textit{diffraction (spreading) time}, defining the space-time correlation of the photon detection (see below) and $\tilde t_d$ is the special case for $t_d$, when ${\bm u}_p$ and ${\bm u}_k$ are collinear. We will hereafter comment on the so-called turning points where the denominator vanishes and the WKB approximation fails.

\begin{figure*}\center
\includegraphics[scale=0.8]{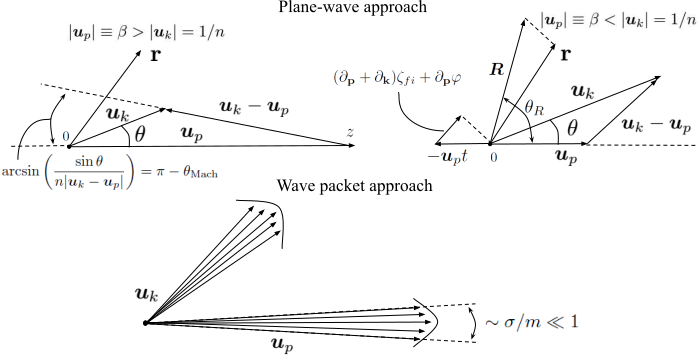}
    \caption{Generation of Cherenkov radiation within the plane-wave approach when the Cherenkov condition, $|{\bm u}_p| > |{\bm u}_k|=1/n$, is met (left upper panel) and when it is not (right upper panel). The photon field is {\it not} vanishing within the formation zone even for $|\bm{u}_p|<|\bm{u}_k|$. The lower panel: the wavepacket approach, where the real electron packet has a momentum distribution with a width $\sigma \ll m_e$ according to Eq.(\ref{WPm}). The Mach angle is defined in Eq.(\ref{Machth}). The vector $\br$ coincides with ${\bm R} = \br - {\bm u}_p t + (\partial_{\bp} + \partial_{\bk})\zeta_{fi} + \partial_{\bp}\varphi$ from Eq.(\ref{Rr}) only for the Gaussian electron packet with $\varphi=0$ at $t=0$ and neglecting the phase $\zeta_{fi} = \text{arg}\,M_{fi}$ of the amplitude, and the difference ${\bm R} - \br$ is due to the distribution of an electric dipole moment induced in the medium.}
    \label{FigGen}
\end{figure*}
Taking the symmetries of all the functions of $t'$ in Eqs. \eqref{MII} and \eqref{A_B} into account, one can write the master integral over $t'$ of the Wigner function as follows
\bw
\bea
\label{Wigner}
& \dst \mathcal W_p(\br, \bp, \bk, t) = \left(\frac{2\sqrt{\pi}}{\sigma}\right)^3 \,\frac{\sqrt{4\pi}}{(2n^2(\bk))^2 2\varepsilon' 2\varepsilon(\bp' + \bk)}\left(1+n^2(\omega(\bk))\right)\,\sum\limits_{\lambda_\gamma} |M_{fi}(\bp' + \bk,\lambda_e,\bk,\lambda_\gamma)|^2 \times \cr
& \dst \exp\left\{- \frac{(\bp' + \bk - \la\bp\ra)^2}{\sigma^2}\right\}\int\limits_{-t_{\rm out}}^{t_{\rm out}}dt'\,\fr{1}{G(t')}\,e^{-R^2/R_{\text{eff}}^2(t')}\exp\left(it'(\varepsilon(\bp) - \varepsilon' - \omega(\bk)) -i\frac{1}{2}\, g(t') + i\frac{R^2}{R_{\text{Im}}^2(t')}\right) \propto \cr
& \dst 2\int\limits_0^{t_{\rm out}}dt'\,\fr{1}{G(t')}\,e^{-R^2/R_{\text{eff}}^2(t')}\cos\left(t'(\varepsilon(\bp) - \varepsilon' - \omega(\bk)) - \frac{1}{2}\, g(t') + \frac{R^2}{R_{\text{Im}}^2(t')}\right)
\label{tpr}
\eea
\ew
where $G(t')$ and $R_{\rm Im}(t')$ are defined in Eqs. \eqref{G2} and \eqref{explhs}, respectively. Here, we changed the infinite limits of the integral over $t'$ to the finite: $\{-t_{\rm out},\,t_{\rm out}\}$ (see for details Appendix \ref{NFA}). This time interval can be interpreted as the window, during which the radiation is generated. Hence, the value $t_{\rm out}$ should satisfy the following condition:
\bea
    t_{\rm out} \lesssim L_\text{bulk}/c
\eea
where $L_{\rm bulk}$ is the length of a medium bulk, which the electron crosses during its time of flight. Here, 
\bea
& \dst {\bm R} = \br - {\bm u}_p t + (\partial_{\bp} + \partial_{\bk})\zeta_{fi}(\bp, \lambda_e, \bk, \lambda_\gamma) + \partial_{\bp}\varphi(\bp) \equiv \cr
& \dst \equiv \{X,Y,Z\} = R\{\sin\theta_R\cos\phi_R, \sin\theta_R\sin\phi_R, \cos\theta_R\},
\label{Rr}
\eea
It can also be noticed that the master integral in Eq. \eqref{tpr} and, therefore, the Wigner function itself are real but \textit{not everywhere positive}.

The integral in Eq. \eqref{tpr} defines the Wigner function of the photon field according to Eq. \eqref{WEE}. The integration over time $t'$ is somewhat similar to that in the retarded potentials, in which the field at the time instant $t$ is defined by a source taken at the retarded moment of time.

The quantity $R_{\text{eff}}(t')$ in the exponent of Eq. \eqref{tpr} is proportional to the size of the electron packet $\sigma_{\perp}^2$, as seen in Eq. \eqref{explhs}, but it is also anisotropic. At $t' \gg t_d$, it can be approximated as
\bea
R_{\text{eff}}^2(t') \propto \sigma^2_{\perp}(0)\, t'^2/t_d^2,
\eea
and therefore at $t'/t_d \to \infty$ we have $R_{\text{eff}} \to \infty$. When $R_{\text{eff}}(t') \gg R$, dependence of the Wigner function and of the photon energy density on the detection coordinates $\br$ and time $t$ vanishes, whereas in the other limit, $R \gg R_{\text{eff}}(t')$, the Wigner function is exponentially suppressed. Thus, an effective region of the time integral (\ref{tpr}) for which the space-time correlation effects exist is limited from above, $t' \lesssim t_d$, and it is the region of coordinates $R \sim R_{\text{eff}}(t')$ where the correlation is most pronounced. At $t' \gg t_d$ the photon field spreads such that there is no longer space-time correlation within the region $R < R_{\text{eff}}(t')$. Thus, $R_{\text{eff}}(t')$ can be called {\it the correlation radius}, which can be seen as the distance from the emitting electron, over which the radiation spreads during its formation.

\begin{figure*}\center
\includegraphics[width=1\textwidth] {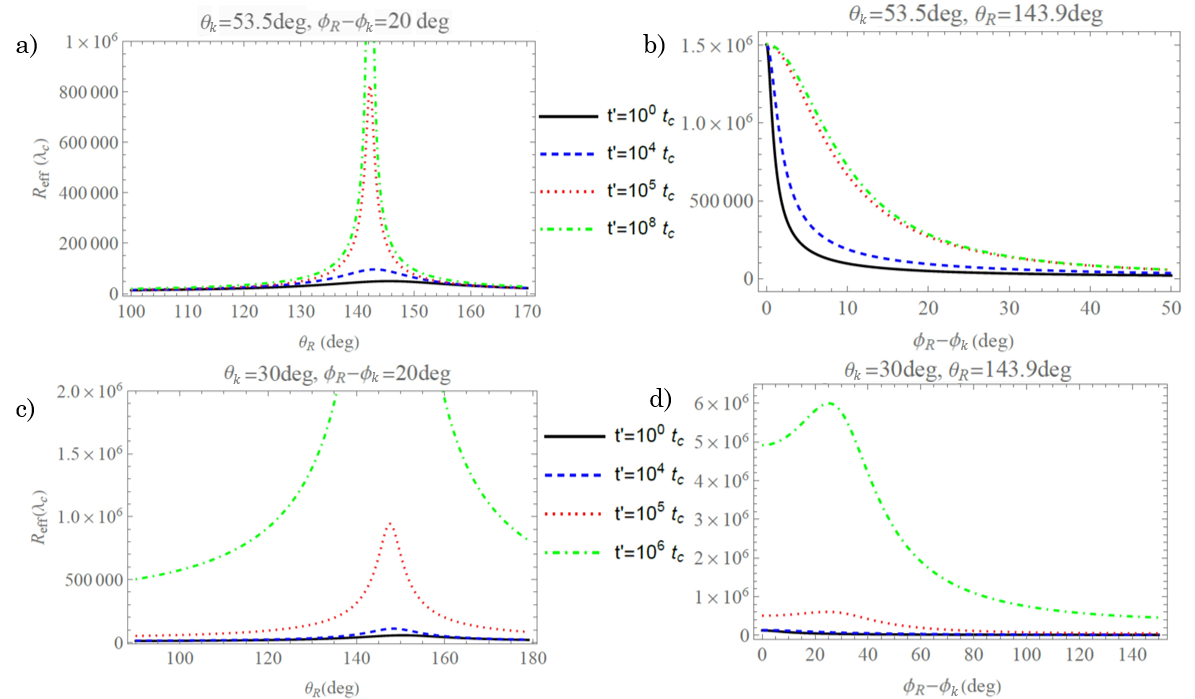}
    \caption{The effective correlation radius, $R_{\rm eff}$, from Eq.(\ref{tpr}) for $\beta = 0.99\,(\gamma \approx 7)$, $n=1.7$, $\omega = 10^{-5}m_e$, $\sigma = 10^{-4}m_e$, the corresponding Cherenkov angle is $\theta_{\text{Ch.cl.}} = \arccos 1/\beta n \approx 53.5^\circ$, the Mach angle is $\theta_{\text{Mach}} \approx 143.9^\circ$ (see Eq.(\ref{Machth})). When $R_{\text{eff}}(t') \to \infty$ at $t' \gg t_d$, the space-time dependence of the Wigner function vanishes, whereas in the other limit, $R \gg R_{\text{eff}}(t')$, the Wigner function is exponentially suppressed. The upper panels, (a) and (b), are for $\theta_k = \theta_{\text{Ch.cl.}}\approx 53.5^\circ$, the lower panels, (c) and (d) - $\theta_k = 30^\circ$. The left panels, (a) and (c), show the dependence of $R_{\rm eff}$ on the polar angle $\theta_R$ of ${\bm R}$ with the fixed difference in azimuthal angles $\phi_R - \phi_k = 20^\circ$. The right panels, (b) and (d), in turn, are for the fixed $\theta_R = 143.9^\circ$ showing the dependence on $\phi_R - \phi_k$. Clearly, when the photon is emitted {\it not} at the Cherenkov angle, the correlation radius is still concentrated in a vicinity of the Mach cone, however it grows much faster with time $t'$ at the angles different from $\theta_{\text{Mach}}$. Here, $\lambda_c = \hbar/m_ec \approx 3.9\cdot 10^{-11}$ cm - the electron Compton wavelength, and $t_c = \lambda_c/c \approx 1.3\cdot 10^{-21}$ s - electron Compton time. Note that $10^6\,t_c \sim 1$ fs, $10^6\,\lambda_c \sim 0.38\,\mu$m.}
    \label{FigReff}
\end{figure*}

\begin{figure*}\center
\includegraphics[width=1\textwidth] {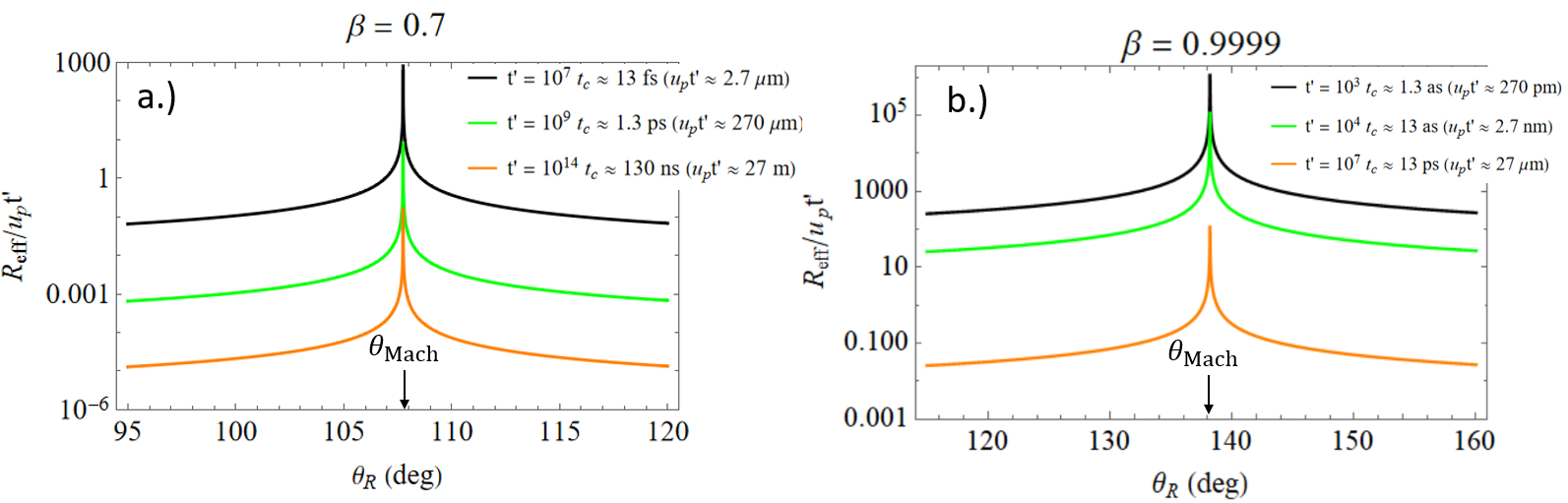}
    \caption{The effective correlation radius from Eq.(\ref{tpr}) divided by the distance $u_p t'$. Parameters: $n = 1.5$, $\theta_k = \theta_{\rm Ch.cl.}$. \textbf{(a)} $\beta = 0.7\,(\gamma = 1.4),\, \theta_{\text{Ch.cl.}} \approx 17.8^\circ$, $\theta_{\text{Mach}} \approx 107.8^\circ$. \textbf{(b)} $\beta = 0.9999\,(\gamma = 70.7),\, \theta_{\text{Ch.cl.}} \approx 48.2^\circ$, $\theta_{\text{Mach}} \approx 138.2^\circ$. Nearby the Mach angle $\theta_{\text{Mach}}$, the space-time dependence of the photon energy density quickly vanishes within the correlation radius $R < R_{\text{eff}}(t')$. For macroscopic targets of the length $u_p t' > 1$ mm, the correlation radius is always much smaller than the length except for the vicinity of the Mach angle where $R_{\text{eff}} \lesssim u_p t'$. An observer placed at an edge of a target at the distance $R \sim u_p t'$ would only see the photon field at $\theta_R \approx \theta_{\text{Mach}}$, otherwise the Wigner function is exponentially suppressed, see Eq.(\ref{tpr}).
    }
    \label{FigReffLog}
\end{figure*}

When the condition of Cherenkov radiation is met
\bea
u_p > u_k,
\eea
the vector ${\bm u}_k - {\bm u}_p$ is directed backwards with respect to the electron velocity ${\bm u}_p$, thus, the dependence of the Wigner function on ${\bm R}$ vanishes along the vector ${\bm u}_k - {\bm u}_p$, defining {\it the Mach cone} with the opening angle (see Fig.\ref{FigGen})
\bea
\theta_{\text{Mach}} = \pi - \arcsin\left(\fr{\sin\theta}{n|{\bm u}_k - {\bm u}_p|}\right).
\label{Machth}
\eea
As the final electron is supposed to be detected in a plane-wave state, scattered at a certain azimuthal angle, the Wigner function itself and the correlation radius $R_{\text{eff}}(t')$ depend on the difference $\phi_R - \phi_k$ between the azimuthal angle of the vector ${\bm R}$ and that of the photon wave vector $\bk$, see Figs. \ref{FigReff},\ref{FigReffLog}. This means that the azimuthal symmetry of the photon energy distribution \eqref{SED_finite} can be restored by integrating its Wigner function over $d^3\bk$. As seen in Fig.\ref{FigReffLog}, this radius as a function of time $t'$ is orders of magnitude smaller than the distance traveled by the electron during that time, $u_p t'$, everywhere except for the Mach angle, $\theta_R \approx \theta_{\text{Mach}}$. So, taking a target of the length $u_p t'$ and measuring the field at its edge, i.e. at the distance $R \sim u_p t'$, we would find that the photon field is {\it not} significantly suppressed only in the vicinity of the Mach angle.

\begin{figure*}\center
\includegraphics[width=1\textwidth] {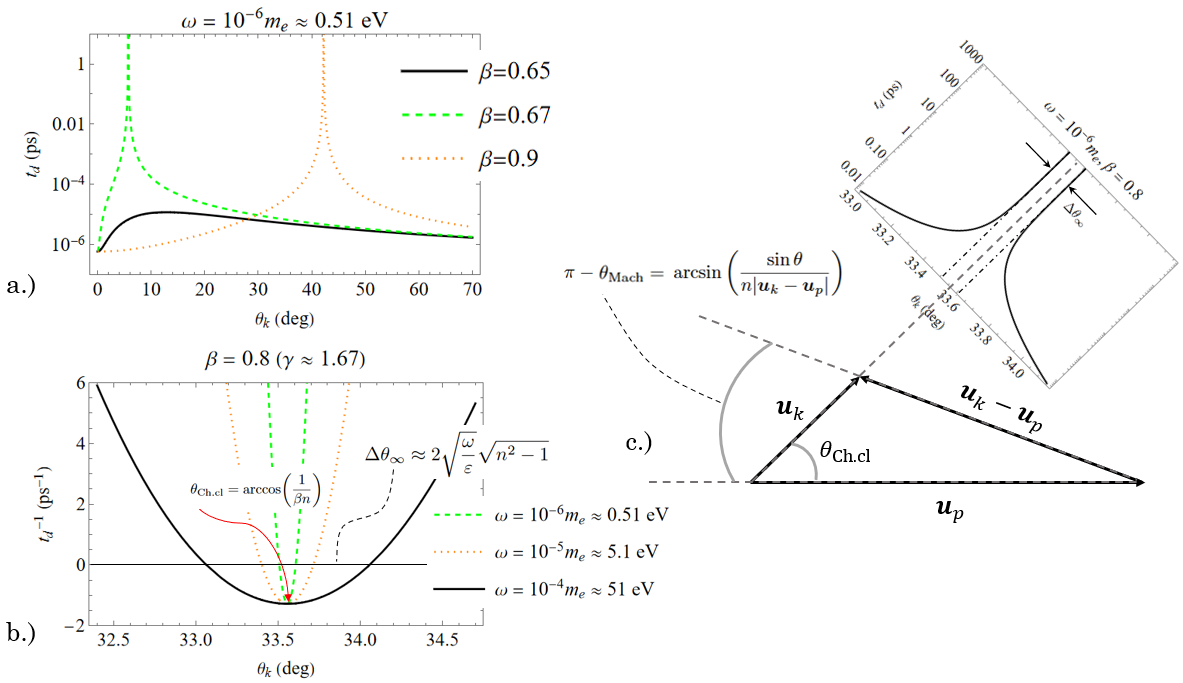}
    \caption{The spreading time $t_d$ from Eq. \eqref{gtp} of the photon field measured in picoseconds (a) and its inverse (b). The sharp maxima are nearby the Cherenkov angle $\theta_{\text{Ch.cl.}} = \arccos(1/u_p n)$, which is illustrated in panel (c). The Cherenkov condition is {\it not} met for the black line in the panel (a), which is why the field quickly spreads during several attoseconds. The panel (b) shows two singular points \eqref{thetainf}, where $t_d^{-1} = 0$. The width between these points is connected to the quantum recoil ($\omega/\varepsilon \ne 0$, see Eq.\eqref{Dth}), where the spreading time turns negative. Notice also, that the extremum of the spreading time lies precisely at the Cherenkov angle}
    \label{Figtd}
\end{figure*}

The spreading time $t_d$ from Eq.(\ref{gtp}) has an extremum either at the classical Cherenkov angle
\bea
\cos\theta = \fr{1}{u_p n} \le 1,
\eea
when the Cherenkov emission condition $u_p \ge u_k$ is met,
or at
\bea
\cos\theta = u_p n < 1,
\eea
when this condition is {\it not} met and $u_p < u_k$ (see Fig.\ref{Figtd}a). Importantly, photon emission takes place in the intermediate zone even if the Cherenkov condition is not fulfilled (see the right panel in Fig.\ref{FigGen}), but the emitted waves do not constructively interfere to form a cone at a certain angle in the far field. 

Along with the extremum, the diffraction time $t_d$ has {\it two singular points}, in which its denominator turns to zero (Fig.\ref{Figtd}b).
\bw
\bea
& \dst \cos\theta_{\infty} = \fr{1}{u_p n} \fr{1}{1-\omega/\varepsilon}\Bigg(1-n^2\fr{\omega}{\varepsilon} \mp \sqrt{\left(1-n^2\fr{\omega}{\varepsilon}\right)^2 - \left(1-\fr{\omega}{\varepsilon}\right)\left(1-n^2\fr{\omega}{\varepsilon}(1-u_p^2 + n^2u_p^2)\right)}\Bigg) \approx \cr
& \dst \approx \fr{1}{u_p n} \left(1 \mp \sqrt{\fr{\omega}{\varepsilon}}\sqrt{(n^2-1)(u_p^2n^2-1)}+\mathcal O\left(\fr{\omega}{\varepsilon}\right)\right).
\label{thetainf}
\eea \ew
Here, at the last step we have kept the first correction of the order of $\sqrt{\omega/\varepsilon}$, which is usually very small, $\omega/\varepsilon \ll 1$ (see details in \cite{Ivanov2016}). Clearly, the diffraction time can only turn to infinity under the condition of Cherenkov radiation. Within the same accuracy, the angular width between the two singular points is
\bea
\Delta \theta_{\infty} \approx 2\sqrt{\fr{\omega}{\varepsilon}}\sqrt{n^2-1} + \mathcal O\left(\left(\fr{\omega}{\varepsilon}\right)^{3/2}\right),
\label{Dth}
\eea
This width vanishes for classical Cherenkov radiation without the electron recoil, $\omega/\varepsilon \to 0$.

Between this singular points the diffraction time $t_d$ becomes \textit{negative} -- see Fig.\ref{Figtd} -- as well as the shift of the Gouy phase $g_1 = \arctan t'/t_d$ from Eq. \eqref{gouy_phase} during the evolution with respect to the time $t'$. In the classical regime, both singular points merge and $t_d$ turns to infinity at the single point $\cos\theta = 1/u_p n$. Indeed, in a vicinity of the classical Cherenkov condition the spreading time becomes negative for $n>1$:
\bea
& \dst t_d\Big|_{\cos\theta = \frac{1}{n u_p}} = -\frac{2\varepsilon}{\sigma^2}\frac{n^2}{n^2-1} = -2\frac{m_e^2}{\sigma^2}\, \gamma\,t_c\, \frac{n^2}{n^2-1} < 0,\cr
& \dst \sigma_t^2\Big|_{\cos\theta = \frac{1}{n u_p}} = \fr{n^2}{2}\sigma_{\perp}^2(t'),
\label{tdCh}
\eea
where 
\[
\gamma = \varepsilon/m_e \quad \text{and}\quad t_c \approx 1.3\times 10^{-21}\, \text{s}
\]
is the electron Compton time.

To understand this result better, let us compare this spreading time of the emitted photon field with that of a freely propagating Gaussian electron packet at rest in vacuum, which is \cite{Karlovets2021}
\bea
t_d^{(\text{e, rest})} = \frac{m_e}{\sigma^2} = \frac{m_e^2}{\sigma^2}\,t_c.
\eea
In the laboratory frame this time interval is $\gamma$ times larger, which coincides with $|t_d|$ up to the factor $2n^2/(n^2-1) > 2$, the latter being due to the medium. So, in a vicinity of the Cherenkov direction the spreading of the photon field in medium {\it still goes} but only due to spreading of the electron packet itself. 

If we initially neglect the quantum recoil, $\omega/\varepsilon \to 0$, the spreading time $t_d$ 
\bea
\dst t_d\Big|_{\frac{\omega}{\varepsilon} \to 0} = \fr{2}{\sigma^2}\fr{\omega n^2}{u_p^2}\fr{n^{-2} + u_p^2 -2u_p\cos\theta/n}{(\cos\theta-1/u_pn)^2}\to
\infty,
\label{tdcl}
\eea
when $\cos\theta \to \dst \frac{1}{u_p n}$, i.e. it becomes infinite at the Cherenkov angle and the two singular points merge together. Thus, spreading of the photon field, of the electron packet itself, and the quantum recoil during Cherenkov emission are intimately connected. In other words, constructive interference of the partial waves results in the slowest spreading rate around the Cherenkov cone (see Fig. \ref{Figtd}a). 

The spatial size of the electron packets for standard sources like cathodes of the electron guns in accelerators or electron microscopes nearby the generation region usually amounts to \cite{Ehberger, Cho, Latychevskaia, Cho_Oshima, Karlovets2021}
\bea
\sigma_{\perp}(0) \sim 1-10\,\text{nm}.
\eea
These estimates can likewise be obtained from the emission duration of non-relativistic photo-electrons from a tungsten tip \cite{Dienstbier} for which the measured sub-femtosecond duration yields nanometer-sized electron packets. 

For $\gamma\gtrsim 1-2, n \gtrsim 1$ the electron spreading time in the laboratory frame is of the order of
\bea
t_d^{(e)} \sim 10\,\text{as} - 10\, \text{ps},
\eea
whereas the time of flight of an electron through a target of 1 cm length is $\sim 100\,\text{ps}$, which is larger than the spreading time of an incident electron, $t_d^{(e)}$. Therefore -- except for very thin layers and ultrarelativistic electrons with $\gamma \gg 1$ -- spreading of an electron packet during the photon emission {\it can} be important even nearby the Cherenkov angle, especially for large Cherenkov light generators employed, for instance, in neutrino telescopes.


\subsection{Flash duration and the shift of the photon arrival time}
\label{sec:FDaSh}

Turning to momentum distribution of the paraxial Wigner function, we note that whereas the momentum conservation law takes place for every partial wave, $\bp = \bp' + \bk$, where $|\bk| = n\, \omega(\bk)$, there is {\it no} corresponding energy conservation, $\varepsilon(\bp) - \varepsilon' - \omega(\bk) \ne 0$, due to spreading and dependence of the integrand \eqref{MII} on $t'$. As a result, the well-known Cherenkov condition in the momentum space (see, e.g., \cite{Ivanov2016}) does \textit{not} hold within the formation zone,
\bea
\cos\theta \neq \cos\theta_{\text{Ch}} =  \frac{1}{\beta n} + \frac{\omega}{2\varepsilon}\,\frac{n^2-1}{\beta n}.
\label{Changle}
\eea
Likewise, there is \textit{no} sharp spectral cutoff, which for the plane-wave emitting electron is
\bea
\omega_{\text{cut-off}} = 2\varepsilon\,\fr{\beta n - 1}{n^2 - 1},
\eea
even when neglecting the dispersion, i.e., dependence of $n$ on $\omega$. One can see from Eq. \eqref{tpr} that the common features of the far-field Cherenkov radiation are regained when one neglects the dependence on time $t'$. This can be done when $\sigma \to 0$, i.e. the incoming electron represents a plane wave, because the Gouy phase $g(t')$ vanishes both when $t' \ll t_d$ and when $\sigma \ll m_e$, and the limit of the function $G(t')$ is the same for both these limits.

{Dependence of the photon Wigner function on the detection time $t$ is only due to the vector $\bm{R}$ (see Eq. \eqref{Rr}) in first term of the exponent in Eq. \eqref{int_tau}:
{\small\bea
& \dst -\frac{1}{2} \frac{[{\bm R}\times({\bm u}_p - {\bm u}_k)]^2}{\eta(t')({\bm u}_p - {\bm u}_k)^2 + \chi(t') \left[{\bm u}_p\times {\bm u}_k\right]^2} \cr
& \dst \propto -\frac{\left[{\bm u}_p\times {\bm u}_k\right]^2}{2} \frac{(t - t_0)^2}{\eta(t') ({\bm u}_p - {\bm u}_k)^2 + \chi(t') \left[{\bm u}_p\times {\bm u}_k\right]^2},
\label{tdep}
\eea}
while the second term is proportional to ${\bm u}_p\cdot[{\bm u}_p\times {\bm u}_k] = 0$.
The real part of Eq. \eqref{tdep} defines the Wigner function with respect to the detection time $t$ as follows
\bea
& \dst \text{Re} \left\{-\frac{\left[{\bm u}_p\times {\bm u}_k\right]^2}{2} \frac{(t - t_0)^2}{\eta({\bm u}_p - {\bm u}_k)^2 + \chi \left[{\bm u}_p\times {\bm u}_k\right]^2}\right\} = \cr
& \dst = -\frac{(t - t_0)^2}{2\sigma_t^2(t')}.
\label{t-t0}
\eea
where the center of the distribution 
\bea
& \dst t_0 \equiv t_0(\br) = \left(\br + (\partial_{\bp} + \partial_{\bk})\zeta_{fi} + \partial_{\bp}\varphi\right)\cdot{\bm l}_0,\cr
& \dst {\bm l}_0 = \frac{\left[({\bm u}_p - {\bm u}_k)\times\left[{\bm u}_k \times {\bm u}_p\right]\right]}{\left[{\bm u}_p\times {\bm u}_k\right]^2},
\label{l0}
\eea 
is the time at which the probability of the photon detection is maximized for the given detector coordinate $\br$, and the width
\bea
& \dst \sigma_t^2(t') = \frac{\sigma^2_{\perp}(t')}{2}\frac{({\bm u}_p - {\bm u}_k)^2}{\left[{\bm u}_p\times {\bm u}_k\right]^2},
\label{stp}
\eea
is \textit{the Cherenkov flash duration}. Here, $\sigma_\perp(t')$ 
\bea
\dst \sigma_\perp^2(t') = \sigma^{-2}\left(1 + (t'/t_d)^2\right)
\label{sigma_perp}
\eea
is the transverse size of the electron packet in a medium, and $t_d$ is the same as in Eq. \eqref{gtp}. 

The exponent \eqref{t-t0} also defines space-time correlation of the energy density and this correlation \textit{vanishes} when $\sigma_t^2 \to \infty$, i.e., either the electron packet is very wide ($\sigma \to 0$) or the spreading is essential ($t'\gg t_d$), which takes place in the far field.}


From the classical considerations, this duration was estimated by Frank in 1956 \cite{Frank,NIM} to be of the order of $1/\Delta \omega$, where $\Delta \omega$ is a frequency interval for which the Cherenkov effect takes place. The microscopical origin of this phenomenon is due to the AC Stark effect \cite{Delone}, in which the atoms of the weakly dispersive medium are off-resonantly polarized by a broadband spectrum of pseudo-photons \cite{Mono_DR}. Clearly, quantum estimates from the uncertainty principle for the off-resonant AC Stark effect yield the same result \cite{Delone}, which is
\bea
\fr{1}{\Delta \omega} \sim 0.1-100\, \text{fs}
\eea
for $\Delta \omega \sim (10^{-2}-10)$ eV (Frank's original estimate was $\Delta\omega^{-1} \leq 1$ ps \cite{Frank,NIM}). 

The flash duration reduces to
\bea
\sigma_t(t') = \frac{\sigma_{\perp}(t')}{\sqrt{2}}\sqrt{\frac{({\bm u}_p - {\bm u}_k)^2}{\left[{\bm u}_p\times {\bm u}_k\right]^2}} \to n\,\frac{\sigma_{\perp}(0)}{\sqrt{2}},
\label{sigma_t_Ch}
\eea
when $\dst \cos\theta_k \to u_k/u_p$ because $t_d \to \infty$ (see Fig.\ref{Figtd}). In other words, in a vicinity of the classical Cherenkov angle only the electron packet size contributes to the finite duration of the Cherenkov flash. For realistic electron packets with $\sigma = 1/\sigma_{\perp}(0) \sim \left(10^{-7} - 10^{-4}\right)\,m_e$, we have
\bea
\displaystyle & \sigma_t(0) \sim \left (10^{4} - 10^{7}\right)\,t_c \sim 10\,\text{as} - 10\,\text{fs}.
\label{durt0}
\eea

Hence, the exponent term 
\[
\exp(-\frac{R^2}{R_{\rm{eff}}^2(t')})
\]
of the master integral in Eq. \eqref{tpr} defines space-time correlation of the energy density and this correlation \textit{vanishes} when $\sigma_t^2 \to \infty$, i.e., either the electron packet is wide enough, $\sigma \ll |\bp - \la\bp\ra|$, or the spreading is essential, $t'\gg t_d$, which takes place in the far field.

Let us direct the $z$ axis along $\bp$, so that 
\bea
\bp = \{0,0,|\bp|\},
\eea
and ($u_p \equiv \beta = u_p/c$)
\bea
& \dst ({\bm u}_p - {\bm u}_k)^2 = n^{-2}+u_p^2-2u_p\cos\theta/n,\cr
& \dst \left[{\bm u}_p\times {\bm u}_k\right]^2 = u_p^2\sin^2\theta/n^2.
\eea
A turning point of $g_2(t')$ in Eq. \eqref{gouy_phase}, where the WKB approximation fails to work for the Wigner function, arises when $\left|{\bm u}_p - {\bm u}_k\right| \to 0$ or
\bea
\cos\theta_k \to \fr{1}{2}\left(\fr{1}{u_p n} + u_p n\right).
\eea
As the r.h.s. of the last expression is equal to or larger than 1, it means that $\cos\theta_k = 1$ and $u_p n = 1/u_pn = 1$. When the Cherenkov condition in met, then the radiation with $u_p n \to 1$ implies vanishingly small emission angle, $\cos\theta_{\text{Ch.cl.}} = 1/u_pn \to 1$. Due to the finite $\sigma$, transverse momentum spread of the initial electron, the turning point will {\it not} be strictly at $\theta_k = 0$ but at a finite angle $\theta_k \sim \sigma/m_e \ll 1$, which remains less than $10^{-3}$ rad for realistic accelerator parameters. Thus, the paraxial approximation is violated when the electron packet becomes non-paraxial with $\sigma/m_e \sim 1$, which implies focusing of the packet to the Compton wavelength, $1/\sigma \sim \lambda_c$.

The finite angular width (\ref{Dth}) of the Cherenkov cone along with the negative spreading time and, therefore, the \textit{negative Gouy phase shift} are purely quantum effects, enhanced for more energetic photons. For instance, X-ray Cherenkov emission is known to take place in a vicinity of atomic absorption edges where \cite{Bazylev, Uglov-2, Uglov-1, Konkov, Mono}
\bea
& \dst \text{Re}\left(n^2(\omega) - 1\right)>0, \cr
& \dst \text{Re}\left(n^2(\omega) - 1\right) \gg \text{Im}\left(n^2(\omega) - 1\right).
\eea
For experimentally relevant materials like Al, Si, Be, and Ti, Cherenkov radiation is observed at the L-edges of their absorption frequencies, which are $\omega \approx 72.5,\, 100,\, 110,\, 453.8$ eV, respectively, for the fixed initial electron energy $\varepsilon \approx 5.7$ MeV (see \cite{Bazylev, Uglov-1, Uglov-2, Konkov}). For nearly this electron energy it yields 
\bea
\Delta \theta_{\infty} < 1^\circ - 2^\circ.
\eea
So, the Cherenkov cone becomes of finite angular width due to quantum recoil, and this width grows larger for harder photons. The finite momentum uncertainty of the electron packet also broadens this cone, but at the much smaller scale, $\sim \sigma/m_e \lesssim 10^{-3}$ rad.

One can also define {\it the formation length} of Cherenkov radiation as following 
\bea
L_{f} = u_p t_d.
\eea
In the classical Tamm problem, Cherenkov energy emitted by a point-like electron in the distance of $L$ in the far field is \cite{Pafomov, Mono, Konkov}

\bea
& \dst \fr{d^2W}{d\omega d\Omega} \propto \fr{\sin^2\left(\fr{L}{2}\,\omega n (\cos\theta - 1/\beta n)\right)}{(\cos\theta - 1/\beta n)^2} \equiv \cr
& \dst \equiv \fr{\sin^2\left(L/L_{\text{cl}.}\right)}{(\cos\theta - 1/\beta n)^2},
\eea
and the classical formation length (also called coherence length by Frank in \cite{NIM})
\bea
L_{\text{cl}.} = \fr{2}{\omega n}\fr{1}{|\cos\theta -1/\beta n|}
\eea
turns to infinity at the Cherenkov angle, $\cos\theta \to 1/\beta n$. Within the current quantum model, this classical singular behavior is reproduced when the electron packet is a plane wave with $\sigma \to 0$ or when neglecting the recoil effects, $\omega/\varepsilon \to 0$, (see Eq.(\ref{tdcl})). Thus, at a finite distance from an electron packet the radiation formation length in quantum theory {\it stays finite} everywhere except for the above two points nearby the Cherenkov angle in which $t_d$ and $L_f$ diverge (see Fig. \ref{Figtd}).

Given that the photon group velocity is $u_k = 1/n$ in a weakly dispersive medium, a detector registers a plane-wave photon in the far field emitted at $t=0, r_0=0$ by a classical point-like electron at the time instant
\bea
t_{\text{cl.}} = \fr{r}{u_k} = r\,n,
\label{tcl}
\eea
which is hereafter called the classical arrival time. Here $r$ is the distance between the origin where the point-like electron was at $t=0$ and the far-field detector. Let us compare this simplest classical prediction with $t_0(\br)$ derived in the framework of quantum mechanics (Eq. \eqref{l0}), which means the time instant when the probability to register the photon is maximized. To this end, we again direct the $z$ axis along the initial electron momentum $\bp$,
\bea
& \dst \bp = \{0,0,|\bp|\},\ \bk = \bp-\bp'= n \omega {\bm l},\cr
& \dst {\bm l}=\{\sin\theta_k\cos\phi_k,\sin\theta_k \sin\phi_k,\cos\theta_k\},\cr
& \dst r \equiv \br\cdot{\bm l},\
t_{\text{cl.}} = \br\cdot{\bm l}\,n,
\label{meanm}
\eea
and find (recall Eq.(\ref{l0}))
\bea
& \dst {\bm l}_0 = \frac{1}{\sin^2\theta}\, \left(\left( \frac{1}{u_p} - n \cos{\theta} \right) \hat{\bm z} + \left(n - \frac{\cos\theta}{u_p}\right) {\bm l}\right).
\eea
In a vicinity of the classical Cherenkov angle, $\cos\theta \to \cos\theta_{\text{Ch.cl.}} = 1/(u_p n)$, one comes to
\bea
{\bm l}_0 \to {\bm l}\,n,
\eea
and so for $\zeta_{fi} = \varphi = 0$
\bea
t_0 \to t_{\text{cl.}} = \br\cdot{\bm l}\,n,
\eea
which is in agreement with Eq.(\ref{tcl}). As was mentioned before, the phase $\zeta_{fi} = \text{arg}\,M_{fi}$ is {\it not} vanishing even at the tree-level (see the Appendix \ref{Amp}). Therefore, in a vicinity of the same classical Cherenkov condition one obtains
\bea
t_0 \to t_{\text{cl.}} + \Delta t,
\eea
where 
\bea
\Delta t = n\,{\bm l}\cdot \left((\partial_{\bp} + \partial_{\bk})\zeta_{fi} + \partial_{\bp}\varphi\right)
\label{Dt}
\eea
is a quantum temporal shift. For arbitrary emission angles we find (see Eq.\eqref{l0})
\bea
\Delta t = t_0 - \br\cdot{\bm l}_0 = {\bm l}_0\cdot \left((\partial_{\bp} + \partial_{\bk})\zeta_{fi} + \partial_{\bp}\varphi\right).
\label{Dtg}
\eea

One can explain physical origin of this shift as follows: the second term $-\partial_{\bp}\varphi$ represents an electric dipole moment volume density in momentum representation of the electron packet due to its non-Gaussianity (its intrinsic mean value vanishes) \cite{PRA2019}, which is why the first term $(\partial_{\bp} + \partial_{\bk})\zeta_{fi}$ can be attributed -- up to the sign -- to the dipole moment density \textit{induced in a medium} by the electric field of the moving electron or, speaking in quantum terms, by its virtual photon. Clearly, the latter dipole distribution can occupy a much larger volume than the former. 

Microscopically, we deal with the mentioned above AC Stark effect \cite{Delone} with the atoms of a weakly dispersive meduim being off-resonantly polarized by a broadband spectrum of pseudo-photons \cite{Mono_DR}. Analogously to time delays observed when a laser wave packet travels in a dielectric medium \cite{Jordan,Sommer,Negtime}, here we encounter similar delays induced by the virtual photons that are reemitted by atoms as the photons of Cherenkov emission.

Classically, one can look at this shift as if a monochromatic photon was emitted not from a point-like electron on the $z$ axis, but from a point shifted laterally to the distance
\bea
\Delta \rho \sim \beta \gamma \lambda/2\pi
\label{emit_size}
\eea
from the particle trajectory (see Fig. \ref{IllustCher}), which is an effective mean free path of the virtual photon with the frequency $\omega$ \cite{Jackson,Mono_DR}. That is why
\bea
\Delta t_{\text{cl.}} \sim \beta \gamma \lambda/2\pi c = \beta\gamma/\omega,
\eea
which yields the rough classical estimate of 
\bea
\Delta t_{\text{cl.}} \sim (10^6-10^8)\,t_c \sim 1\,\text{fs}-100\,\text{fs}.
\label{Dtcl}
\eea
for photons from IR to UV ranges, $\omega \sim (10^{-7}-10^{-5})\, m_e$, and $\gamma = \varepsilon/m_e \sim 10$.

\begin{figure*}\center
\includegraphics[width=0.85\textwidth] {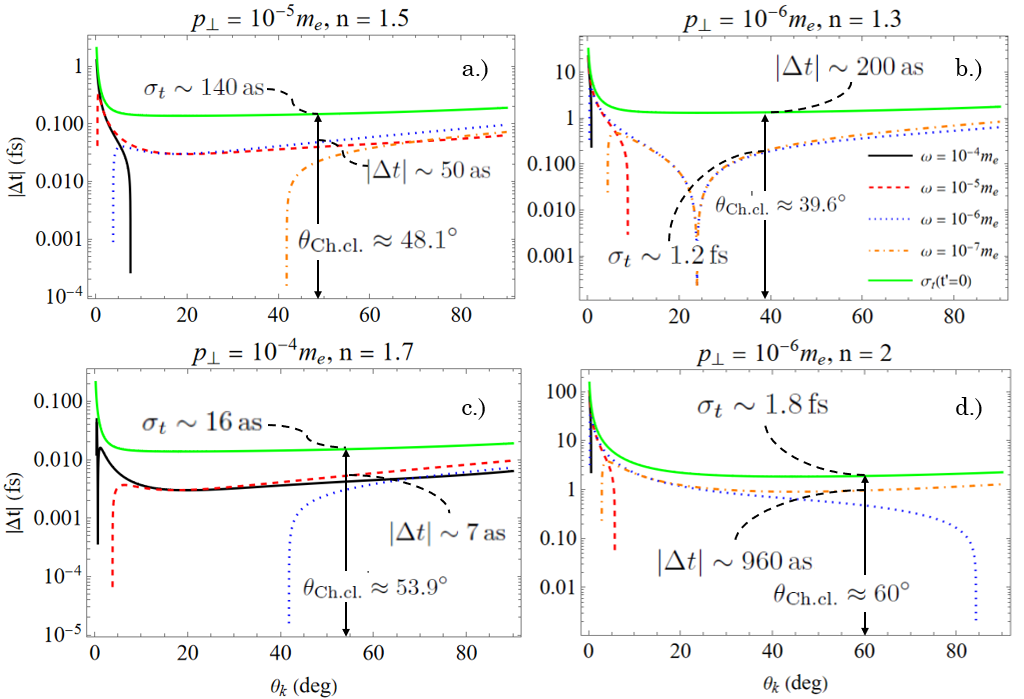}
    \caption{The temporal shift $\Delta t = t_0 - \br\cdot{\bm l}_0 = {\bm l}_0\cdot(\partial_{\bp} + \partial_{\bk})\zeta_{fi}$ from Eq.\eqref{Dtg} of the photon arrival time. An incident electron packet is Gaussian, which implies $\varphi=0$. The shift is measured in the electron Compton times, $t_c = \lambda_c/c \approx 1.3\cdot 10^{-21}\, $s. The parameters: $\beta = 0.999\, (\gamma \approx 22), p'_{\perp} = 0.99\, p_{\perp}, p'_z = 0.99\, \beta m_e$. \textbf{(a)} $p_{\perp} = 10^{-5}\, m_e,\, n=1.5$. \textbf{(b)} $p_{\perp} = 10^{-6}\, m_e,\, n=1.3$. \textbf{(c)} $p_{\perp} = 10^{-4}\, m_e,\, n=1.7$ (no emission for $\omega = 10^{-7}\,m_e$). \textbf{(d)} $p_{\perp} = 10^{-6}\, m_e,\, n=2$. The green line represents the flash duration, $\sigma_t(t')$ from Eq. \eqref{stp} at $t'=0$ with $\sigma = p_{\perp}$. The signs of the helicities can influence these results only at the very small angles, which are of no practical interest because the paraxial approximation may break down there. The typical shifts are of the order of $(10^4-10^6)\,t_c \sim 10^{-2} -1$ fs for large emission angles, which correspond to the distances $u_p\,(10^4-10^6)\,t_c \sim (10^4-10^6)\,\lambda_c \sim 1 - 100\,\text{nm}$.
    }
    \label{FigDt}
\end{figure*}

\begin{figure*}\center
\includegraphics[scale=0.55] {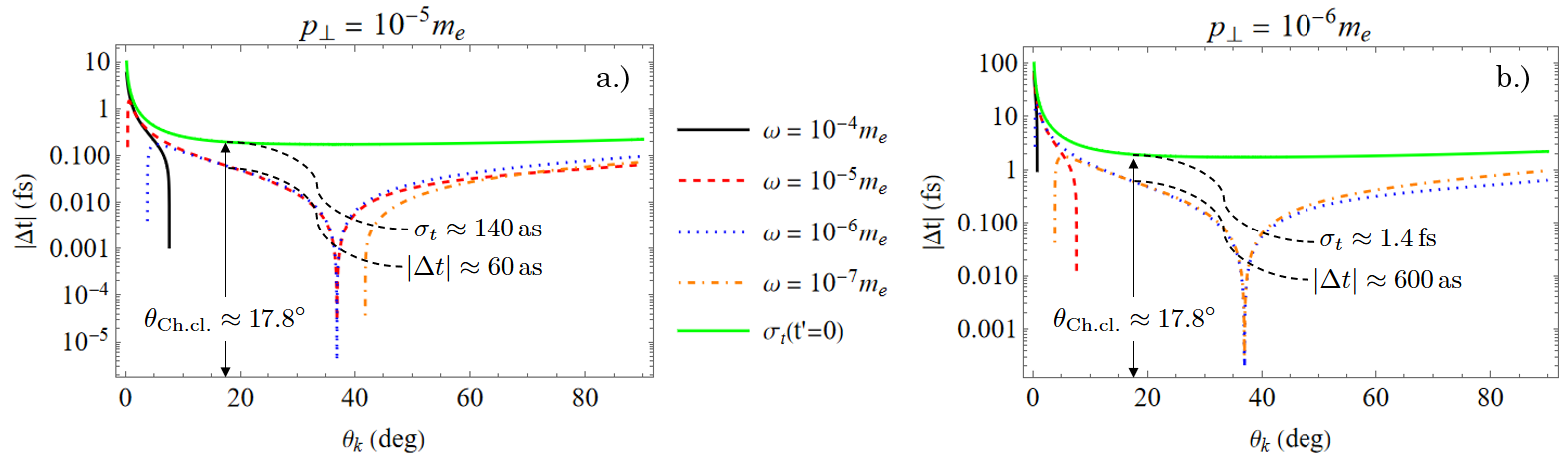}
    \caption{The same as in Fig.\ref{FigDt} but for lower electron energy, typical for TEM: $\beta = 0.7, p'_{\perp} = 0.99\, p_{\perp}, p'_z = 0.9\, \beta m_e, n=1.5$. \textbf{(a)} $p_{\perp} = 10^{-5}\, m_e,\, 1/p_{\perp} \gtrsim 10\,\text{nm}$. \textbf{(b)} $p_{\perp} = 10^{-6}\, m,\, 1/p_{\perp} \gtrsim 100\,\text{nm}$.}
    \label{FigDtNonr}
\end{figure*}

\begin{figure*}\center
\includegraphics[width=1\textwidth] {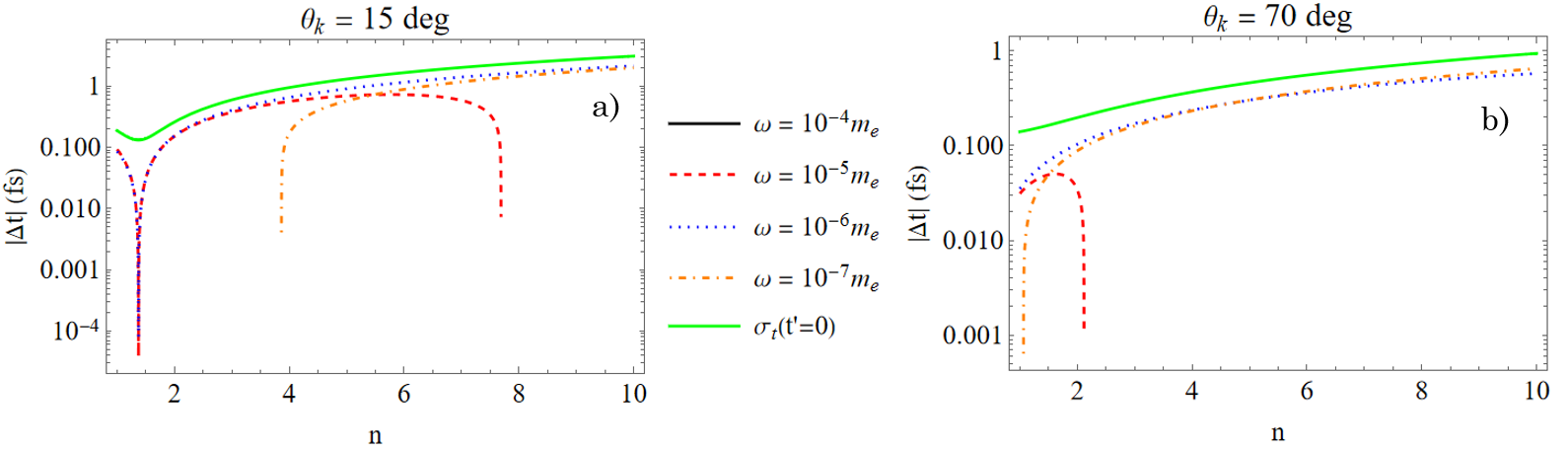}
    \caption{The temporal shift $\Delta t = t_0 - \br\cdot{\bm l}_0 = {\bm l}_0\cdot(\partial_{\bp} + \partial_{\bk})\zeta_{fi}$ from Eq.\eqref{Dtg} of the photon arrival time as a function of the refractive index $n$ at a certain emission angle: \textbf{(a)} $\theta_k = 15^\circ$, \textbf{(b)} $\theta_k = 70^\circ$ for a Gaussian electron packet with $\varphi=0$. Parameters: $\beta = 0.999\, (\gamma \approx 22)$, $p'_{\perp} = 0.99\, p_{\perp}$, $p'_z = 0.99\, \beta m_e$, $p_{\perp} = 10^{-5}\, m_e$, $1/p_{\perp} \gtrsim 10\, \text{nm}$. For the TEM energies, $\beta = 0.7$, $p'_z = 0.9\, \beta m_e$, these lines are nearly the same. The green line shows the flash duration $\sigma_t$ from Eq.\eqref{stp} at $t'=0$. There is no emission of THz photons with $\omega \sim 10^{-8}\,m_e$ at small angles $\theta_k$. The signs of the helicities are not important. The lines for $p_{\perp} = 10^{-6}\, m_e, 1/p_{\perp} \gtrsim 100\, \text{nm}$ are the same, however the shifts are of one order of magnitude larger.}
    \label{FigN}
\end{figure*}

\begin{figure*}\center
\includegraphics[width=1\textwidth] {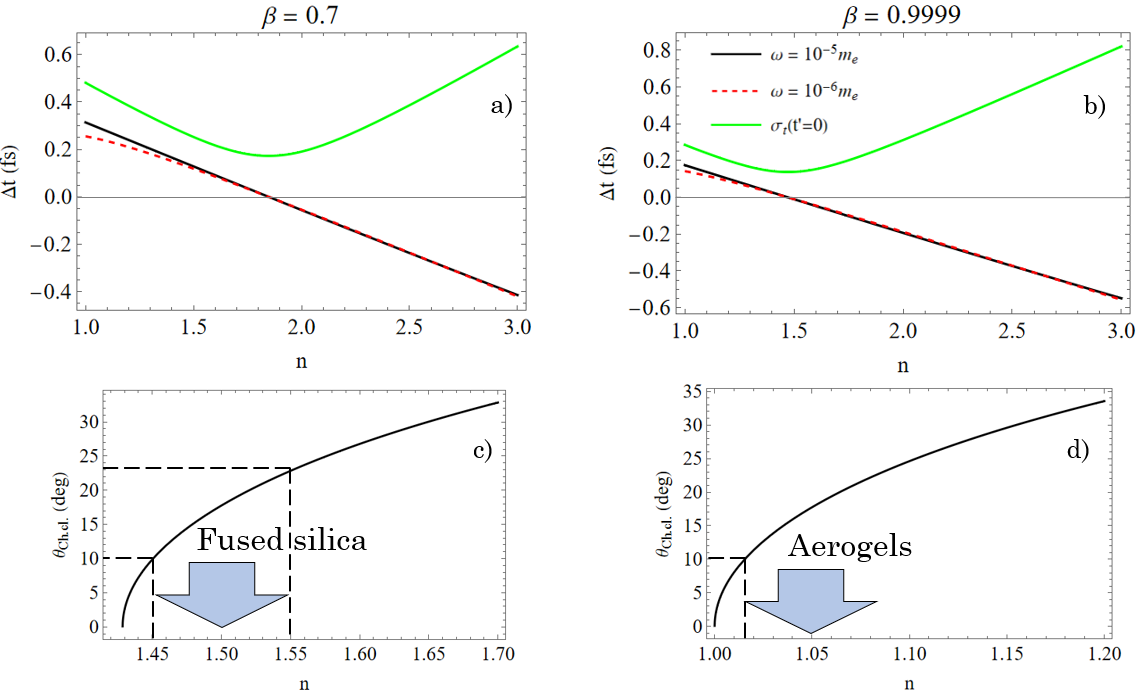}
    \caption{The same as in Fig. \ref{FigN}, but for the photon emission angle $\theta_k = 10^\circ$. \textbf{(a)} $\beta = 0.7, p'_z = 0.9\, \beta m_e$ (a TEM regime; the minimal refractive index, for which the Cherenkov condition is fulfilled, is $n_{\rm min}\approx 1.429$, whereas $n=1.45-1.55$ for fused silica in the given spectral range), \textbf{(b)} $\beta = 0.9999, p'_z = 0.99\, \beta m_e$ (an accelerator regime; here $n_{\rm min} \approx 1.0001$, which is suitable for the study of Cherenkov radiation in aerogels). The classical Cherenkov angle $\theta_{\text{Ch.cl.}} = \arccos(1/\beta n)$ for (a) and (b) is illustrated in panels (c.) and (d.), respectively.}
    \label{FigN2}
\end{figure*}

When measuring those small deviations of the photon arrival time from its classical value, the overall width of the temporal distribution $\sigma_t$ from Eq.(\ref{stp}) is crucial because the deviations can hardly be discerned when $\sigma_t \gg \Delta t$. In Figures \ref{FigDt}, \ref{FigDtNonr}, \ref{FigN}, and \ref{FigN2} we show that our quantum model yields smaller values of the temporal shift than the above classical estimate (\ref{Dtcl}), as it takes the finite width of the spectrum into account that contributes to the emission, and that the flash duration $\sigma_t(0)$ at $t'=0$, shown with the green lines, is always larger than the shift. We see that the shift is non-vanishing only in those regions of the momentum space that are allowed by the conservation laws and that the region of medium polarization grows for soft photons with $\omega \ll m_e$, thus $|\Delta t| \propto 1/(\omega\,\sin\theta_k) = 1/k_\perp$. For large emission angles, $\theta_k \lesssim \pi/2$, we have $|\Delta t| \propto 1/k_{\perp} \sim 1/p_{\perp}$ according to the triangle rules. Formally, the shift goes to infinity for a plane-wave electron with $p_{\perp} \to 0$, as the region of polarization grows, but the shift itself makes sense only for a packet with finite $p_{\perp}$. The transverse momenta of the incoming electron packet were chosen to be $p_{\perp} \sim (10^{-7}-10^{-4})\,m_e$ because they correspond to realistic spatial widths of the electron packet
\bea
1/p_{\perp} \gtrsim 1\, \text{nm} - 1\,\mu\text{m},
\eea
respectively. 

It is also implied that spreading of the electron packet during emission can be neglected, which always holds in a vicinity of the Cherenkov angle where $t_d \to \infty$ and so $\sigma_{\perp}(t') \to \sigma_{\perp}(0)$ (see Fig.\ref{Figtd}). We would like to emphasize that within the formation zone the photon emission does {\it not} take place only at the Cherenkov angle, which is why one can fix the detector at the angle $\theta_k$ different from $\theta_{\text{Ch.cl.}}$. Here, the spreading of the emitter wave function can still be neglected when it is a relativistic electron or when a target is much shorter than the electron Rayleigh length, $|\bm{u}_p|t_d^{(\text{e})} = \gamma |\bm{u}_p|t_d^{(\text{e, rest})} \sim 1-10$ cm for $\gamma \sim 10-100$ (see, e.g. \cite{Karlovets2021}). For the green lines in Figures \ref{FigDt}, \ref{FigDtNonr}, \ref{FigN}, and \ref{FigN2} we take $\sigma = p_{\perp}$ because the mean value of $p_{\perp}$, averaged over the packet momenta, equals $\sigma$ up to a factor of $\sqrt{2}$.

The sign of the shift {\it swaps} between the two points of the triangle (see Appendix \ref{ApTri}). Such swapping points can be seen in Fig. \ref{FigDt}b for $\omega = 10^{-6}m_e$ and $\omega = 10^{-7}m_e$ at $\theta_k \approx 24^\circ$, in Fig. \ref{FigDtNonr} at $\theta_k \approx 37^\circ$, in Fig. \ref{FigN}a for $\omega = 10^{-5}m_e$ and $\omega = 10^{-6}m_e$ at $n\approx 1.4$ and in Fig. \ref{FigN2} at $n \approx 1.85$ for the panel (a) and at $n \approx 1.5$ for the panel (b). Fixing the photon detector at the certain angles $\theta^{(\rm det)}, \phi^{(\rm det)}$ and the distance $r^{(\rm det)}$, one would see that the Cherenkov photons can arrive to the detector either later (time delay) than the classical time $t_{\text{cl.}}$ or earlier than that (negative time delay) with equal probablity. In contrast to transmission of the photons through medium, here, the negative time delays do not require any resonant behaviour of the refractive index and they hold even when the dispersion is neglected (see, e.g., \cite{Dienstbier}).

If one wishes to catch only the shifts with {\it one particular sign}, one should detect the photon and the electron in coincidence when the electron is post-selected to a state with a certain azimuthal angle $\phi'$ of the momentum and so the photon azimuthal angle is thereby fixed as well (see the Appendix \ref{ApTri}) together with the shift sign. 
Although detecting the two particles in coincidence is technically more demanding, the electron scattering angle is usually very small, whereas the photon emission angle is much larger, which is why such an experiment can in principle be carried out.

As seen in Figures \ref{FigDt}, \ref{FigDtNonr}, \ref{FigN}, and \ref{FigN2}, the shift is larger for wider electron packets of the width $1/p_{\perp} \gtrsim 0.1\,-1\,\mu\text{m}$. Whereas the typical widths nearby the cathodes do not exceed a few nanometers \cite{Cho,Cho_Oshima,Latychevskaia,Ehberger,Karlovets2021}, such electron packets can spread to the required values on a distance of a few millimetres for moderately relativistic energies of the transmission electron microscopes (TEM) with sub-nanoamper currents and no space charge influence. For such wide packets, the shift mostly takes place in the infrared and even sub-THz regions, so a target with $n \gg 1$ can in principle be created artificially as a meta-medium. In an accelerator chamber with $\beta \sim 0.999, \gamma \gg 1$ one needs to have a distance of at least several centimeters after the cathode to reach those widths. Importantly, the values of the shift stay of the same order of magnitude for lower electron energies of $\beta \sim 0.7-0.8$, as seen in comparison of Figs. \ref{FigDt} and \ref{FigDtNonr}.

As seen in Figs. \ref{FigN} and \ref{FigN2}, the temporal shift lies within the region from {\it attoseconds to a few femtoseconds} for realistic parameters, which are typical time scales of the atomic excitation processes, studied by the attosecond spectroscopy and metrology. For the TEM energies, a target made of fused silica with $n=1.458$ in the optical range can do the job, whereas for ultrarelativistic electrons aerogels with $n \sim 1.01-1.3$ can be employed, which are already used as Cherenkov generators for GeV electrons \cite{Adachi,sol-gel}. In a vicinity of the Cherenkov angle, the flash duration is governed by the size of the electron packet according to Eq. \eqref{sigma_t_Ch}, and therefore one can control duration of the emission by taking the electron packets with the needed size. For example, the electrons emitted from either a photo-cathode or a field-emitter have the sizes of a few nanometers \cite{Cho,Cho_Oshima,Latychevskaia,Ehberger,Karlovets2021} which yields the flash duration of the order of $10$ attoseconds. Conversely, if one measures duration of the Cherenkov radiation pulses with attosecond precision, one could retrieve the length of the emitting electron packet with nanometer accuracy.

\subsection{Medium with arbitrary dispersion}
\label{ArbDisp}
When the weak dispersion condition (Eq.(\ref{Disp_cond})) is not satisfied, the photon group velocity and the corresponding derivations should be generalized for the arbitrary dispersion as shown below:
\bea
& \dst {\bm u}_k = \frac{\partial\omega}{\partial\bk} = \frac{\bk}{n^2(\omega)\omega\left(1 + \dst \frac{\omega}{n}n'(\omega)\right)}, \label{uk_disp} \cr
& \dst \partial^2_{ij}\omega \equiv \frac{\partial^2\omega}{\partial k_i\partial k_j} = \frac{\delta_{ij} - (\bm u _k)_i({\bm u}_k)_j \, \xi (\omega, n,n',n'')}{n^2\omega \left(1 + \dst \frac{\omega}{n}n'(\omega)\right)},
\eea
where $\xi(\omega, n,n',n'') = n \omega^2 n'' + (n')^2\omega^2 + n^2 + 4nn'\omega$, and $n'(\omega)$ and $n''(\omega)$ are the first and second derivatives of the refractive index, respectively. To take the arbitrary $n(\omega)$ into account let us introduce two parameters, characterising the dispersion strength of a medium:
\bea
& \dst \mathcal{D}(\omega) = \frac{\omega}{n(\omega)} \,n'(\omega),
\label{D_par}
\eea
\bea
& \dst  \mathcal{E} (\omega) = n(\omega)\,\omega^2n''(\omega).
\label{E_par}
\eea
Thus, the group velocity of the photon has changed to
\bea
\displaystyle |{\bm u}_k| = \frac{1}{n(\omega) \left(1 + \mathcal{D}(\omega) \right)},
\eea
and the previous formulas for the correlation radius in Eq.\eqref{explhs}, diffraction time Eq.\eqref{gtp}, formation length Eq.\eqref{stp} and shift of the photon arrival time Eq.\eqref{Dtg} have to be modified. When $\mathcal{D} \ll 1$ and $\mathcal{E} \ll1 $, i.e. the dispersion is feeble, the expressions in Eq. (\ref{uk_disp}) are reduced to those in Eq. \eqref{uk_nodisp}.

Accordingly, the paraxial approximation for a weakly dispersive medium can be similarly generalized to the arbitrary dispersion law with the corresponding correction to the Gouy phase:
\bea
g^{(d)}(t') = \arctan\fr{t'}{t^{(d)}_d} + \arctan\fr{t'}{\tilde{t}\,^{(d)}_d},
\eea
with the diffraction times:

{\small \bea
& \dst t_d^{(d)} = \frac{2}{\sigma^2}\frac{1}{\dst \left(\frac{|\bm{u}_k|}{|\bk|} - \frac{1}{\varepsilon(\bp)}\right) + \left(\frac{1}{\varepsilon(\bp)} - \xi \frac{|\bm{u}_k|}{|\bk|
}\right)\frac{[\bm{u}_p\times\bm{u}_k]^2}{(\bm{u}_p - \bm{u}_k)^2}},
\label{td_disp}
\eea}
where $|\bk| = n(\omega)\, \omega$ and the upper index $(d)$ stands for dispersion. The time $\tilde t_d$ correspondingly is

\bea
& \dst \tilde{t}\,^{(d)}_d = \frac{2}{\sigma^2}\left(\frac{|\bm{u}_k|}{|\bk|} - \frac{1}{\varepsilon(\bp)}\right)^{-1} .
\eea

\begin{figure*}\center
\includegraphics[width=1\textwidth]{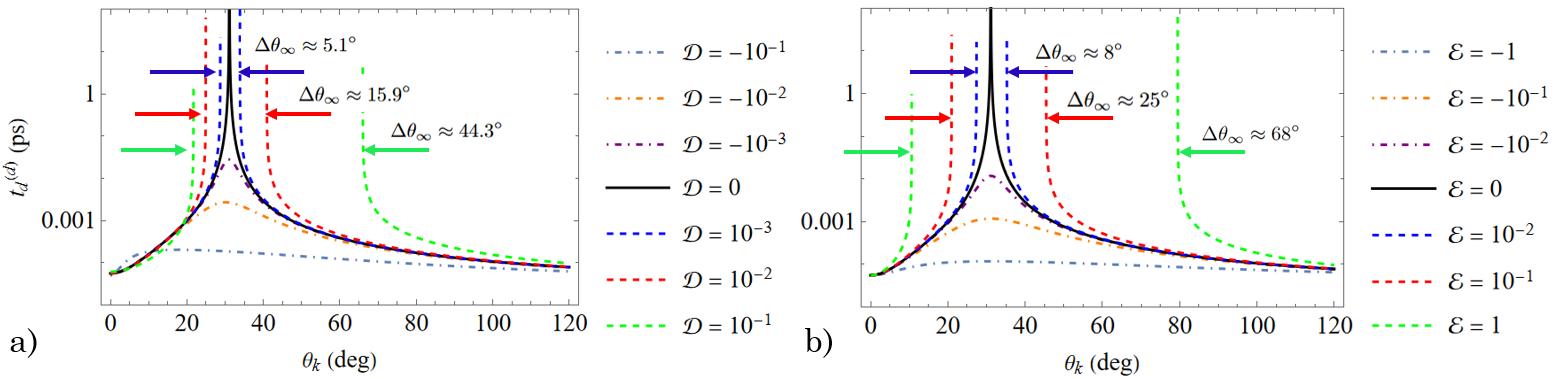}
    \caption{The diffraction times from Eq. \eqref{td_disp} as functions of the emission angle $\theta_k$ for different values of the first, $n'(\omega)$ and second $n''(\omega)$ derivatives of the refractive index. Discontinuous $t_d^{(d)}$ indicates $|\bm{u}_p|/|\bm{u}_k| \ge 1$. Parameters: $\omega_0 = 10^{-6}m_e \approx 0.51$ eV, $n = 1.5$, $\sigma = 10^{-5}\,m_e$, $\beta = 0.8$, $\theta_{\rm Ch.cl.} \approx 31.1$. \textbf{(a)} Lines are for $\mathcal{D} \equiv \omega n'/n = \{-10^{-1},\,-10^{-2},\,-10^{-3},\,0,\,10^{-3},\,10^{-2},\,10^{-1}\}$ and $\mathcal{E} = 0$. \textbf{(b)} Lines are for different $\mathcal{E} \equiv n\omega^2n'' = \{-10^{-1},\,-10^{-2},\,-10^{-3},\,0,\,10^{-3},\,10^{-2},\,10^{-1}\}$ and $\mathcal{D} = 0$. As seen, $\mathcal{D}$ and $\mathcal{E}$ grow, the angular width of $\Delta \theta_{\infty}$ increases and it can reach $\sim 10^\circ$.}
    \label{FigtdDisp}
\end{figure*}

In Fig. \ref{FigtdDisp} the effects of dispersion are shown for the spreading time of a photon for different values of $\mathcal{D}$ and $\mathcal{E}$. As seen, for $\mathcal{D} \ge 0$ and $\mathcal{E} \ge 0$ the interval $\Delta \theta_{\infty}$ increases with these parameters, reaching several tens of degrees when $\mathcal{D} = 10^{-1}$ or for $\mathcal{E} = 10^{-1}$. However, the diffraction time is continuous for negative $\mathcal{D}$ and $\mathcal{E}$, indicating the absence of a far photon field in such a medium for the given velocity of the incident electron. The further increase of this velocity affects only the Cherenkov angle and, therefore, the positions of the maximum and negative interval region, while the interval $\Delta\theta_\infty$ does not change (see Eq.\eqref{Dth}). The effective correlation radius $R_{\rm eff}$ from Eq. \eqref{explhs} in this scenario is modified as follows:
\bw
\bea
& \dst \frac{R^2}{\left(R_{\rm{eff}}^{(d)}\right)^2} = \frac{1}{\left(\sigma_\perp^{(d)}(t')\right)^2}\Bigg[\fr{[{\bm R}\times (\bm{u}_p - \bm{u}_k)]^2}{(\bm{u}_p - \bm{u}_k)^2} + \frac{t'^2}{\left(\tau^{(d)}_d\right)^2\left(1 + \left(t'/\tilde t^{(d)}_d\right)^2\right)}\fr{({\bm R}\cdot \left[{\bm u}_p \times {\bm u}_k\right])^2}{(\bm{u}_p - \bm{u}_k)^2}\Bigg]\cr
& \dst \left(\sigma_\perp^{(d)}(t')\right)^2 = \sigma^{-2}\left(1 + \frac{(t')^2}{\left(t_d^{(d)}\right)^2}\right),\quad \left(\tau_d^{(d)}\right)^2 = \frac{2}{\sigma^2}\frac{t_d^{(d)}\tilde t_d^{(d)}}{\left(\dst \frac{1}{\varepsilon(\bp)} - \xi\frac{|\bm{u}_k|}{|\bk|}\right)\left(t_d^{(d)} + \tilde t_d^{(d)}\right)}
\eea
\ew

The shift of the photon arrival time in Eq.\eqref{Dtg} changes only due to the different photon group velocity, $\bm{u}_k$. Thus, $\Delta t$ depends only on $\mathcal{D}$ and not on $\mathcal{E}$. One can ensure the linear dependence of the shift on $\mathcal{D}$:
\bea
    & \dst \Delta t \propto \frac{[(\bk\cdot \bm{\Phi}) u_p^2 - (\bm{u}_p\cdot \bm{\Phi})(\bk\cdot\bm{u}_p)]n^2(\omega)\omega}{[\bk\times \bm{u}_p]^2}\mathcal{D}, \cr
    & \bm{\Phi} = (\partial_\bp + \partial_\bk)\zeta_{fi} + \partial_\bp\varphi.
\eea

\begin{figure*}[h!]\center
\includegraphics[width=1\linewidth]{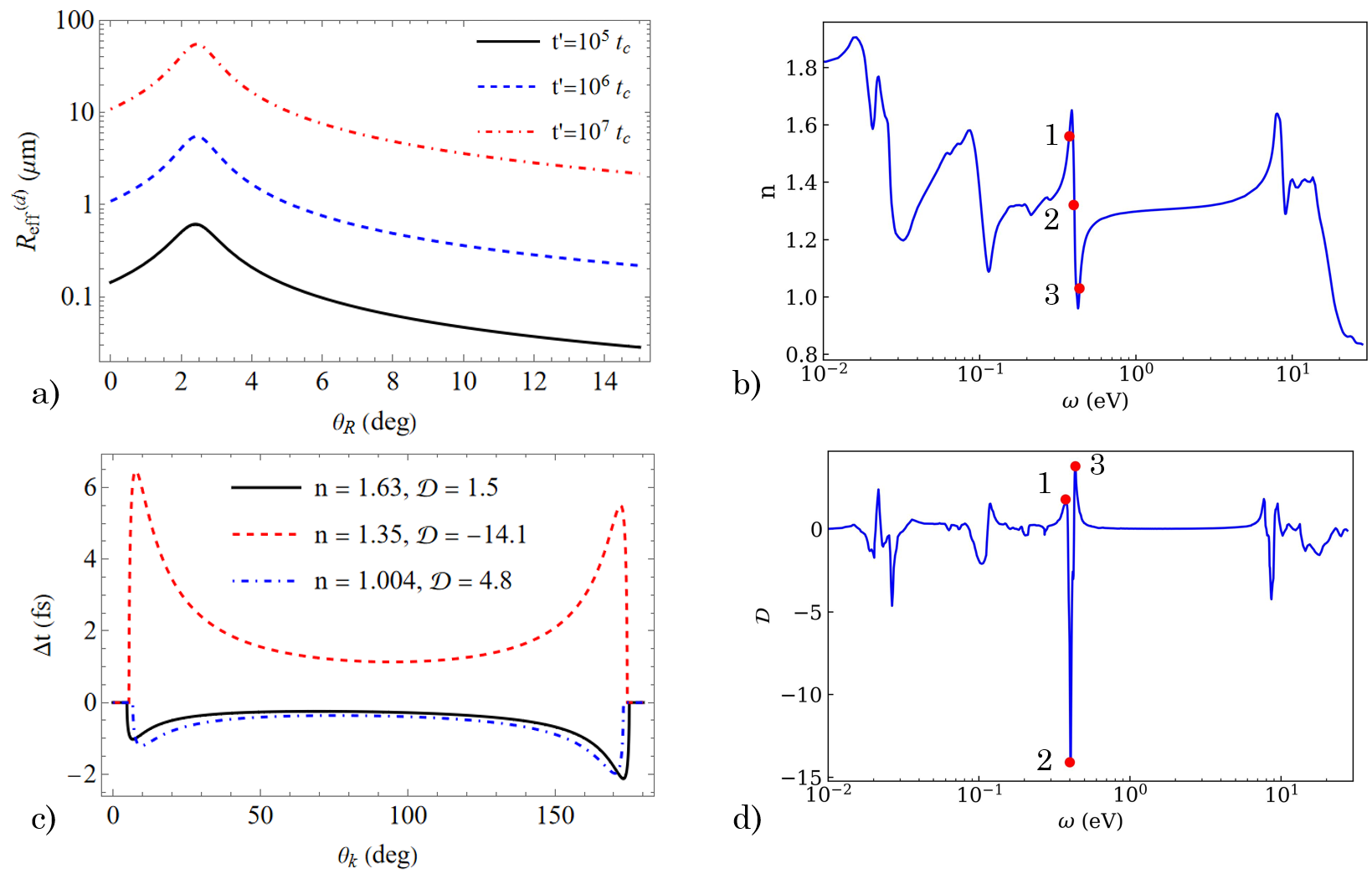}
    \caption{The dispersion characteristics for ice at the temperature of $T = -7^\circ$. \textbf{(a).} The correlation radius $R_{\rm eff}$ from Eq.\eqref{explhs} as a function of the polar angle $\theta_R$ of $\bm R$ for $t' = 10^5\, t_c,\, 10^6\,t_c,\, 10^7\,t_c$. Parameters: $\omega = 4$ eV, $n = 1.35$, $\theta_k = \theta_{\rm Ch.cl.} \approx 41.5^\circ$, $\sigma = 10^{-4}m_e$, $\beta = 0.99$, $\phi_R - \phi_k = 20^\circ$, and the dimensionless parameters from Eqs. \eqref{D_par} and \eqref{E_par} are $\mathcal{D} \equiv \omega n'/n = -13.65$ and $\mathcal{E} \equiv n\omega^2n'' \approx 0$, respectively (the second red dot in panels (b) and (d)). \textbf{(b).} The refractive index for ice as a function of the radiation frequency taken from \cite{Warren1984}. \textbf{(d).} The values of $\mathcal{D} \equiv \omega n'/n$ for ice as a function of the radiation frequency. \textbf{(c).} The temporal shift, $\Delta t$ from Eq.\eqref{Dtg}, in ice for different $\mathcal{D}(\omega)$, corresponding to the red points in panels (b) and (d) as follows: the solid black line is for $\omega = 0.37$ eV, $n = 1.63, \mathcal{D} = 1.5$ (red dot 1), the blue dotted line is for $\omega = 4$ eV, $n = 1.35,\, \mathcal{D} = -14.1$ (red dot 2) and the red dashed line is for $\omega = 4.3$ eV, $n = 1.004,\, \mathcal{D} = 4.8$ (red dot 3). Other parameters: $p_\perp = 10^{-5}m_e$, $p'_\perp = 0.99\,p_\perp$, $p'_z = 0.99\,\beta m_e$, $\sigma = 10^{-4}m_e$}
    \label{FigIceDisp}
\end{figure*}

The effects of dispersion can be demonstrated experimentally in the materials having sharp jumps of their refractive index, which are often used for experiments with the slow light \cite{khurgin2018slow,bigelow2003}. This sharp behaviour of the refractive index can be found even in water around the absorption lines (see, e.g., \cite{Warren1984}). In Fig. \ref{FigIceDisp} the correlation radius and time shift are illustrated for ice at the temperature of $-7^\circ$C for $\lambda \approx 3\, \mu m$. Fig. \ref{FigIceDisp}a demonstrates that the maximum of the correlation radius for the strong dispersion of ice ($|\mathcal{D}| \sim 10$) lies no longer at the vicinity of the classical Mach angle (Eq. \eqref{Machth}), which is $\theta_{\rm Mach} \approx 131^\circ$, but near the electron propagation direction ($\theta_R = 0$). 

\begin{figure*}[h!]\center
\includegraphics[width=1\textwidth]{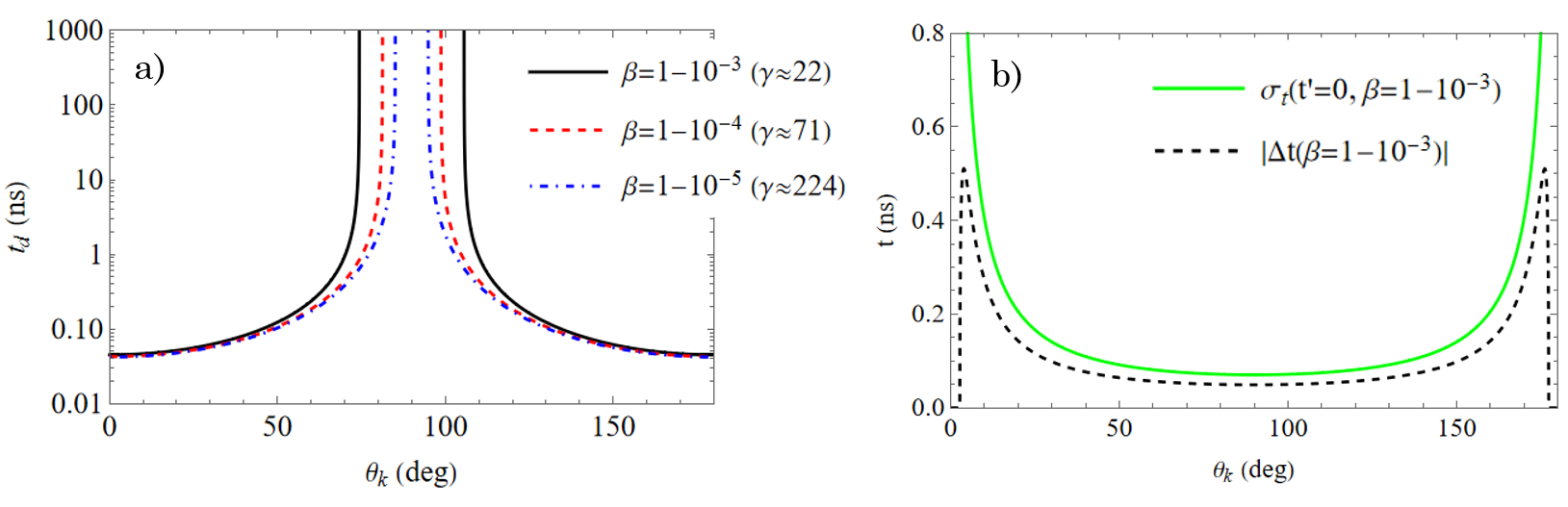}
    \caption{The spreading times (a) and the shift of the photon arrival time (b) for the ultracold gas of sodium atoms, for which $n = 1.001$ and dimensionless dispersion parameters from Eqs. \eqref{D_par} and \eqref{E_par} are $\mathcal{D}(\omega = 2\cdot 10^{-5}m_e) \approx 5\cdot 10^6$, $\mathcal{E} = 0$, respectively. This gas is used for the experiments with the slow light (see, for instance \cite{Hau1999Feb}). Parameters: $p_\perp = 10^{-5}m_e$, $p_\perp' = 0.99\, p_\perp$, $p_z' = 0.99\, \beta m_e$, $\sigma = 10^{-5}m_e$ (the transverse momentum spread of the incoming electron). The green line in the panel (b) represents the flash duration $\sigma_t$ from Eq.\eqref{stp} taken at $\beta = 0.999$ ($\gamma \approx 22$)}
    \label{FigNaDisp}
\end{figure*}

It is also well-known that an extremely large dispersion ($|\mathcal{D}| \gg 1$) can be obtained using the electromagnetically induced transparency in vapor cells of atoms (see, e.g., \cite{Harris1992Jul}). For example, the gas of sodium atoms at the temperature close to absolute zero and at a specific frequency is described by the value $\mathcal{D} \sim 10^6$, leaving $n \approx 1$ and $\mathcal{E} \approx 0$ \cite{Hau1999Feb}. The diffraction time, flash duration and time shift for these conditions are shown in Fig. \ref{FigNaDisp}. As seen, due to extremely large $\mathcal{D}$ the time shift reaches \textit{several hundreds of picoseconds} at the angles $\theta_k \approx 5^\circ$ and $\theta_k \approx 175^\circ$, which in turn can be easier to measure experimentally than the shift of $\Delta t \sim 10\,\text{as} - 1\,\text{fs}$ which is typical for the materials with weak dispersion (see Section \ref{sec:FDaSh}).

\section{Images of the photon Wigner function}
\label{sec:Images}

\begin{figure*}[h!]
\includegraphics[width=\textwidth]{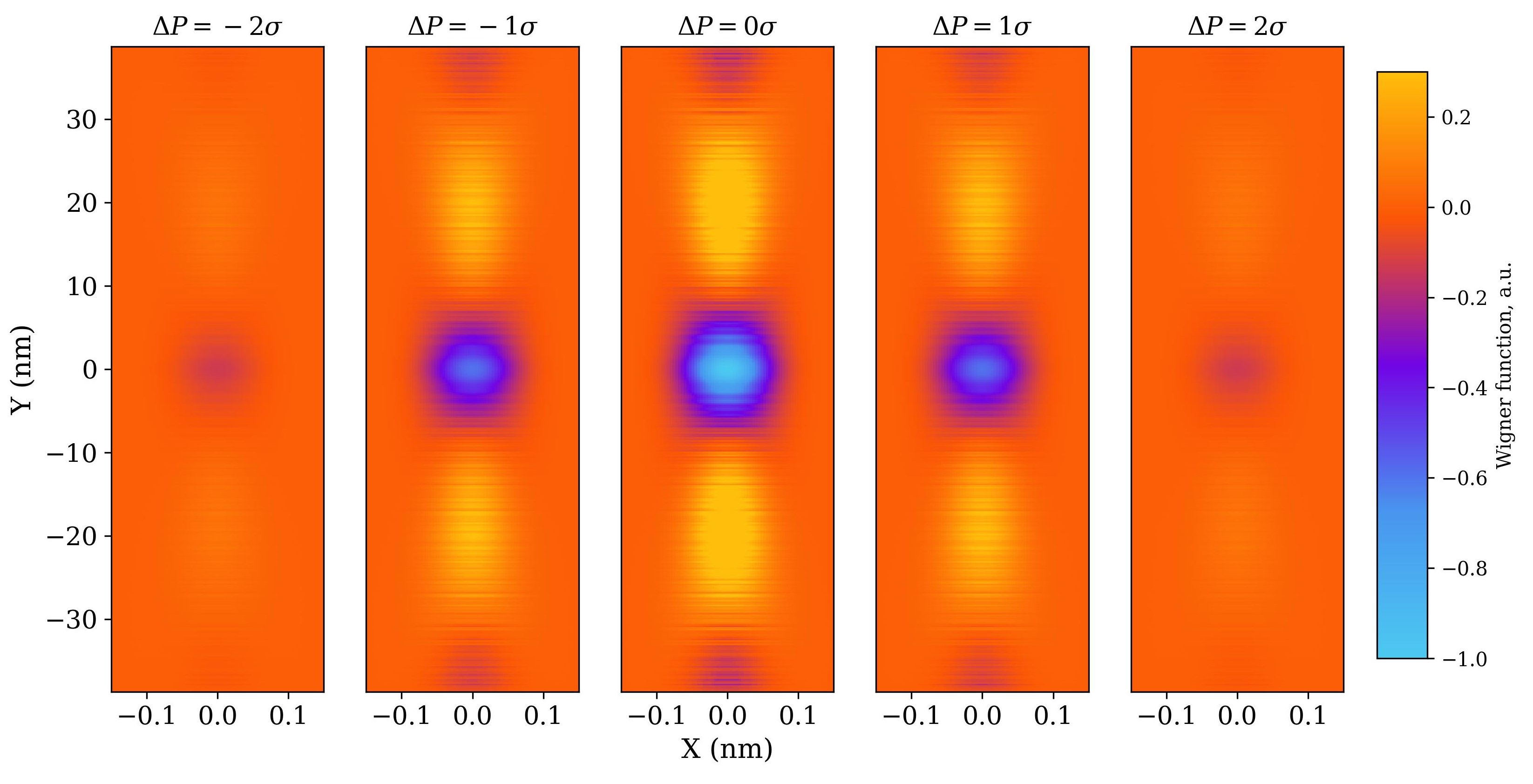}
	\caption{The coordinate dependence of the Wigner function according to Eq.(\ref{tpr}) where ${\bm R} = (X,Y,Z)$ in the plane $Z=0$. Parameters: $\beta = 0.99\, (\gamma \approx 7)$, $n=1.4$, $\theta_k = \theta_{\rm Ch.cl} = \arccos(1/\beta n) \approx 43.82^\circ$, $\phi_k = \pi/2$, $\omega = 10^{-5}\,m_e $, $\sigma = 10^{-2}\,m_e$ (transverse momentum spread of the incoming electron). The panels from left to right are for the different values of $\Delta P \equiv |\bp' + \bk - \la \bp\ra|$, which are: $-2\sigma$, $-\sigma$, $0$, $\sigma$, $2\sigma$. This illustrates an approximate momentum conservation law in the photon emission within the employed WKB-approach}
	\label{Wig_plot_ppx}
\end{figure*}

\begin{figure*}[h!]
\includegraphics[width=1\linewidth]{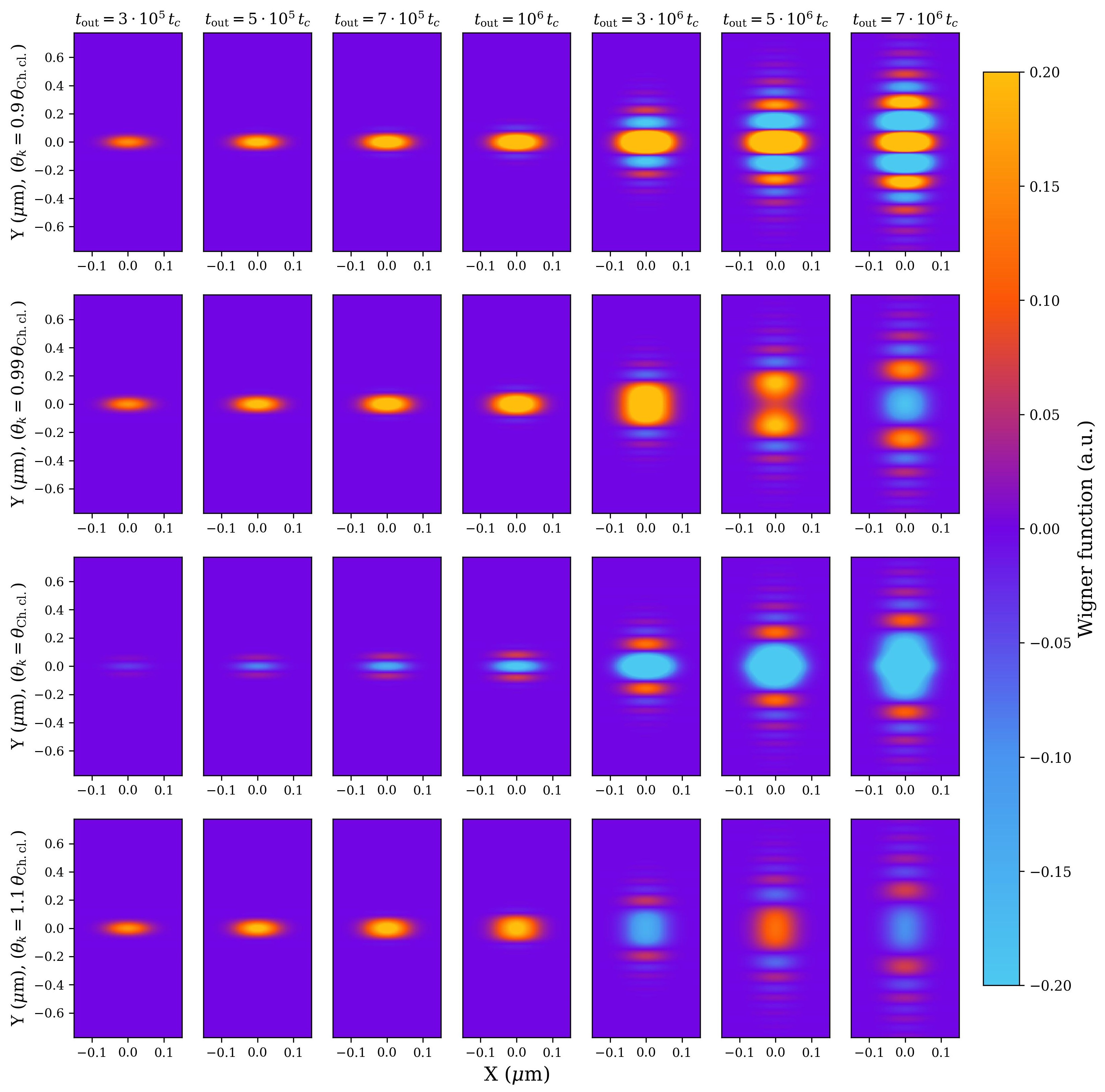}
    \caption{The coordinate dependence of the Wigner function according to Eq.(\ref{tpr}) where ${\bm R} = (X,Y,Z)$ in the plane $Z=0$. Parameters: $\beta = 0.99\, (\gamma \approx 7)$, $\theta_p = \la\theta_p\ra = 0, \,\phi_p = \la\phi_p\ra = \pi/2$, $n = 1.4$, $\phi_k = \pi/2$, $\omega = 10^{-5}\,\text{m}_e $, $\sigma = 10^{-5}\,m_e$, where, $\la\theta_p\ra,\, \la\phi_p\ra$ are the polar and azimutal angles of the mean momentum $\la\bp\ra$ of the initial electron, introduced in Eq.\eqref{WPm}. From left to right: $t_{\text{out}} = 3\cdot 10^6\, t_c$,  $5\cdot 10^6\, t_c$,  $7\cdot 10^6\, t_c$,  $10^7\, t_c$,  $3\cdot 10^7\, t_c$,  $5\cdot 10^7\, t_c$,  $7\cdot 10^7\, t_c$. Upper row - $\theta_k = 0.9\, \theta_{\rm Ch.cl.}$, second row - $\theta_k = 0.99\,\theta_{\rm Ch.cl.}$, third row - $\theta_k = \theta_{\rm Ch.cl.}$ and lower row - $\theta_k = 1.1\, \theta_{\rm Ch.cl.}$}
    \label{Wig_plot_theta=0p99_1_1p1Ch}
\end{figure*}

\begin{figure*}[h!]
\includegraphics[width=1\textwidth]{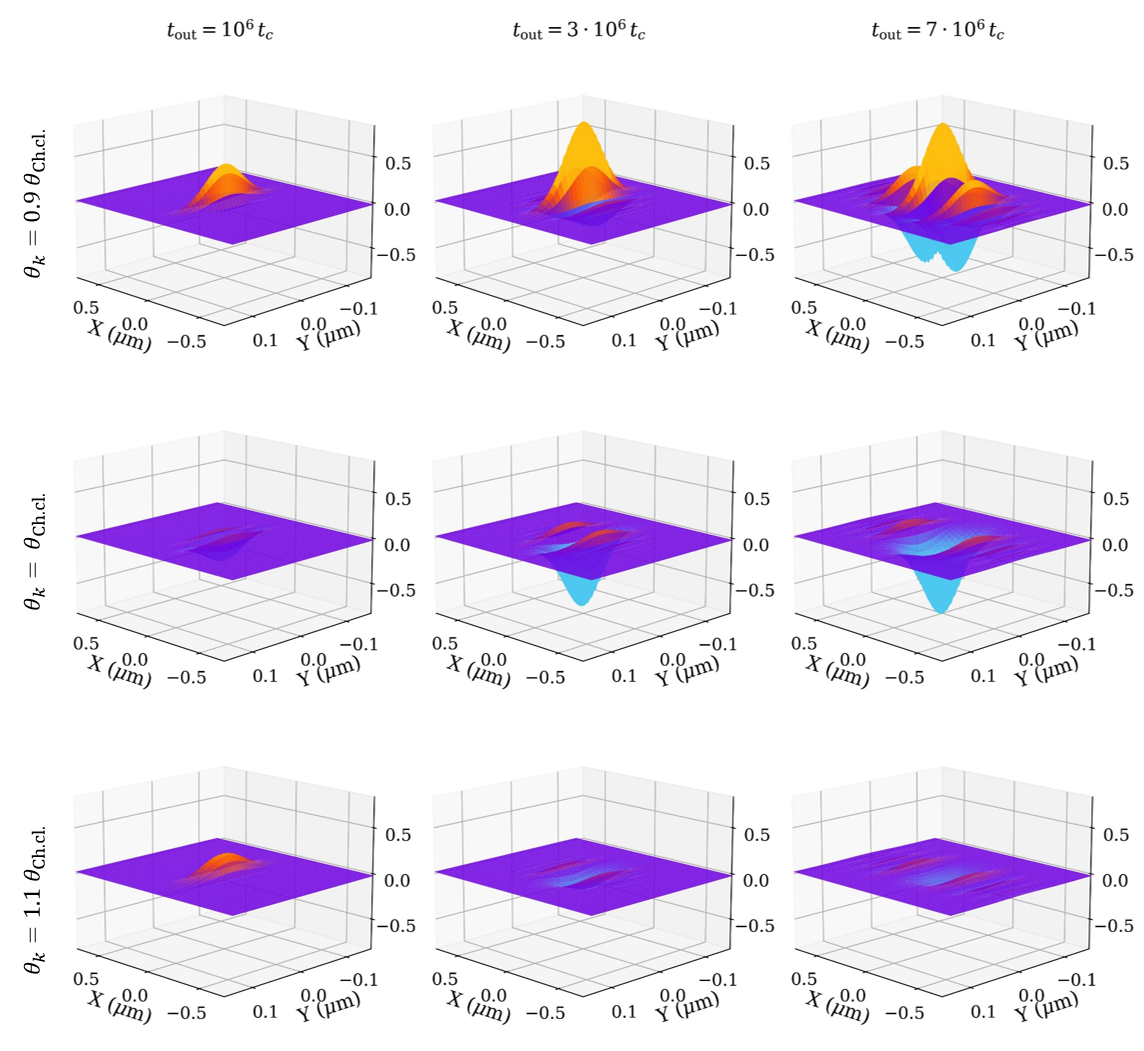}
    \caption{The coordinate dependence of the Wigner function according to Eq.(\ref{tpr}) where ${\bm R} = (X,Y,Z)$ in the plane $Z=0$. Parameters: $\beta = 0.99\, (\gamma \approx 7)$, $\theta_p = \la\theta_p\ra = 0$ (the polar angles of the detected, $\bp$ and the mean, $\la\bp\ra$, electron momenta from Eq.\eqref{WPm}), $\phi_p = \la\phi_p\ra = \pi/2$ (the azimuthal angles of the detected, $\bp$ and the mean, $\la\bp\ra$, electron momenta from Eq.\eqref{WPm}), $n = 1.4$, $\phi_k = \pi/2$, $\omega = 10^{-5}\, m_e$, $\sigma = 10^{-5}\,m_e$ (transverse momentum spread of the incoming electron). Upper row - $\theta_k = 0.9\,\theta_{\rm Ch.cl}  \approx 39.44^\circ$, middle row - $\theta_k = \theta_{\rm Ch.cl}  \approx 43.82^\circ$, lower row - $\theta_k = 1.1\,\theta_{\rm Ch.cl}  \approx 48.20^\circ$. Left column - $t_{\rm out} = 10^6\,t_c \approx 1.3$ fs, middle column - $t_{\rm out} = 3\cdot 10^6\,t_c \approx 3.9$ fs, right column - $t_{\rm out} = 7\cdot 10^6\,t_c \approx 9.1$ fs}
    \label{Wig_plot_3D}
\end{figure*}

{To see how the Wigner function Eq.\eqref{tpr} evolves at different emission angles, one can visualize it as a function of $\bm{R}$. The corresponding images in $X,Y$-coordinates, evaluated in the plane $Z = 0$, are shown in Fig. \ref{Wig_plot_theta=0p99_1_1p1Ch}. 

It is important to emphasize that the structure of the Wigner function in Eq.\eqref{tpr} supposes that the \textit{final electron state is measured by projecting it onto a plane wave with the momentum $\bp'$}. This scenario does not show the physical evolution of the emitted photon state because this measurement collapses the full two-particle quantum state. Thus, to investigate the photon independently of the final electron state, i.e. without the measurement of its quantum numbers, one should integrate the Wigner function Eq. \eqref{tpr} over the final electron momentum, $\bp'$
\bea
\mathcal{W}(\br,\bk, t) = \int d^3\bp' \mathcal{W}_p(\br,\bp'+\bk,\bk,t).
\label{Wigner_intd}
\eea
However, within the same WKB approximation the exponential factor can be well approximated by the delta-function $\delta^{(3)}(\bp' + \bk - \la\bp\ra)$. Fig.\ref{Wig_plot_ppx} shows that, as the value $\Delta P \equiv |\bp' + \bk - \la\bp\ra|$ increases, the Wigner function only changes its intensity but not the shape, which confirms the validity of this simplification. Hence, we can employ the following approximate equality
\bea
\mathcal{W}(\br,\bk,t) \approx \mathcal{W}_p(\br,\la\bp\ra,\bk,t)
\eea 

As seen in Fig.\ref{Wig_plot_theta=0p99_1_1p1Ch}, at $\theta_k = 0.99\,\theta_{\rm Ch.cl.}$ the middle part of the Wigner function exhibits oscillatory behavior: it is positive at $t_{\rm out} = 3\cdot 10^5$ and becomes negative by $t_{\rm out} = 7\cdot 10^6\,t_c$. A similar trend is observed for $\theta_k = 1.1\,\theta_{\rm Ch.cl.}$ and $\theta_k = 0.9\theta_{\rm Ch.cl.}$, as shown in Fig. \ref{Wig_plot_3D}. In contrast, this central region remains negative for $\theta_k = \theta_{\rm Ch.cl.}$ for all $t_{\rm out}$, which implies that the totally evolved Wigner function ($t_{\rm out} \rightarrow \infty$) must be also negative in this region of ($X,Y$)-plane. 

In fact, the Wigner function becomes negative not only exactly at $\theta_k = \theta_{\rm Ch.cl.}$ but \textit{throughout the finite interval of angles around it}, lying within the range of angles corresponding to the negative spreading time (see Eq. \eqref{thetainf}). Physically, this indicates destructive interference of the near field within this angular sector. This behavior contrasts with the classical emission scenario, where at $\theta_k = \theta_{\rm Ch.cl.}$ there is always a constructive interference in the far field. The deviation from classical theory is connected to the quantum recoil of the emitting electron.

The negativity of the Wigner function is experimentally accessible through the generalized quantum-state measurement techniques (see, e.g., \cite{Nogues2000_Wig_meas}). These regions can also indicate that the corresponding quantum state possesses the features with no counterparts in classical realm.}

\section{Conclusion}
\label{sec:conc}

In this work, we have developed a general quantum phase-space description of a single-photon emission generated by an electron wave packet and illustrated it with Cherenkov radiation. This approach is based on the Wigner function of the emitted photon field, which explicitly incorporates the finite spatial structure and coherence of the incoming electron as well as the details of the measurement procedure. The method has been constructed not only from the conventional projective measurements but also from the generalized measurements framework, including positive operator valued measurements.

Using this generalized-measurement formulation, we have derived the evolved photon Wigner function and demonstrated how the spatial coherence of the electron packet imprints itself onto the emitted radiation. Several effects with no direct classical analogue emerge naturally from this treatment:
\begin{itemize}
    \item Negative spreading time and negative formation length in a finite angular interval around Cherenkov direction.
    \item Finite flash duration of the emitted radiation.
    \item Quantum shift of photon arrival time produced by the medium-induced dipole response and AC-Stark-type phases.
    \item Near-field ``snapshots'' of the electron wave function, in which the photon spatial profile reproduces the coherent structure of the emitter.
\end{itemize}

We have further extended the phase-space formalism of Cherenkov radiation for media with arbitrary dispersion. Strongly dispersive materials, such as those near absorption edges or exhibiting electromagnetically induced transparency, substantially modify the spreading time, correlation radius, flash duration, and arrival-time shift. This modification can increase the angular width of negative spreading time, and the temporal shift of the photon arrival can reach experimentally resolvable femtosecond-nanosecond scales. The results also demonstrate that the quantum recoil after the emission, electron-packet coherence, and medium dispersion create a structure in space-time that cannot be captured by the classical Tamm-Frank picture.

We have analyzed the phase-space images of the emitted radiation with respect to the emission angle and the evolution time of the photon quantum state. The Wigner function images possess the regions of persistent negativity, particularly in the vicinity of the Cherenkov angle, lying within the angular range of the negative spreading time. These negative regions of the Wigner function arise from quantum interference. Their persistence over a finite interval of emission angles signifies the quantum features of the photon state, which, in principle, can be probed experimentally using generalized-state tomography techniques.

Beyond photon generation, this approach opens a path toward diagnostic applications: by measuring photon pulse durations with subpicosecond accuracy, one can infer the longitudinal size of the emitting electron packet. Such techniques could enable in situ control of wavepacket coherence in electron microscopy, accelerator facilities, and even astrophysical contexts involving ultrarelativistic particles. Consequently, Cherenkov radiation can be an outstanding representative of medium-induced emission processes where phase-space methods uncover attosecond-scale temporal phenomena inaccessible to conventional momentum-space analysis.

\section{Acknowledgements}

We are grateful to T.\,Panfilov and I.\,Pavlov for fruitful discussions and suggestions, and also to M. Zhuravleva for the help with the illustrating. The studies of the temporal characterictics of Cherenkov radiation are supported by the Ministry of Science and Higher Education of the Russian Federation (Project FSER-2025-0012). Those in a dispersive medium are supported by the Russian Science Foundation (Project No. 23-62-10026 https://rscf.ru/en/project/23-62-10026/). The work on the visualization of the Wigner function is supported by the Government of the Russian Federation through the ITMO Fellowship and Professorship Program.

\section{Data availability}

The data that support the findings of this article are openly available at \cite{Warren1984} and at \cite{iceref}.

\appendix

\section{Derivation of the photon spatial energy density}
\label{SpatEnerDens}
Let us define the photon field energy density in the evolved state for the real space-time. The Hermitian vector potential operator is 
\bea
&  \displaystyle \hat{\bm A}({\br}, t) = \sum\limits_{\lambda_\gamma=\pm 1} \int\bar{d}^3\bk \left({\bm A}_{{\bk}\lambda_\gamma}({\br}, t)\, \hat{c}_{{\bk}\lambda_\gamma} + \text{h.c.}\right),
\label{FOp}
\eea
where $\hat{c}_{{\bk}\lambda_\gamma}$ are the annihilation operators, h.c. means hermitian conjugate. The Fourier component of this operator is
\bea
&  \displaystyle {\bm A}_{{\bk}\lambda_\gamma}({\br}, t) = \frac{\sqrt{4\pi}}{\sqrt{2\omega}}\, {\bm e}_{{\bk}\lambda_\gamma}\, e^{-i\omega t + i{\bk}\cdot{\br}}, 
\label{A_k_lamb}
\eea
where ${\bm e}_{{\bk}\lambda_\gamma}\cdot\bk = 0,\ {\bm e}_{{\bk}\lambda_\gamma}\cdot {\bm e}_{{\bk}\lambda'_{\gamma}}^* = \delta_{\lambda_\gamma\lambda'_{\gamma}}$, and the Coulomb gauge is implied. The normalization condition for $A_{\bk\lambda_\gamma}$ is
\bea
{\bm A}^*_{{\bk}\lambda_\gamma}({\br}, t)\,{\bm A}_{{\bk}\lambda_\gamma}({\br}, t) = \frac{2\pi}{\omega}
\label{A_norm_cond}
\eea
The field average $\hat{\bm A}({\br}, t)$ can be defined with the generalized measurement S-matrix (\ref{gammaGM}) and the Fourier amplitude (\ref{A_k_lamb}) (see \cite{Scully}) as following:
\bea
&  \displaystyle \la 0|\hat{\bm A}({\br}, t)|\gamma\ra = \sum\limits_{\lambda_\gamma} \int\bar{d}^3\bk {\bm A}_{{\bk}\lambda_\gamma}({\br}, t)\,S_{fi}^{\text{GM}}(\bk,\lambda_\gamma) = \cr
& \dst = \int\bar{d}^3\bk \bm{\tilde{A}}({\bk}, \omega)\, e^{-ikx}, \cr
& \dst \bm{\tilde{A}}({\bk}, \omega) = \frac{\sqrt{4\pi}}{\sqrt{2\omega}} \sum\limits_{\lambda_\gamma} {\bm e}_{\bk\lambda_\gamma} S_{fi}^{\text{GM}}(\bk,\lambda_\gamma)
\label{Fav}
\eea
Clearly, they can have imaginary parts:
\bea
& \dst \la 0|\hat{\bm A}({\bm r}, t)|\gamma\ra^* = \int\bar{d}^3\bk \bm{\tilde{A}}^*({\bk}, \omega)\, e^{ikx} \cr
& \dst = \int\bar{d}^3\bk \bm{\tilde{A}}^*(-{\bk}, \omega)\, e^{i\omega t + i\bk\cdot\br}.
\label{Aevrt}
\eea
$\bm{\tilde{A}}({\bk}, \omega)$ can be interpreted as ``the photon wave function'' in momentum space. The sign $\bk \to -\bk$ has been modified in the second line of Eq. \eqref{Aevrt}. Comparing this to $\la 0|\hat{\bm A}({\bm r}, t)|\gamma\ra$, we see that
\bea
\bm{\tilde{A}}^*(-{\bk}, \omega)\, e^{i\omega t} \ne \bm{\tilde{A}}({\bk}, \omega)\, e^{-i\omega t}
\eea
whatever symmetries ${\bm \tilde{A}}({\bk}, \omega)$ has. So at a time instant $t\ne 0$ the average \eqref{Aevrt} may be complex. Note that the diagonal part of the matrix element $\hat{S} - \hat{S}^{(1)} = \hat{1}$ does not contribute to Eq.(\ref{Fav}).

Then, the Hermitian electric and magnetic field operators are 
\bea
& \dst \hat{\bm E}({\br}, t) = - \frac{\partial \hat{\bm A}({\br}, t)}{\partial t}\cr
& \dst \hat{\bm H}({\br}, t) = \nabla\times \hat{\bm A}({\br}, t),
\label{field_oper}
\eea
and the field averages are defined analogously as Eq. \eqref{Fav} (see for details Appendix \ref{Amp}).

A real spatial observable in this problem is \textit{the energy density}. Electric and magnetic parts of it are:
\bea
&  \displaystyle \la\gamma|\hat{\bm E}^\dagger({\br}, t)\hat{\bm E}({\br}, t)|\gamma\ra \equiv \la\gamma|\hat{\bm E}^2({\br}, t)|\gamma\ra = \cr
& \dst \varepsilon_0 + 2\left|\la 0|\hat{\bm E}({\br}, t)|\gamma\ra\right|^2,
\label{E_part}
\eea
\bea
& \dst \la\gamma|\hat{\bm H}^2({\br}, t)|\gamma\ra = \varepsilon_0 + 2\left|\la 0|\hat{\bm H}({\br}, t)|\gamma\ra\right|^2,
\label{H_part}
\eea
respectively, where the first term in Eqs. \eqref{E_part} and \eqref{H_part} is
\bea
& \dst \varepsilon_0 = \la\gamma|\gamma\ra \sum\limits_{\lambda_\gamma}\int\bar{d}^3\bk\, \omega^2 |{\bm A}_{{\bk}\lambda_\gamma}(\br,t)|^2 = \cr
& \dst \la\gamma|\gamma\ra \sum\limits_{\lambda_\gamma}\int\bar{d}^3\bk\, 2\pi\omega
\eea
It \textit{diverges} and, therefore, should be associated with the vacuum energy. The diagonal term $\hat{S} - \hat{S}^{(1)} = \hat{1}$ only contributes to the factor $\la\gamma|\gamma\ra$. The finite contributions are
\bea
&  \displaystyle \la\gamma|\hat{\bm E}^2({\br}, t)|\gamma\ra - \varepsilon_0 = 2 \left|\la 0|\hat{\bm E}({\br}, t)|\gamma\ra\right|^2,\cr
& \dst \la\gamma|\hat{\bm H}^2({\br}, t)|\gamma\ra - \varepsilon_0 = 2 \left|\la 0|\hat{\bm H}({\br}, t)|\gamma\ra\right|^2.
\eea
In a transparent non-magnetic medium with a refractive index $n(\omega)$ we have $\bk^2 = n^2\omega^2$, and the photon normalization factor in \eqref{A_norm_cond} changes from $\sqrt{4\pi/2\omega}$ to $\sqrt{4\pi/2\omega n^2}$ (see e.g. \cite{Ivanov2016}). As a result, $\bk^2 |{\bm A}_{{\bk}\lambda_\gamma}({\br}, t)|^2 = 2\pi\omega$ does not change. 

Thus, a finite part of the spatial energy density is
\bea
& \dst W(\br, t) = \frac{1}{8\pi}\la\gamma|\hat{\bm E}^2({\br}, t)+\hat{\bm H}^2({\br}, t)|\gamma\ra -\frac{\varepsilon_0}{4\pi} = \cr 
& \dst \frac{1}{4\pi} \left(\left|\la 0|\hat{\bm E}({\br}, t)|\gamma\ra\right|^2+\left|\la 0|\hat{\bm H}({\br}, t)|\gamma\ra\right|^2\right),
\label{E2H2}
\eea
One can also speak of it as a probability to detect the emitted photon in a region of space-time centered at the point $(\br, t)$, whereas the electron is jointly detected in the generalized measurement scheme, described in momentum space with the complex function $f_e^{(\text{det})}(\bp',\lambda')$. The electric and magnetic fields averages in \eqref{E2H2} can be expanded as follows
\bw
\bea
& \dst \frac{1}{4\pi}\Big|\la 0|\hat{\bm E}(\br,t)|\gamma\ra\Big|^2 = \frac{1}{4\pi}\Bigg|\sum_{\lambda_\gamma}\int \bar{d}^3\bk\,{\bm E}_{\lambda_\gamma}(\bk)e^{-ikx}\Bigg|^2 = \frac{1}{4\pi}\sum_{\lambda_\gamma, \tilde\lambda_\gamma}\int \bar{d}^3\bk \,\bar{d}^3\tilde\bk \,{\bm E}_{\lambda_\gamma}(\bk)\,{\bm E}_{\tilde\lambda_\gamma}^*(\tilde\bk) e^{-i(k - \tilde k)x} = \cr
& \dst = \left/\begin{matrix}\bk \rightarrow \bk + \tilde\bk/2,\\ \tilde\bk \rightarrow \bk - \tilde\bk/2 \end{matrix}\right/ = \frac{1}{4\pi}\sum_{\lambda_\gamma,\tilde\lambda_\gamma}\int \bar{d}^3\bk\,\bar{d}^3\tilde\bk\, {\bm E}_{\tilde\lambda_\gamma}^*(\bk - \tilde\bk/2)\,{\bm E}_{\lambda_\gamma}(\bk + \tilde\bk/2)e^{-it\left(\omega(\bk + \tilde\bk/2) - \omega(\bk - \tilde\bk/2)\right) + i(\br\cdot\bk)},
\label{E_to_expand}
\eea
\bea
& \dst \frac{1}{4\pi}\Big|\la 0|\hat{\bm H}(\br,t)|\gamma\ra\Big|^2 = \frac{1}{4\pi}\sum_{\lambda_\gamma,\tilde\lambda_\gamma}\int \bar{d}^3\bk\,\bar{d}^3\tilde\bk\, {\bm H}_{\tilde\lambda_\gamma}^*(\bk - \tilde\bk/2)\,{\bm H}_{\lambda_\gamma}(\bk + \tilde\bk/2)e^{-it\left(\omega(\bk + \tilde\bk/2) - \omega(\bk - \tilde\bk/2)\right) + i(\br\cdot\bk)}.
\label{H_to_expand}
\eea
\ew
\section{{Details on the near-field and WKB approximations for the photon Wigner function}}
\label{NFA}

\paragraph{Near-field approximation.}

The near-field approximation of the Wigner function in Eq. \eqref{WEE} implies that $\tilde\bk$ is set to zero everywhere except for the electron wavepacket function Eq. \eqref{WPm} and the exponent in Eq. \eqref{WEE}. It results in the following contraction:
\bw
\bea
& \dst \left({\bm e}_{\bk + {\tilde \bk}/2,\lambda_\gamma}{\bm e}^*_{\bk - {\tilde \bk}/2,\tilde{\lambda}_{\gamma}} + \frac{1}{\omega(\bk+{\tilde \bk}/2)\omega(\bk-{\tilde \bk}/2)}\left[\bk - \frac{\tilde\bk}{2}\times{\bm e}^*_{\bk-{\tilde \bk}/2,\tilde{\lambda}_{\gamma}}\right]\cdot\left[\bk + \frac{\tilde\bk}{2}\times {\bm e}_{\bk+{\tilde \bk}/2,\lambda_\gamma}\right]\right)\times \cr
& \dst e^{-it(\omega(\bk + {\tilde \bk}/2)-\omega(\bk - {\tilde \bk}/2)) + i\br\cdot{{\tilde \bk}}}\delta\left(\varepsilon(\bp + {\tilde \bk}/2) - \varepsilon' - \omega(\bk + {\tilde \bk}/2)\right)\delta\left(\varepsilon(\bp-{\tilde \bk}/2)-\varepsilon' - \omega(\bk - {\tilde \bk}/2)\right) \to \cr
  & \dst \to 
\delta_{\lambda_\gamma,\tilde{\lambda}_{\gamma}}\left(1 + n^2(\omega(\bk))\right)\frac{T}{2\pi}\, \delta\left(\varepsilon(\bp) - \varepsilon' - \omega(\bk)\right) e^{-it(\varepsilon(\bp + {\tilde \bk}/2) - \varepsilon(\bp - {\tilde \bk}/2)) + i\br\cdot{{\tilde \bk}}},
\label{NFA_E}
\eea
\ew
where $T$ is a normalization time emerging from the squared energy delta-function:
\bea
& \dst \delta(\varepsilon(\bp) - \varepsilon' - \omega(\bk))\,\delta(\varepsilon(\bp) - \varepsilon' - \omega(\bk)) \approx \cr
& \dst \approx \delta(\varepsilon(\bp) - \varepsilon' - \omega(\bk)) \frac{1}{2\pi}\int\limits_{-T/2}^{T/2}dt' e^{-it'(\varepsilon(\bp) - \varepsilon' - \omega(\bk))} = \cr 
& \dst \delta(\varepsilon(\bp) - \varepsilon' - \omega(\bk))\, \frac{1}{2\pi}\int\limits_{-T/2}^{T/2}dt' = \cr
& \dst \frac{T}{2\pi}\delta(\varepsilon(\bp) - \varepsilon' - \omega(\bk))
\label{nrm_cnst_T}
\eea
The Wigner function within this approximation then reads
\bw
\bea
& \dst \mathcal W(\br, \bk, t) = (2\pi)^5 \int \frac{\bar{d}^3\bp}{4n^4\left(\omega(\bk)\right)\varepsilon'\varepsilon(\bp)}\frac{T}{2\pi}\delta^{(4)}(p - p' - k) \left(1 + n^2(\omega(\bk))\right) |M_{fi}\left(\bp,\lambda,\bk,\lambda_\gamma\right)|^2\times\cr
& \dst \underbrace{\int\bar{d}^3{\tilde \bk}\, \left(f_e^{(\text{in})}(\bp-{\tilde \bk}/2)\right)^* f_e^{(\text{in})}(\bp+{\tilde \bk}/2)\, e^{-it(\varepsilon(\bp + {\tilde \bk}/2) - \varepsilon(\bp - {\tilde \bk}/2)) + i\br\cdot{{\tilde \bk}}}}_{w_e^{(\rm in)}(\br,\bp,t)},
\label{WF_NFA}
\eea
\ew
where one can see exactly the Wigner function of an incoming electron (Eq. \eqref{WWd3k}).
\paragraph{Paraxial (WKB) approximation}

The more accurate WKB-approximation implies the neglection of all the terms $\mathcal O(\tilde\bk)$ leaving unchanged those of $\mathcal O(\tilde\bk^2)$ in the space-time dependent phase. This leads to the following reductions in Eq. \eqref{WEE} 
\bw
\bea
& \dst {\bm e}^*_{\bk-{\tilde \bk}/2,\tilde{\lambda}_{\gamma}}\cdot{\bm e}_{\bk+{\tilde \bk}/2,\lambda_\gamma} + \frac{1}{\omega(\bk+{\tilde \bk}/2)\omega(\bk-{\tilde \bk}/2)} \left[\left(\bk - \frac{\tilde\bk}{2}\right)\times{\bm e}^*_{\bk-{\tilde \bk}/2,\tilde{\lambda}_{\gamma}}\right]\cdot\left[\left(\bk + \frac{\tilde\bk}{2}\right)\times {\bm e}_{\bk+{\tilde \bk}/2,\lambda_\gamma}\right] \approx \cr
& \dst \approx (1+n^2(\omega(\bk)))\left(\delta_{\tilde{\lambda}_{\gamma}\lambda_\gamma} + \frac{{\tilde k}_i}{2}\left({\bm e}^*_{\bk,\tilde{\lambda}_{\gamma}}\cdot\frac{\partial {\bm e}_{\bk,\lambda_\gamma}}{\partial k_i} - {\bm e}_{\bk,\lambda_\gamma}\cdot\frac{\partial {\bm e}^*_{\bk,\tilde{\lambda}_{\gamma}}}{\partial k_i}\right) + \mathcal O({\tilde \bk}^2)\right)
\label{eexpan}
\eea
\bea
& \dst M_{fi}(\bp + {\tilde \bk}/2, \bk + {\tilde \bk}/2,\lambda_e,\lambda_\gamma)  M^*_{fi}(\bp - {\tilde \bk}/2, \bk -{\tilde \bk}/2,\lambda_e,\lambda_\gamma) = \cr 
& \dst \left(\left|M_{fi}(\bp, \bk,\lambda_e,\lambda_\gamma)\right|^2 + \mathcal O ({\tilde \bk}^2) \right) \exp\left\{i{\tilde \bk}\cdot \left(\partial_{\bp} + \partial_{\bk}\right)\zeta_{fi}(\bp, \bk,\lambda_e,\lambda_\gamma) + \mathcal O ({\tilde \bk}^3)\right\}.
\label{Mexpan}
\eea
\bea
& \dst \delta\left(\varepsilon(\bp + {\tilde \bk}/2) - \varepsilon' - \omega(\bk + {\tilde \bk}/2)\right)\delta\left(\varepsilon(\bp - {\tilde \bk}/2) - \varepsilon' - \omega(\bk - {\tilde \bk}/2)\right) = \cr
& \dst = \frac{1}{(2\pi)^2}\int\limits_{-\infty}^{\infty} dt_1\,dt_2 \exp{it_1\left(\varepsilon(\bp + {\tilde \bk}/2) - \varepsilon' - \omega(\bk + {\tilde \bk}/2)\right) + it_2\left(\varepsilon(\bp - {\tilde \bk}/2) - \varepsilon' - \omega(\bk - {\tilde \bk}/2)\right)} \approx \cr 
& \dst \left/\varepsilon(\bp \pm {\tilde \bk}/2) \approx \varepsilon(\bp) \pm \frac{\tilde\bk\cdot{\bm u}_p}{2} + \frac{1}{2}\frac{\tilde k_i}{2}\frac{\tilde k_j}{2}\partial_{ij}^2\varepsilon,\quad \omega(\bk \pm {\tilde \bk}/2) \approx \omega(\bk) \pm \frac{\tilde\bk\cdot {\bm u}_k}{2} + \frac{1}{2}\frac{\tilde k_i}{2}\frac{\tilde k_j}{2}\partial_{ij}^2\omega\right/\approx \cr
& \dst \approx \frac{1}{(2\pi)^2}\int\limits_{t_{\rm in}}^{t_{\rm out}} dt' \int\limits_{-\infty}^{\infty} d\tau \exp{it'(\varepsilon(\bp) - \varepsilon' - \omega(\bk)) + it'\frac{1}{2}\frac{\tilde k_i}{2}\frac{\tilde k_j}{2}(\partial_{ij}^2\varepsilon - \partial_{ij}^2\omega) + i\tilde\bk\,\tau({\bm u}_p - {\bm u}_k)}
\label{delta_expan}
\eea
\ew
where in the last row we have done the substitution $t_1 + t_2 = t'$ and $t_1 - t_2 = \tau$. As seen the integral over $t'$ reminds the expansion of delta-function $\delta(\varepsilon(\bp) - \varepsilon' - \omega(\bk))$, which is the energy conservation as if the particles were the plane waves with well-defined momenta $\bp$, $\bp'$ and $\bk$ for initial, final electrons and emitted photon, respectively. Thus, the variable $t'$ can be interpreted as the time of the photon state evolution during its propagation. Therefore, we can investigate formation of the state, taking this integral in finite limits (let it be symmetric from $t_{\rm in} = -t_{\rm out}$, to $t_{\rm out}$).

The integral over $d^3{\tilde\bk}$ in Eq.\eqref{MII} is a Gaussian one and can be evaluated analytically as
\bea
& \dst \int\bar{d}^3{\tilde \bk}\, \exp\left\{-\mathcal {\bm A}\cdot {\tilde \bk} - \frac{1}{2}{\tilde k}_i{\tilde k}_jB_{ij}\right\} = \cr 
& \dst (2\pi)^{-3/2}\frac{1}{\sqrt{\det B}}\,\exp\left\{\frac{1}{2} B_{ij}^{-1}\mathcal A_i \mathcal A_j\right\},
\label{Gauss}
\eea
where
\bea
& \dst \det B = \eta^2 (\eta + \chi\, {\bm u}_p^2),\cr
& \dst B_{ij}^{-1} = \eta^{-1}\delta_{ij} - \frac{\chi}{\eta(\eta + \chi\, {\bm u}_p^2)}({\bm u}_p)_i({\bm u}_p)_j,\cr
& \dst \eta(t') = \frac{1}{2\sigma^2} + \frac{it'}{4}\Big(\frac{1}{\omega n^2} - \frac{1}{\varepsilon}\Big),\ \chi(t') = \frac{it'}{4}\Big(\frac{1}{\varepsilon} - \frac{1}{\omega}\Big).
\eea
The integral over $\tau$ is also Gaussian and evaluated as follows:
{\small\bea
& \dst \int\limits_{-\infty}^{+\infty}\frac{d\tau}{2\pi}\,\exp\left\{\frac{1}{2}B^{-1}_{ij}\mathcal A_i \mathcal A_j\right\} = \cr
& \dst = \frac{1}{\sqrt{2\pi}}\, \sqrt{\frac{\eta + \chi {\bm u}_p^2}{({\bm u}_p - {\bm u}_k)^2 + \frac{\chi}{\eta} [{\bm u}_p\times {\bm u}_k]^2}}\times \cr
& \dst \exp\left\{-\frac{1}{2\eta} \frac{[{\bm R}\times({\bm u}_p - {\bm u}_k)]^2 + \frac{\chi}{\eta} ({\bm u}_p\cdot \left[{\bm R} \times {\bm u}_k\right])^2}{({\bm u}_p - {\bm u}_k)^2 + \frac{\chi}{\eta} \left[{\bm u}_p\times {\bm u}_k\right]^2}\right\},
\label{int_tau}
\eea}
where the vector ${\bm R}$ together with $\br$ are defined in Eq. \eqref{Rr} and is illustrated in Fig.\ref{FigGen}. We can rewrite the pre-exponential factor in Eq.\eqref{int_tau} as follows: 
\bea
& \dst  \sqrt{\frac{\eta + \chi\, {\bm u}_p^2}{\det B_{ij} \left(({\bm u}_p - {\bm u}_k)^2 + \frac{\chi}{\eta} [{\bm u}_p\times {\bm u}_k]^2\right)}} = \cr
& \dst \sqrt{\frac{1}{\eta \left(\eta({\bm u}_p - {\bm u}_k)^2 + \chi [{\bm u}_p\times {\bm u}_k]^2\right)}} \equiv \cr
& \dst \frac{1}{G(t')}\, \exp\left\{-\frac{i}{2}\, g(t')\right\},\cr 
& \dst G(t') = \fr{\left|{\bm u}_p - {\bm u}_k\right|}{2\sigma^2}\left[\left(1+(t'/t_d)^2\right)\left(1+(t'/{\tilde t}_d)^2\right)\right]^{1/4}, \cr
& \dst g(t') = g_1(t') + g_2(t') = \arctan\fr{t'}{t_d} + \arctan\fr{t'}{\tilde t_d}.
\label{G2}
\eea
and the exponent in Eq. \eqref{int_tau} can be presented in terms of the diffraction times $\tilde{t}_d, t_d$ from Eq.\eqref{gtp}:
\bw
\bea
& \dst - \frac{1}{2\eta(t')} \frac{\eta(t')[{\bm R}\times({\bm u}_p - {\bm u}_k)]^2 + \chi(t') ({\bm R}\cdot \left[{\bm u}_p \times {\bm u}_k\right])^2}{\eta(t')({\bm u}_p - {\bm u}_k)^2 + \chi(t') \left[{\bm u}_p\times {\bm u}_k\right]^2} = \cr
& \dst = -\sigma^2\fr{1-it'/\tilde{t}_d}{1+(t'/\tilde{t}_d)^2}\fr{1-it'/t_d}{1+(t'/t_d)^2} \left(\fr{[{\bm R}\times ({\bm u}_p - {\bm u}_k)]^2}{({\bm u}_p - {\bm u}_k)^2} \left(1 + \fr{it'}{\tilde{t}_d}\right) + it' \fr{\sigma^2}{2}\Big(\frac{1}{\varepsilon} - \frac{1}{\omega}\Big)\fr{({\bm R}\cdot \left[{\bm u}_p \times {\bm u}_k\right])^2}{({\bm u}_p - {\bm u}_k)^2}\right) = \cr
& \dst = -\fr{1}{\sigma_{\perp}^2(t')}\left(\underbrace{\fr{[{\bm R}\times ({\bm u}_p - {\bm u}_k)]^2}{({\bm u}_p - {\bm u}_k)^2}}_{\text{finite at $t'=0$}} + \underbrace{\fr{(t')^2}{\tau_d^2 (1 + (t'/\tilde{t}_d)^2)}\fr{({\bm R}\cdot \left[{\bm u}_p \times {\bm u}_k\right])^2}{({\bm u}_p - {\bm u}_k)^2}}_{\text{due to spreading at $t'\ne 0$}}\right) + \cr
& \dst + i\frac{t'}{\sigma_\perp^2(t')}\Bigg(\frac{1}{t_d}\frac{[{\bm R}\times ({\bm u}_p - {\bm u}_k)]^2}{({\bm u}_p - {\bm u}_k)^2} - \frac{\sigma^2}{2}\left(\frac{1}{\varepsilon}-\frac{1}{\omega}\right)\frac{1-t'^2/(t_d\tilde{t}_d)}{1+t'^2/\tilde{t}_d^2}\frac{({\bm R}\cdot \left[{\bm u}_p \times {\bm u}_k\right])^2}{({\bm u}_p - {\bm u}_k)^2}\Bigg) \equiv \cr
& \dst \equiv - \fr{R^2}{R_{\text{eff}}^2(t')} + i\frac{R^2}{R_{\text{Im}}^2(t')}
\label{explhs}
\eea
\ew
where we have introduced the correlation radius $R_{\rm eff}(t')$ and the characteristic radius of the imaginary part, $R_{\rm Im}$. 

When the electron packet is wide and $\sigma$ can be treated as a small parameter compared to the electron mass $m_e$, the first term of the real part of Eq. \eqref{explhs} is finite at $t' = 0$ and it is $\mathcal O(\sigma^2)$, whereas the second one vanishes at $t' = 0$ due to spreading and it is $\mathcal O(\sigma^6)$. The imaginary part also vanishes at $t' = 0$ and is $\mathcal O(\sigma^4)$. At small $t' \ll t_d, \tilde{t}_d$, we get
\bea
& \dst - \frac{1}{2\eta(t')} \frac{\eta(t')[{\bm R}\times({\bm u}_p - {\bm u}_k)]^2 + \chi(t') ({\bm R}\cdot \left[{\bm u}_p \times {\bm u}_k\right])^2}{\eta(t')({\bm u}_p - {\bm u}_k)^2 + \chi(t') \left[{\bm u}_p\times {\bm u}_k\right]^2} \cr
& \dst \approx -\sigma^2 \fr{[{\bm R}\times ({\bm u}_p - {\bm u}_k)]^2}{({\bm u}_p - {\bm u}_k)^2} = -\sigma^2 {\bm R}^2 \sin^2\Delta\Theta,
\eea
where $\Delta\Theta$ is the angle between $\bm R$ and $\bm{u}_k - \bm{u}_p$.

\section{Transverse momentum conservation}
\label{ApTri}

In the sections above we used the following expansion of the delta function for the transverse momentum conservation in cylindrical coordinates
\bw
\bea
& \displaystyle \delta^{(2)}(\bp_{\perp} - \bp'_{\perp} - \bk_{\perp}) = \cr 
& \displaystyle = \frac{\Theta(p_{\perp}, k_{\perp}, p'_{\perp})}{2\Delta} \Big(\delta\left(\phi' - \phi + \alpha\right)\delta\left(\phi_k - \phi - \gamma\right) + \delta\left(\phi' - \phi - \alpha\right)\delta\left(\phi_k - \phi + \gamma\right)\Big) = \cr
& \displaystyle = \frac{\Theta(p_{\perp}, k_{\perp}, p'_{\perp})}{2\Delta} \Big(\delta\left(\phi' - \phi_k - (\vartheta-\pi)\right)\delta\left(\phi_k - \phi - \gamma\right) + \delta\left(\phi' - \phi_k + (\vartheta-\pi)\right)\delta\left(\phi_k - \phi + \gamma\right)\Big),
\label{delta2}
\eea
\ew
where $\Delta$ is an area of the triangle with the legs $p_{\perp}, p'_{\perp}, k_{\perp}$ and the angles $\alpha, \vartheta, \gamma\, (\alpha + \vartheta + \gamma = \pi)$,
\bea
& \displaystyle\Delta = \frac{1}{2} p_{\perp}p'_{\perp}\sin\alpha = \frac{1}{2} k_{\perp}p'_{\perp}\sin\vartheta = \frac{1}{2} p_{\perp}k_{\perp}\sin\gamma,\cr
& \dst \alpha = \arccos\left\{\frac{p_{\perp}^2 + (p'_{\perp})^2 - k_{\perp}^2}{2p_{\perp}p'_{\perp}}\right\},\cr 
& \dst \vartheta = \arccos\left\{\frac{(p'_{\perp})^2 + k_{\perp}^2 - p_{\perp}^2}{2k_{\perp}p'_{\perp}}\right\},\cr 
& \dst \gamma = \arccos\left\{\frac{p_{\perp}^2 + k_{\perp}^2 - (p'_{\perp})^2}{2k_{\perp}p_{\perp}}\right\},
\label{Deltatri}
\eea
and the legs satisfy the triangle rules,
\bea
p_{\perp} \leq k_{\perp} + p'_{\perp},\ p'_{\perp} \leq k_{\perp} + p_{\perp},\ k_{\perp} \leq p_{\perp} + p'_{\perp},
\label{trirules}
\eea
so that the function $\Theta(p_{\perp}, k_{\perp}, p'_{\perp})$ in Eq.(\ref{delta2}) equals 1 when these inequalities are simultaneously satisfied and $0$ otherwise. 

\section{Helicity amplitudes}\label{Amp}
The first-order amplitude of the photon emission by an electron is
\bea
M_{fi} = \sqrt{4\pi \alpha_{\rm em}}\, \bar{u}_{\bp'\lambda'}\gamma^{\mu}e^*_{\mu}u_{\bp\lambda} = |M_{fi}|\,e^{i\zeta_{fi}},
\label{Mfi}
\eea
where $\alpha_{\rm em} = e^2/\hbar c$ is the fine structure constant, $\gamma^{\mu}e^*_{\mu} = - {\bm \gamma}\cdot{\bm e}^*_{\bk\lambda_\gamma}$ in the Coulomb gauge, and the bispinors for initial and final electron state as well as the  polarization vector are expanded in the following series:
\bea
& \displaystyle u_{\bp\lambda} = \sum\limits_{\sigma=\pm 1/2} u_{\varepsilon\lambda}^{(\sigma)}\, d^{(1/2)}_{\sigma\lambda}(\theta)\, e^{-i\sigma\phi},\cr
& \displaystyle \bar{u}_{\bp'\lambda'} = \sum\limits_{\sigma'=\pm 1/2} \bar{u}_{\varepsilon'\lambda'}^{(\sigma')}\, d^{(1/2)}_{\sigma'\lambda'}(\theta')\, e^{i\sigma'\phi'},\cr
& \displaystyle {\bm e}_{\bk\lambda_\gamma} = \sum\limits_{\sigma_{\gamma}=0,\pm 1} {\bm \chi}^{(\sigma_{\gamma})} \, d^{(1)}_{\sigma_{\gamma}\lambda_\gamma}(\theta_k)\, e^{-i\sigma_{\gamma}\phi_k}.
\label{uexp_second}
\eea
Here, $\theta, \phi$ are the polar and azimuthal angles of $\bp$, while $\theta'$, $\phi'$ are those for $\bp'$ and $\theta_k, \phi_k$ -- the same for $\bk$, and also 
\bea
& \dst\hat{s}_z u_{\varepsilon\lambda}^{(\sigma)} =\sigma u_{\varepsilon\lambda}^{(\sigma)}, \cr
& \dst \left({\bm \chi}^{(\sigma_{\gamma})}\right)^*\cdot {\bm \chi}^{(\sigma'_{\gamma})} = \delta_{\sigma_{\gamma}\sigma'_{\gamma}}.
\eea
We employ the phase convention of Ref.\cite{BLP}, so that $\hat{j}_z u_{\bp\lambda} = 0$ (see details in \cite{EPJC}). The small Wigner functions are 
\bw
\bea
& \displaystyle d^{(1/2)}_{\sigma\lambda}(\theta) = \delta_{\sigma\lambda} \cos(\theta/2) - 2\sigma\, \delta_{\sigma,-\lambda} \sin(\theta/2), \, \sigma,\lambda = \pm \frac{1}{2}\cr
& \displaystyle  d^{(1)}_{\sigma_{\gamma}\lambda_\gamma}(\theta_k) = \left\{d^{(1)}_{11} = \cos^2(\theta_k/2), d^{(1)}_{-11} = \sin^2(\theta_k/2), d^{(1)}_{0\lambda_\gamma} = -d^{(1)}_{\lambda_\gamma 0} = \frac{\lambda_\gamma}{\sqrt{2}}\sin(\theta_k)\right\},\, \lambda_\gamma = \pm 1.
\label{uexp}
\eea
\ew
We find the following equalities for the spatial component of the 4-transition current $j_{fi}^\mu$:
{\small\bea
& \displaystyle \bar{u}_{\varepsilon'\lambda'}^{(\sigma')}\,{\bm \gamma}\, u_{\varepsilon\lambda}^{(\sigma)} = \cr
& \dst \left(2\lambda \sqrt{\varepsilon - m_e} \sqrt{\varepsilon' + m_e} + 2\lambda' \sqrt{\varepsilon' - m_e} \sqrt{\varepsilon + m_e}\right) \left(\omega^{(\sigma')}\right)^{\dagger} {\bm \sigma} \omega^{(\sigma)} = \cr
& \displaystyle = \left(2\lambda \sqrt{\varepsilon - m_e} \sqrt{\varepsilon' + m_e} + 2\lambda' \sqrt{\varepsilon' - m_e} \sqrt{\varepsilon + m_e}\right) \times \cr
& \dst 2\sigma \left({\bm \chi}^{(0)} \, \delta_{\sigma\sigma'} - {\bm \chi}^{(2\sigma)} \sqrt{2}\, \delta_{\sigma,-\sigma'}\right).
\eea}
and then
\bea
& \displaystyle \bar{u}_{\varepsilon'\lambda'}^{(\sigma')}\,\left({\bm \chi}^{(\sigma_{\gamma})}\right)^*\cdot{\bm \gamma}\, u_{\varepsilon\lambda}^{(\sigma)} = \cr
& \dst \left(2\lambda \sqrt{\varepsilon - m_e} \sqrt{\varepsilon' + m_e} + 2\lambda' \sqrt{\varepsilon' - m_e} \sqrt{\varepsilon + m_e}\right) \times \cr
& \dst 2\sigma \left(\delta_{\sigma_{\gamma}0} \, \delta_{\sigma\sigma'} - \sqrt{2}\,\delta_{\sigma_{\gamma},2\sigma}\,  \delta_{\sigma,-\sigma'}\right).
\eea
Summing over $\sigma,\sigma',\sigma_{\gamma}$, we notice that only the terms obeying $\sigma = \sigma'+\sigma_{\gamma}$ contribute, and so there are four of them
\bea
& \displaystyle M_{fi} = g_{\lambda\lambda'}\sum\limits_{\sigma,\sigma',\sigma_{\gamma}}\delta_{\sigma, \sigma'+\sigma_{\gamma}}\, M_{fi}^{(\sigma\sigma'\sigma_{\gamma})}\, e^{i\zeta_{fi}^{(\sigma\sigma'\sigma_{\gamma})}} = \cr
& \dst g_{\lambda\lambda'}\Bigg(M_{fi}^{(\frac{1}{2},-\frac{1}{2},1)}\, e^{i\zeta_{fi}^{(\frac{1}{2},-\frac{1}{2},1)}} + 
M_{fi}^{(\frac{1}{2},\frac{1}{2},0)}\, e^{i\zeta_{fi}^{(\frac{1}{2},\frac{1}{2},0)}} +\cr
& \dst M_{fi}^{(-\frac{1}{2},\frac{1}{2},-1)}\, e^{i\zeta_{fi}^{(-\frac{1}{2},\frac{1}{2},-1)}} + M_{fi}^{(-\frac{1}{2},-\frac{1}{2},0)}\, e^{i\zeta_{fi}^{(-\frac{1}{2},-\frac{1}{2},0)}}\Bigg),
\eea
where 
\bea
\displaystyle g_{\lambda\lambda'} = \sqrt{4\pi\alpha_{\rm em}}\,\left(2\lambda \sqrt{\varepsilon - m_e} \sqrt{\varepsilon' + m_e} + 2\lambda' \sqrt{\varepsilon' - m_e} \sqrt{\varepsilon + m_e}\right),
\eea
and the helicity amplitudes, which are real but not necessarily positive, are
\bw
\bea 
&& \displaystyle M_{fi}^{(\frac{1}{2},-\frac{1}{2},1)} = \sqrt{2}\,\,  d^{(1/2)}_{1/2,\lambda}(\theta)\, d^{(1/2)}_{-1/2,\lambda'}(\theta')\, d^{(1)}_{1\lambda_\gamma}(\theta_k),\quad \zeta_{fi}^{(\frac{1}{2},-\frac{1}{2},1)} = -\frac{1}{2}(\phi+\phi') + \phi_k,\cr 
&& \displaystyle 
M_{fi}^{(\frac{1}{2},\frac{1}{2},0)} = - d^{(1/2)}_{1/2,\lambda}(\theta)\, d^{(1/2)}_{1/2,\lambda'}(\theta')\, d^{(1)}_{0\lambda_\gamma}(\theta_k),\quad 
\zeta_{fi}^{(\frac{1}{2},\frac{1}{2},0)} = \frac{1}{2}(\phi'-\phi), \cr 
&& \displaystyle
M_{fi}^{(-\frac{1}{2},\frac{1}{2},-1)} = -\sqrt{2}\,\, d^{(1/2)}_{-1/2,\lambda}(\theta)\, d^{(1/2)}_{1/2,\lambda'}(\theta')\, d^{(1)}_{-1\lambda_\gamma}(\theta_k),\quad \zeta_{fi}^{(-\frac{1}{2},\frac{1}{2},-1)} =  -\zeta_{fi}^{(\frac{1}{2},-\frac{1}{2},1)},\cr 
&& \displaystyle
M_{fi}^{(-\frac{1}{2},-\frac{1}{2},0)} = d^{(1/2)}_{-1/2,\lambda}(\theta)\, d^{(1/2)}_{-1/2,\lambda'}(\theta')\, d^{(1)}_{0\lambda_\gamma}(\theta_k),\quad
\zeta_{fi}^{(-\frac{1}{2},-\frac{1}{2},0)} = -\zeta_{fi}^{(\frac{1}{2},\frac{1}{2},0)}.
\eea
Finally,

\bea
& \displaystyle |M_{fi}|^2/g^2_{\lambda\lambda'} = \sum\limits_{\sigma,\sigma',\sigma_{\gamma}}\delta_{\sigma, \sigma'+\sigma_{\gamma}}\, \left(M_{fi}^{(\sigma\sigma'\sigma_{\gamma})}\right)^2 + 2 M_{fi}^{(\frac{1}{2},-\frac{1}{2},1)} M_{fi}^{(\frac{1}{2},\frac{1}{2},0)}\,\cos\left(\zeta_{fi}^{(\frac{1}{2},-\frac{1}{2},1)} - \zeta_{fi}^{(\frac{1}{2},\frac{1}{2},0)} \right) +\cr 
& \displaystyle +  2 M_{fi}^{(\frac{1}{2},-\frac{1}{2},1)} M_{fi}^{(-\frac{1}{2},\frac{1}{2},-1)}\,\cos\left(\zeta_{fi}^{(\frac{1}{2},-\frac{1}{2},1)} - \zeta_{fi}^{(-\frac{1}{2},\frac{1}{2},-1)} \right) + 2 M_{fi}^{(\frac{1}{2},-\frac{1}{2},1)} M_{fi}^{(-\frac{1}{2},-\frac{1}{2},0)}\,\cos\left(\zeta_{fi}^{(\frac{1}{2},-\frac{1}{2},1)} - \zeta_{fi}^{(-\frac{1}{2},-\frac{1}{2},0)} \right) +\cr 
& \displaystyle +  2 M_{fi}^{(\frac{1}{2},\frac{1}{2},0)} M_{fi}^{(-\frac{1}{2},\frac{1}{2},-1)}\,\cos\left(\zeta_{fi}^{(\frac{1}{2},\frac{1}{2},0)} - \zeta_{fi}^{(-\frac{1}{2},\frac{1}{2},-1)}\right) + 2 M_{fi}^{(\frac{1}{2},\frac{1}{2},0)} M_{fi}^{(-\frac{1}{2},-\frac{1}{2},0)}\,\cos\left(\zeta_{fi}^{(\frac{1}{2},\frac{1}{2},0)} - \zeta_{fi}^{(-\frac{1}{2},-\frac{1}{2},0)}\right) +\cr 
& \displaystyle + 2 M_{fi}^{(-\frac{1}{2},\frac{1}{2},-1)} M_{fi}^{(-\frac{1}{2},-\frac{1}{2},0)}\,\cos\left(\zeta_{fi}^{(-\frac{1}{2},\frac{1}{2},-1)} - \zeta_{fi}^{(-\frac{1}{2},-\frac{1}{2},0)}\right),\cr
\label{MDz}
\eea
\ew
where
{\small\bea
& \displaystyle \zeta_{fi} = \arctan\frac{\sum\limits_{\sigma,\sigma',\sigma_{\gamma}}\delta_{\sigma, \sigma'+\sigma_{\gamma}}\, M_{fi}^{(\sigma\sigma'\sigma_{\gamma})} \sin\left(\zeta_{fi}^{(\sigma\sigma'\sigma_{\gamma})}\right)}{\sum\limits_{\sigma,\sigma',\sigma_{\gamma}}\delta_{\sigma, \sigma'+\sigma_{\gamma}}\, M_{fi}^{(\sigma\sigma'\sigma_{\gamma})} \cos\left(\zeta_{fi}^{(\sigma\sigma'\sigma_{\gamma})}\right)},
\label{zeta_fi}
\eea}
Here the sums include only four above terms obeying $\sigma = \sigma' + \sigma_{\gamma}$.

On the triangle point from Eq.(\ref{delta2}) $\phi = \phi'+\alpha, \phi_k = \phi'+\alpha+\gamma$, we have 
\bea
 && \displaystyle \zeta_{fi}^{(\frac{1}{2},-\frac{1}{2},1)} = -\frac{1}{2}(\phi+\phi') + \phi_k \to \gamma +\alpha/2,\cr 
&& \displaystyle 
\zeta_{fi}^{(\frac{1}{2},\frac{1}{2},0)} = \frac{1}{2}(\phi'-\phi) \to -\alpha/2, \cr 
&& \displaystyle
\zeta_{fi}^{(-\frac{1}{2},\frac{1}{2},-1)} = \pi+\frac{1}{2}(\phi+\phi') - \phi_k \to -\gamma-\alpha/2,\cr 
&& \displaystyle
\zeta_{fi}^{(-\frac{1}{2},-\frac{1}{2},0)} = -\frac{1}{2}(\phi'-\phi) \to \alpha/2,
\eea
where on the second point $\phi = \phi'-\alpha, \phi_k = \phi'-\alpha-\gamma$, the phases change the signs and so
\bea
\zeta_{fi}\Big|_{\phi = \phi'+\alpha, \phi_k = \phi'+\alpha+\gamma} = - \zeta_{fi}\Big|_{\phi = \phi'-\alpha, \phi_k = \phi'-\alpha-\gamma},
\eea
whereas
\bea
|M_{fi}|^2\Big|_{\phi = \phi'+\alpha, \phi_k = \phi'+\alpha+\gamma} =  |M_{fi}|^2\Big|_{\phi = \phi'-\alpha, \phi_k = \phi'-\alpha-\gamma},
\eea

As seen, the phase takes opposite signs for two momentum configurations from Eq.\eqref{delta2} and it also determines the quantum shift in the photon arrival time through its derivative. In contrast, the squared modulus of the matrix element $|M_{fi}|^2$ stays independent of the momentum configuration choice. Note that the phases - and, therefore, the matrix element - do not depend on the azimuthal angle, $\phi'$, of the final electron momentum and the transverse momenta in the amplitudes must satisfy the triangle rules (Eq.\eqref{trirules}).

\putbib  
\end{bibunit}

\clearpage
\onecolumngrid
\begingroup
\begin{bibunit}  
\setcounter{section}{0}
\setcounter{subsection}{0}
\setcounter{equation}{0}
\setcounter{figure}{0}
\setcounter{table}{0}
\renewcommand{\appendixname}{Supplement}
\renewcommand{\thesection}{S\arabic{section}}
\renewcommand{\thesubsection}{\thesection.\arabic{subsection}}
\renewcommand{\theequation}{S\arabic{equation}}
\renewcommand{\thefigure}{S\arabic{figure}}
\renewcommand{\thetable}{S\Roman{table}}

\begin{center}
  {\Large\bfseries Supplemental information to phase-space description of photon emission\par}
  \vspace{1.0em}
  {D.V.~Karlovets$^{1,2}$, A.A.~Shchepkin$^{1,2}$, A.D.~Chaikovskaya$^{1}$,
  D.V.~Grosman$^1$, D.A.~Kargina$^{1}$, U.G.~Rybak$^3$ and G.K.~Sizykh$^{1,2}$\par}
  \vspace{0.7em}
  {\small
  $^1$School of Physics and Engineering, ITMO University, 197101 St. Petersburg, Russia\\
  $^2$Petersburg Nuclear Physics Institute named by B.P. Konstantinov of National
  Research Centre ``Kurchatov Institute'', 188300, Gatchina, Russia\\
  $^3$Peter the Great St. Petersburg Polytechnic University, 194021, St. Petersburg, Russia\par}
\end{center}
\vspace{1.2em}

\section{Detection in the scheme of positive-operator-valued measurements}\label{supp:POVM}
\subsection{Generalized measurements with loss of information}

The simplest model of a realistic detection scheme implies that one performs generalized measurements with a positive-operator-valued measurement (POVM) of an electron quantum numbers. Conceptually, POVM can be described by projective measurements (PM) which undergo a so-called \textit{scrambling}. This is a process performed by a classical device that for each output of PM randomly chooses the state from the detector output basis with a certain probability. Thus, in this scheme the information about the quantum system \textit{is} lost and the final photon can only be described by a density matrix, defined as \cite{Lvovsky}:
\bea
\hat{\rho}_{\gamma}^{(\text{POVM})} = \text{Tr}\{\hat{F}^{(\text{det})}_e\,\hat{\rho}_{e\gamma}\},
\label{supp:rhoPOVM}
\eea
where $\hat{\rho}_{e\gamma}$ is a density matrix of the final bipartite system, which for the pure state becomes $\hat{\rho}_{e\gamma} \to |e',\gamma\ra \la e',\gamma|$, and the trace is taken over the electron component. The operator $\hat{F}^{(\text{det})}_e$ describes detection of the electron and it can be written as an \textit{incoherent} mixture of states from an orthogonal set that we use. For instance, for plane waves it is
\bea
\hat{F}^{(\text{det})}_e = \sum\limits_{\lambda'}\int\frac{d^3p'}{(2\pi)^3}\, f_e^{(\text{det})}(\bp',\lambda')\, |\bp',\lambda'\ra\la\bp',\lambda'|,
\label{supp:POVMpw}
\eea
where $f_e^{(\text{det})}(\bp',\lambda')$ is a normalized probability to detect the electron with the momentum $\bp'$ and the helicity $\lambda'$.

Let us first assume that one performs the POVM detection of the electron state in the two-particle evolved state as
\bea
|e',\gamma\ra = |\text{in}\ra + \sum_{\lambda', \lambda_\gamma} \int\bar{d}^3\bk\,\bar{d}^3\bp'\, |\bp',\lambda'\ra\otimes|\bk,\lambda_\gamma\ra\, S_{fi}^{(1)}. \quad
\label{supp:2partWFPW}
\eea
(see Eq. (4) in the article) with $\hat{\rho}_{e\gamma} = |e',\gamma\ra \la e',\gamma|$ and with the operator (\ref{supp:POVMpw}). Then we note that $\la \bp,\lambda|\bp',\lambda'\ra= (2\pi)^3\delta_{\lambda\lambda'}\delta(\bp-\bp')$ and obtain the following photon density matrix from Eq.(\ref{supp:rhoPOVM}):
\bea
& \displaystyle \hat{\rho}_{\gamma}^{(\text{POVM})} = \sum\limits_{\lambda'}\int\frac{d^3p'}{(2\pi)^3} f_e^{(\text{det})}(\bp',\lambda')\sum\limits_{\lambda_{\gamma},\lambda'_{\gamma}}\int\frac{d^3k}{(2\pi)^3}\frac{d^3k'}{(2\pi)^3} |\bk,\lambda_{\gamma}\ra\la \bk',\lambda'_{\gamma}|\cr 
& \displaystyle \times S_{fi}^{(1)}(\bp',\lambda',\bk,\lambda_{\gamma})\left(S_{fi}^{(1)}(\bp',\lambda',\bk',\lambda'_{\gamma})\right)^*,
\label{supp:rhog}
\eea
where $S_{fi}^{(1)}(\bp',\lambda',\bk,\lambda_{\gamma}) = \la \bp',\lambda';\bk,\lambda_{\gamma}|\hat{S}^{(1)}|\text{in}\ra$.
Let us further assume that the incoming electron state \textit{is a plane wave} with the four-momentum $p = \{\varepsilon(\bp),\bp\}, \varepsilon(\bp) = \sqrt{m_e^2 + \bp^2}$ (say, during the Cherenkov emission), so that  
\bea
S_{fi}^{(1)}(\bp',\lambda',\bk,\lambda_{\gamma}) = i(2\pi)^4 \delta^{(4)}(p - p' - k)\, T_{fi}^{(\lambda'\lambda_{\gamma})}(\bp',\bk),
\label{supp:Tfi}
\eea
and
\bea
& \displaystyle S_{fi}^{(1)}(\bp',\lambda',\bk,\lambda_{\gamma})\left(S_{fi}^{(1)}(\bp',\lambda',\bk',\lambda'_{\gamma})\right)^* \propto \delta(\varepsilon(\bp) - \varepsilon(\bp') - \omega(\bk))\delta(\bp - \bp' - \bk)\cr & \displaystyle \times \delta(\varepsilon(\bp) - \varepsilon(\bp') - \omega(\bk')) \delta(\bp - \bp' - \bk') = [\delta(\varepsilon(\bp) - \varepsilon(\bp') - \omega(\bk))]^2 \delta(\bp - \bp' - \bk)\delta(\bk - \bk') = \cr 
& \displaystyle = \frac{T}{2\pi}\, \delta(\varepsilon(\bp) - \varepsilon(\bp') - \omega(\bk)) \delta(\bp - \bp' - \bk)\delta(\bk - \bk')
\label{supp:SSp}
\eea
with $T \to \infty$. 
As a result, the matrix (\ref{supp:rhog}) becomes diagonal in $\bp'$ and $\bk$, but not in the photon helicity, 
\bea
\hat{\rho}_{\gamma} = T \sum\limits_{\lambda',\lambda_{\gamma},\lambda'_{\gamma}} \int d\Gamma\, f_e^{(\text{det})}(\bp',\lambda')\, T_{fi}^{(\lambda'\lambda_{\gamma})} \left(T_{fi}^{(\lambda'\lambda'_{\gamma})}\right)^*\, |\bk,\lambda_{\gamma}\ra\la \bk,\lambda'_{\gamma}|,
\label{supp:rhog2}
\eea
where $T_{fi}^{(\lambda'\lambda_{\gamma})} = T_{fi}^{(\lambda'\lambda_{\gamma})}(\bp',\bk)$ and we have denoted
\bea
d\Gamma = (2\pi)^4\,\delta^{(4)}(p - p' - k)\,\frac{d^3p'}{(2\pi)^3}\frac{d^3k}{(2\pi)^3}. 
\label{supp:Gamma}
\eea

Let us estimate the vorticity of the photon, that is, its mean total angular momentum (TAM) projection onto the direction of the incoming electron ($z$ axis),
\bea
\label{supp:AvTAM}
\la\hat{J}_z\ra = \text{Tr} \left\{\hat{J}_z \hat{\rho}_{\gamma}\right\}/\text{Tr} \left\{ \hat{\rho}_{\gamma}\right\}.
\eea
We need to calculate $\la \bk,\lambda'_{\gamma}|\hat{J}_z|\bk,\lambda_{\gamma}\ra$, and as the plane wave is not an eigenfunction of $\hat{J}_z$ operator, we expand the unity operator over the twisted states,
\bea
\label{supp:Unity}
\hat{1}_{\gamma} = \sum\limits_{m=-\infty}^{\infty}\sum\limits_{\Lambda=\pm 1}\int\frac{dk'_z}{2\pi}\frac{k'_{\perp} dk'_{\perp}}{2\pi}\, |k'_z,k'_{\perp},m,\Lambda\ra\la k'_z,k'_{\perp},m,\Lambda|,
\eea
where $\Lambda$ is the helicity, $m$ is the total angular momentum projection, and $\hat{J}_z |k'_z,k'_{\perp},m,\Lambda\ra = m |k'_z,k'_{\perp},m,\Lambda\ra$. As a result, we have
\bea
\la \bk,\lambda'_{\gamma}|\hat{J}_z|\bk,\lambda_{\gamma}\ra = \sum\limits_{m=-\infty}^{\infty} m \sum\limits_{\lambda=\pm 1}\int\frac{dk'_z}{2\pi}\frac{k'_{\perp} dk'_{\perp}}{2\pi} \la \bk,\lambda'_{\gamma}|k'_z,k'_{\perp},m,\lambda\ra\la k'_z,k'_{\perp},m,\lambda|\bk,\lambda_{\gamma}\ra,
\eea
where in the cylindrical coordinates $\bk = \{k_{\perp},\phi_{\gamma},k_z\}$ 
\bea
\label{supp:ScProd}
\la \bk,\lambda'_{\gamma}|k'_z,k'_{\perp},m,\Lambda\ra \propto \delta(k_z - k'_z)\delta(k_{\perp} - k'_{\perp})\delta_{\lambda'_{\gamma} \Lambda}\, e^{i m \phi_{\gamma}}, 
\eea
which is a scalar part of the twisted photon wave function in momentum representation, and so the product $\la \bk,\lambda'_{\gamma}|k'_z,k'_{\perp},m,\Lambda\ra\la k'_z,k'_{\perp},m,\Lambda|\bk,\lambda_{\gamma}\ra$ does \textit{not} depend on $m$. As a result, 
\bea
& \displaystyle \la \bk,\lambda'_{\gamma}|\hat{J}_z|\bk,\lambda_{\gamma}\ra \propto \sum\limits_{m=-\infty}^{\infty} m = ... -2 -1 +0 +1 +2+... = 0,\cr
& \displaystyle \la\hat{J}_z\ra = 0. 
\label{supp:Jz0}
\eea
Meanwhile
\bea
& \displaystyle \la \bk,\lambda'_{\gamma}|\hat{J}^2_z|\bk,\lambda_{\gamma}\ra \propto \sum\limits_{m=-\infty}^{\infty} m^2 = \infty,\cr
& \displaystyle \la\hat{J}_z^2\ra = \infty. 
\label{supp:Jz20}
\eea 

The photon has thus the vanishing mean vorticity and the infinitely wide TAM spread, which is merely a consequence of the plane-wave basis for the POVM-operator, however it does not depend on the specific form of the detector function $f_e^{(\text{det})}(\bp',\lambda')$. In particular, this means that whatever errors of the electron measurement in the plane-wave basis are, both the equations (\ref{supp:Jz0}) and (\ref{supp:Jz20}) still hold. An important consequence one can draw is that even if the azimuthal angle of the electron momentum is measured with a large error, $\sigma_{\phi} \to 2\pi$, this does \textit{not} automatically put the photon TAM dispersion $\la\hat{J}_z^2\ra$ to a finite value because the conjugate variable -- the TAM projection -- is not measured by the plane-wave detector. \textbf{To obtain a finite} TAM dispersion, one needs to perform the electron measurements in the \textbf{cylindrical basis}, even if with the finite errors. 

Importantly, this is \textit{not} the feature of the POVM scheme itself, but rather that of the plane-wave approximation for the incoming electron \textit{or} the detector state. Let us first examine a scenario with an OAM-resolved detector described by POVM
\bea
\hat{F}^{(\text{det})}_{e,tw} = \sum\limits_{m' = -\infty}^{\infty} \sum\limits_{\lambda'} \int\frac{dp'_z}{2\pi}\frac{p'_{\perp} dp'_{\perp}}{2\pi} \, f_{e,tw}^{(\text{det})}(p'_z, p'_{\perp}, m', \lambda') \, |p'_z, p'_{\perp}, m', \lambda' \ra \la p'_z, p'_{\perp}, m', \lambda'|.
\eea
Further we use the expression for the scalar product of a plane-wave and a twisted electron states analogous to Eq. \eqref{supp:ScProd}:
\bea
\label{supp:ScProdEqEl}
\la \bp,\lambda | p'_z, p'_{\perp}, m', \lambda' \ra = \left( N_{PW}^{(e)} \right)^* N_{tw}^{(e)} \frac{(2 \pi)^2}{p_{\perp}} e^{i m' \varphi_p} \delta(p_z - p'_z) \delta(p_{\perp} - p'_{\perp}) \delta_{\lambda \lambda'}, 
\eea
where $N_{PW}^{(e)}$ and $ N_{tw}^{(e)}$ are the normalization factors of the plane-wave and the twisted electron states, respectively.

In accordance with Eq. \eqref{supp:rhoPOVM} the photon state is given by the density matrix
\bea
& \displaystyle \hat{\rho}_{\gamma}^{(\text{POVM})} = \sum\limits_{m' = -\infty}^{\infty} \sum\limits_{\lambda',\lambda_{\gamma}, \lambda'_{\gamma}} \int\frac{dp'_z}{2\pi}\frac{p'_{\perp} dp'_{\perp}}{2\pi} \, f_{e,tw}^{(\text{det})}(p'_z, p'_{\perp}, m', \lambda') \int\frac{d^3k}{(2\pi)^3}\frac{d^3k'}{(2\pi)^3} |N_{PW}^{(e)}|^2 |N_{tw}^{(e)}|^2 \cr 
& \displaystyle \times S_{fi,tw}^{(1)}(p'_z, p'_{\perp}, m', \lambda',\bk, \lambda_{\gamma}) \left(S_{fi,tw}^{(1)}(p'_z, p'_{\perp}, m', \lambda', \bk',\lambda'_{\gamma})\right)^* |\bk,\lambda_{\gamma}\ra\la \bk',\lambda'_{\gamma}|,
\eea
where the twisted amplitudes may be obtained from their plane-wave counterparts by the relation.
\be
S_{fi,tw}^{(1)}(p'_z, p'_{\perp}, m', \lambda',\bk, \lambda_{\gamma}) = \int_0^{2 \pi} \frac{d \phi_{p'}}{2 \pi} e^{- i m' \phi_{p'}} S_{fi}^{(1)}(\bp', \lambda', \bk, \lambda_{\gamma}).
\ee
For the twisted amplitudes the following equality takes place (see Appendix E in the article):
\bea
\nonumber & S_{fi,tw}^{(1)}(p'_z, p'_{\perp}, m', \lambda',\bk, \lambda_{\gamma}) = \\
& = \dst i (2 \pi)^4 \delta(\varepsilon - \varepsilon' - \omega) \delta(p_z - p'_z - k_z) \frac{1}{2 \Delta(k_{\perp})}\Bigg\{e^{-i m' (\phi_p - \alpha(k_{\perp}))} \delta(\phi_k - \phi_p - \gamma) T_{fi}^{\lambda' \lambda_{\gamma}}(p'_z, p'_{\perp}, \phi_p - \alpha(k_{\perp}), \bk) + \cr 
& e^{-i m' (\phi_p + \alpha(k_{\perp}))} \delta(\phi_k - \phi_p + \gamma) T_{fi}^{\lambda' \lambda_{\gamma}}(p'_z, p'_{\perp}, \phi_p + \alpha(k_{\perp}), \bk)\Bigg\}.
\eea
Finally, the photon density matrix becomes
\bea
\label{supp:rhogTw}
\nonumber & \displaystyle \hat{\rho}_{\gamma}^{(\text{POVM})} = \sum\limits_{m' = -\infty}^{\infty} \sum\limits_{\lambda',\lambda_{\gamma}, \lambda'_{\gamma}} \int d \Gamma_{tw} \, f_{e,tw}^{(\text{det})}(p'_z, p'_{\perp}, m', \lambda') \frac{\omega}{(2 \Delta)^2} |N_{PW}^{(e)}|^2 |N_{tw}^{(e)}|^2 \times \\
\nonumber & \Bigg(\Big\{T_{fi,tw}^{(\lambda' \lambda_{\gamma})}(p'_z, p'_{\perp}, \phi_p - \alpha, k_z, k_{\perp}, \phi_p + \gamma) \left( T_{fi,tw}^{(\lambda' \lambda'_{\gamma})}(p'_z, p'_{\perp}, \phi_p - \alpha, k_z, k_{\perp}, \phi_p + \gamma) \right)^* \times \\
\nonumber & | k_z, k_{\perp}, \phi_p + \gamma, \lambda_{\gamma} \ra \la k_z, k_{\perp}, \phi_p + \gamma, \lambda'_{\gamma}| + \\
\nonumber & e^{i m' 2 \alpha} T_{fi,tw}^{(\lambda' \lambda_{\gamma})}(p'_z, p'_{\perp}, \phi_p - \alpha, k_z, k_{\perp}, \phi_p + \gamma) \left( T_{fi,tw}^{(\lambda' \lambda'_{\gamma})}(p'_z, p'_{\perp}, \phi_p + \alpha, k_z, k_{\perp}, \phi_p - \gamma) \right)^* \times \\
& | k_z, k_{\perp}, \phi_p + \gamma, \lambda_{\gamma} \ra \la k_z, k_{\perp}, \phi_p - \gamma, \lambda'_{\gamma}| \Big\} + \cr
& \Big\{T_{fi,tw}^{(\lambda' \lambda_{\gamma})}(p'_z, p'_{\perp}, \phi_p + \alpha, k_z, k_{\perp}, \phi_p - \gamma) \left(T_{fi,tw}^{(\lambda' \lambda'_{\gamma})}(p'_z, p'_{\perp}, \phi_p + \alpha, k_z, k_{\perp}, \phi_p - \gamma) \right)^* \times \\
\nonumber & | k_z, k_{\perp}, \phi_p - \gamma, \lambda_{\gamma} \ra \la k_z, k_{\perp}, \phi_p - \gamma, \lambda'_{\gamma}| + \\
\nonumber & e^{-i m' 2 \alpha} T_{fi,tw}^{(\lambda' \lambda_{\gamma})}(p'_z, p'_{\perp}, \phi_p + \alpha, k_z, k_{\perp}, \phi_p - \gamma) \left( T_{fi,tw}^{(\lambda' \lambda'_{\gamma})}(p'_z, p'_{\perp}, \phi_p - \alpha, k_z, k_{\perp}, \phi_p + \gamma) \right)^* \times \\
& | k_z, k_{\perp}, \phi_p - \gamma, \lambda_{\gamma} \ra \la k_z, k_{\perp}, \phi_p + \gamma, \lambda'_{\gamma}| \Big\}\Bigg),
\eea
where the measure
\be
d \Gamma_{tw} = (2 \pi)^4 \delta(\varepsilon - \varepsilon' - \omega) \delta(p_z - p'_z - k_z) \frac{d p'_z}{2 \pi} \frac{p'_{\perp} d p'_{\perp}}{2 \pi} \frac{d k_z}{2 \pi} \frac{k_{\perp} d k_{\perp}}{2 \pi},
\ee
and the second expression in parentheses is the first one with $\{\alpha,\gamma\} \rightarrow \{-\alpha,-\gamma\}$. Note that the density matrix \eqref{supp:rhogTw} consists of two parts, one of which is diagonal in $\bk$ (terms, proportional to $| k_z, k_{\perp}, \phi_p - \gamma, \lambda_{\gamma} \ra \la k_z, k_{\perp}, \phi_p - \gamma, \lambda'_{\gamma}|$ and $| k_z, k_{\perp}, \phi_p + \gamma, \lambda_{\gamma} \ra \la k_z, k_{\perp}, \phi_p + \gamma, \lambda'_{\gamma}|$) and the other is not. The second one indicates the presence of the OAM in the photon state.

To calculate the average total angular momentum projection we use the unity operator \eqref{supp:Unity} and the scalar product of the plane-wave and the twisted photon states
\bea
\label{supp:ScProdEqPh}
\la \bk,\lambda_{\gamma} | k'_z, k'_{\perp}, m', \lambda'_{\gamma} \ra = \left( N_{PW}^{(\gamma)} \right)^* N_{tw}^{(\gamma)} \frac{(2 \pi)^2}{k_{\perp}} e^{i m' \phi_k} \delta(k_z - k'_z) \delta(k_{\perp} - k'_{\perp}) \delta_{\lambda_{\gamma} \lambda'_{\gamma}}.
\eea
This, in accordance with Eq. \eqref{supp:AvTAM}, leads to the following expression:
\bea
& \dst \la\hat{J}_z\ra = \frac{\text{Tr} \left\{\hat{J}_z \hat{1} \hat{\rho}_{\gamma}\right\}}{\text{Tr} \left\{ \hat{\rho}_{\gamma} \right\}} = \sum \limits_{M = - \infty}^{\infty} M \sum \limits_{\lambda_{\gamma}} \sum \limits_{m' = - \infty}^{\infty} \sum \limits_{\lambda'} \int d \Gamma_{tw} \, \frac{(2 \pi)^4}{k_{\perp}^2} \frac{\omega}{(2 \Delta)^2} f_{e,tw}^{(\text{det})}(p'_z, p'_{\perp}, m', \lambda') \times \cr 
\nonumber & \dst \frac{|N_{PW}^{(e)}|^2 |N_{tw}^{(e)}|^2 |N_{PW}^{(\gamma)}|^2 |N_{tw}^{(\gamma)}|^2}{\text{Tr} \left\{ \hat{\rho}_{\gamma} \right\}} \delta(k_{\perp} - k_{\perp}) \delta(k_z - k_z) \times \nonumber \cr
& \nonumber \Bigg\{ \left| T_{fi,tw}^{(\lambda' \lambda_{\gamma})}(p'_z, p'_{\perp}, \phi_p - \alpha, k_z, k_{\perp}, \phi_p + \gamma) \right|^2 + e^{- i M 2 \gamma} e^{i m' 2 \alpha} T_{fi,tw}^{(\lambda' \lambda_{\gamma})}(p'_z, p'_{\perp}, \phi_p - \alpha, k_z, k_{\perp}, \phi_p + \gamma) \times \\
& \times \left( T_{fi,tw}^{(\lambda' \lambda_{\gamma})}(p'_z, p'_{\perp}, \phi_p + \alpha, k_z, k_{\perp}, \phi_p - \gamma) \right)^* +  \left|T_{fi,tw}^{(\lambda' \lambda_{\gamma})}(p'_z, p'_{\perp}, \phi_p + \alpha, k_z, k_{\perp}, \phi_p - \gamma) \right|^2 + \cr 
& e^{i M 2 \gamma} e^{-i m' 2 \alpha} T_{fi,tw}^{(\lambda' \lambda_{\gamma})}(p'_z, p'_{\perp}, \phi_p + \alpha, k_z, k_{\perp}, \phi_p - \gamma)\left( T_{fi,tw}^{(\lambda' \lambda_{\gamma})}(p'_z, p'_{\perp}, \phi_p - \alpha, k_z, k_{\perp}, \phi_p + \gamma) \right)^* \Bigg\}.
\eea
The first term in the expression above vanishes due to $\sum M = ... - 2 -1 + 0 + 1 + 2... = 0$, while the second term contributes to the average TAM according to $\sum_M M e^{- i M 2 \gamma} \rightarrow \frac{2 \pi i}{4} \frac{d \delta(\gamma)}{d \gamma}$.

To make use of the arising derivative of the delta function let us change the integration over $dp_{\perp}'dk_{\perp}$ to $d\alpha\, d\gamma$. The electron and photon transverse momentum are related to the angles via the following expressions
\begin{equation}
k_{\perp} = \frac{p_{\perp}\sin\alpha}{\sin(\alpha \pm \gamma)}, \; p'_{\perp} = \frac{p_{\perp}\sin\gamma}{\sin(\alpha \pm \gamma)}
\end{equation}
Notice that that the sign ``$+$'' should be chosen as the angles $\alpha$ and $\gamma$ are angles of a triangle, therefore, $0<\alpha+\gamma<\pi$ always. Thus we can integrate over $d\alpha\, d\gamma$ in the region $[0,\pi]\times[0,\pi-\alpha]$ (notice that $\gamma$ varies in $[0,\pi-\alpha]$ not in $[-\alpha,\pi-\alpha]$ as it has to be in the region $[0,\pi]$ as a proper angle in a triangle.

The Jacobian of this transformation is 
\begin{equation}
\begin{vmatrix}
\dst \frac{\partial p_{\perp}'}{\partial\alpha} & \dst \frac{\partial p_{\perp}'}{\partial\gamma}\\
\dst \frac{\partial k_{\perp}}{\partial\alpha} & \dst \frac{\partial k_{\perp}}{\partial\gamma}
\end{vmatrix} = \frac{p_{\perp}^2\sin\gamma\sin\alpha}{\sin^2(\alpha+\gamma)} = \frac{4\Delta^2}{p_{\perp}'k_{\perp}\sin^2(\alpha+\gamma)} = p_{\perp}'k_{\perp}.
\end{equation}
Thus our measure is now
\begin{equation}
d\Gamma_{tw}' = \frac{(2\pi)^2}{k_{\perp}}\frac{1}{(2\Delta)^2}d\Gamma_{tw} =  (2 \pi)^4 \delta(\varepsilon - \varepsilon' - \omega) \delta(p_z - p'_z - k_z) \frac{d p'_z}{2 \pi} \frac{d k_z}{2 \pi} \frac{1}{p_{\perp}\sin\alpha\sin(\alpha+\gamma)}d\alpha d\gamma.
\end{equation}
and the mean TAM projection is
\begin{equation}
\begin{aligned}
&\langle \hat{J}_z \rangle = (LR)\left(\text{Tr} \left\{ \hat{\rho}_{\gamma} \right\}\right)^{-1}\sum\limits_{\lambda_{\gamma},\lambda'}\sum\limits_{m' = -\infty}^{\infty}\int d\Gamma' f_{e,tw}^{(\text{det})}(p'_z, p'_{\perp}, m', \lambda') |N_{PW}^{(e)}|^2 |N_{tw}^{(e)}|^2 |N_{PW}^{(\gamma)}|^2 |N_{tw}^{(\gamma)}|^2 \times\\
&\left[T_{fi,tw}^{(\lambda' \lambda_{\gamma})}(p'_z, p'_{\perp}, \phi_p - \alpha, k_z, k_{\perp}, \phi_p + \gamma) T_{fi,tw}^{(\lambda' \lambda_{\gamma})*}(p'_z, p'_{\perp}, \phi_p + \alpha, k_z, k_{\perp}, \phi_p - \gamma)e^{2im'\alpha} -\right.\\
&
\left.T_{fi,tw}^{(\lambda' \lambda_{\gamma})}(p'_z, p'_{\perp}, \phi_p + \alpha, k_z, k_{\perp}, \phi_p - \gamma) T_{fi,tw}^{(\lambda' \lambda_{\gamma})*}(p'_z, p'_{\perp}, \phi_p - \alpha, k_z, k_{\perp}, \phi_p + \gamma)e^{-2im'\alpha}\right]\frac{\pi i}{2}\delta'(\gamma).
\end{aligned}
\end{equation}

Now we need to exploit the derivative of the delta function. It gives the derivative of the integrand evaluated at $\gamma = 0$. The only place where $\gamma$ enters are the $T$-matrices and the measure.
Importantly, to get a non-vanishing photon TAM one can still employ the plane-wave basis for the POVM scheme, but with the initial electron state being different from the simplified plane wave. That is why let us consider now the more general problem with the incoming electron being a wave packet,
\bea
& \displaystyle |\text{in}\ra = \sum\limits_{\lambda} \int\frac{d^3p}{(2\pi)^3}\,f_e^{(\text{in})}(\bp,\lambda) |\bp,\lambda\ra,\cr
& \displaystyle \la \text{in}|\text{in}\ra = \sum\limits_{\lambda} \int\frac{d^3p}{(2\pi)^3}\,|f_e^{(\text{in})} (\bp,\lambda)|^2 = 1
\eea
described with a normalized wave function $f_e^{(\text{in})}(\bp,\lambda)$. In Eq.(\ref{supp:rhog}) we have instead of Eq.(\ref{supp:SSp})
\bea
& \displaystyle S_{fi}^{(1)}(\bp',\lambda',\bk,\lambda_{\gamma})\left(S_{fi}^{(1)}(\bp',\lambda',\bk',\lambda'_{\gamma})\right)^* \to \sum\limits_{\lambda,\bar{\lambda}}\int\frac{d^3p}{(2\pi)^3}\frac{d^3\bar{p}}{(2\pi)^3}\,\left(f_e^{(\text{in})}(\bar{\bp},\bar{\lambda})\right)^* f_e^{(\text{in})}(\bp,\lambda)\cr 
& \displaystyle \times (2\pi)^8\,\delta(\varepsilon(\bp) - \varepsilon(\bp') - \omega(\bk))\delta(\bp - \bp' - \bk)\delta(\varepsilon(\bar{\bp}) - \varepsilon(\bp') - \omega(\bk')) \delta(\bar{\bp} - \bp' - \bk')\cr
& \displaystyle \times T_{fi}^{(\lambda\lambda'\lambda_{\gamma})}(\bp,\bp',\bk) \left(T_{fi}^{(\bar{\lambda}\lambda'\lambda'_{\gamma})}(\bar{\bp},\bp',\bk')\right)^*,
\label{supp:SSpg}
\eea
and so $\bk \ne \bk'$. Here we sum over three polarization variables, which is why we indicate three upper indices of the amplitude $T_{fi}^{(\lambda\lambda'\lambda_{\gamma})}$. That is why the mean TAM of the resultant photon, defined by $\la \bk',\lambda'_{\gamma}|\hat{J}_z|\bk,\lambda_{\gamma}\ra$, is \textit{not necessarily vanishing}, as we will see hereafter, and it depends on the detection protocol.

We can reproduce the vanishing result (\ref{supp:Jz0}) in the paraxial approximation when the incoming wave packet behaves nearly like a plane wave. To this end, let us change the variables in Eq.(\ref{supp:SSpg}) as follows:
\bea
\bp_+ = \frac{\bp+\bar{\bp}}{2},\ \bp_- = \bp-\bar{\bp},\ d^3p\,d^3\bar{p} = d^3p_+d^3p_-.
\label{supp:varW0}
\eea
Assume now that the electron wave function $f_e^{(\text{in})}$ has a smooth Gaussian envelope (see, for instance, \cite{JHEP, PRD}),
\bea
& \displaystyle f_e^{(\text{in})}(\bp,\lambda) \propto \exp\left\{-\frac{(\bp - \la\bp\ra)^2}{2\sigma^2}\right\}
\label{supp:feG}
\eea
centered at some mean momentum $\la\bp\ra$, and then
\bea
& \displaystyle f_e^{(\text{in})}(\bp_+ + \bp_-/2,\lambda)(f_e^{(\text{in})}(\bp_+ - \bp_-/2,\bar{\lambda}))^* \propto \exp\left\{-\frac{(\bp_+ - \la\bp\ra)^2}{\sigma^2} -\frac{\bp_-^2}{(2\sigma)^2}\right\}.
\label{supp:feG2}
\eea
In the paraxial regime with a wide packet with $\sigma \ll m$ (or $1/\sigma \gg \lambda_c = 1/m$), one can first expand all the functions under the integral in series over $\bp_-$. The leading term is obtained when $\bp_- = 0$ everywhere except for $f_e^{(\text{in})}(\bp_+ + \bp_-/2,\lambda)(f_e^{(\text{in})}(\bp_+ - \bp_-/2,\bar{\lambda}))^*$. 
The resultant paraxial photon density matrix is
\bea
& \displaystyle \hat{\rho}_{\gamma}^{(\text{POVM})} \approx T
\sum\limits_{\lambda,\bar{\lambda},\lambda'}\sum\limits_{\lambda_{\gamma},\lambda'_{\gamma}}\int \frac{d^3p_+}{(2\pi)^3}\, d\Gamma\,f_e^{(\text{det})}(\bp',\lambda')\, n_{\text{in}}^{(\lambda\bar{\lambda})}(\br=0,\bp_+,t=0)\, \cr 
& \displaystyle \times T_{fi}^{(\lambda\lambda'\lambda_{\gamma})}(\bp_+,\bp',\bk) \left(T_{fi}^{(\bar{\lambda}\lambda'\lambda'_{\gamma})}(\bp_+,\bp',\bk)\right)^*|\bk,\lambda_{\gamma}\ra\la \bk,\lambda'_{\gamma}|,
\label{supp:rhogpar}
\eea
where $d\Gamma = (2\pi)^4\,\delta^{(4)}(p_+ - p' - k)\,\frac{d^3p'}{(2\pi)^3}\frac{d^3k}{(2\pi)^3}$ and
\bea
n_{\text{in}}^{(\lambda\bar{\lambda})}({\br} = 0,\bp_+, t=0) = \int\frac{d^3p_-}{(2\pi)^3}\, f_e^{(\text{in})}(\bp_+ + \bp_-/2,\lambda)(f_e^{(\text{in})}(\bp_+ - \bp_-/2,\bar{\lambda}))^*\, e^{-i\bp_-\cdot {\br}}\Big|_{{\br} = 0}
\label{supp:Wigner}
\eea
is a Wigner function of the electron (see, for instance, \cite{JHEP, PRD}) taken at ${\br} = 0$ in the initial moment of time $t = 0$. This matrix is diagonal in $\bk$, which is why the TAM stays vanishing in the paraxial approximation. Note that depending on the incoming electron function $f_e^{(\text{in})}$ the use of the paraxial approximation can become impossible, which is the case, for instance, for the Bessel beams (see below).

\subsection{Informationally complete generalized measurements}

It is instructive to compare the above results with an analogous expression for the following pure photon state
\bea
& \dst |\gamma\ra = \la f_{e}^{(\text{det})}|e_{\text{in}}\ra|0_{\gamma}\ra + \sum\limits_{\lambda',\lambda_\gamma}\int\bar{d}^3\bk\,\bar{d}^3\bp'\, |\bk,\lambda_\gamma\ra\,\left(f_e^{(\text{det})}(\bp',\lambda')\right)^*\, S_{fi}^{(1)},\cr
& \dst S_{fi}^{(1)} \equiv  S_{fi}^{(1)}(\bp',\lambda',\bk,\lambda_\gamma) = \la \bk,\lambda_\gamma;\bp',\lambda'|\hat{S}^{(1)}|\text{in}\ra,\quad \left(f_e^{(\text{det})}(\bp',\lambda')\right)^* = \la f_{e}^{\text{(det)}}|\bp',\lambda'\ra,
\label{supp:1partPhGenPW}
\eea
when the electron is jointly detected in a pure state, though probably with finite errors, which gives no loss of the information: 
\bea
& \displaystyle |\gamma\ra \la\gamma| = \sum\limits_{\lambda',\bar{\lambda},\lambda_{\gamma},\bar{\lambda}_{\gamma}}\int\frac{d^3p'}{(2\pi)^3}\frac{d^3\bar{p}}{(2\pi)^3}\frac{d^3k}{(2\pi)^3}\frac{d^3k'}{(2\pi)^3}\, f_e^{(\text{det})}(\bar{\bp},\bar{\lambda})(f_e^{(\text{det})}(\bp',\lambda'))^*\, |\bk,\lambda_{\gamma}\ra\la \bk',\bar{\lambda}_{\gamma}|\cr
& \displaystyle \times
S_{fi}^{(1)}(\bp',\lambda',\bk,\lambda_{\gamma}) \left(S_{fi}^{(1)}(\bar{\bp},\bar{\lambda},\bk',\bar{\lambda}_{\gamma})\right)^*,
\label{supp:rhogpu}
\eea
where $f_e^{(\text{det})}(\bp',\lambda')$ is the electron detector function, introduced in Eq.(\ref{supp:1partPhGenPW}). Clearly, the photon matrix is not diagonal in $\bk$ even if the incoming electron is in a plane-wave state, which is assumed below. Thus, the photon TAM can be \textit{non-vanishing}. After changing the variables as
\bea
\bp_+ = \frac{\bp'+\bar{\bp}}{2},\ \bp_- = \bp'-\bar{\bp},\ \bk_+ = \frac{\bk+\bk'}{2},\ \bk_- = \bk-\bk',
\label{supp:varW}
\eea
we take into account that (see Eq.(\ref{supp:Tfi}))
\bea
& \displaystyle S_{fi}^{(1)}(\bp',\lambda',\bk,\lambda_{\gamma}) \left(S_{fi}^{(1)}(\bar{\bp},\bar{\lambda},\bk',\bar{\lambda}_{\gamma})\right)^* = (2\pi)^8 \delta\Big(\varepsilon - \varepsilon(\bp_+ + \bp_-/2) - \omega(\bk_+ - \bp_-/2)\Big)\cr
& \displaystyle \times\delta\Big(\varepsilon(\bp_+ + \bp_-/2) + \omega(\bk_+ - \bp_-/2) - \varepsilon(\bp_+ - \bp_-/2) - \omega(\bk_+ + \bp_-/2)\Big) \delta(\bp - \bp_+ - \bk_+) \delta(\bp_- + \bk_-) \cr
& \displaystyle \times T_{fi}^{(\lambda'\lambda_{\gamma})}(\bp_+ + \bp_-/2, \bk_+ - \bp_-/2)\left(T_{fi}^{(\bar{\lambda}\bar{\lambda}_{\gamma})}(\bp_+ - \bp_-/2, \bk_+ + \bp_-/2)\right)^*,
\eea
and arrive at the following general expression:
\bea
& \displaystyle |\gamma\ra \la\gamma| = \sum\limits_{\lambda',\bar{\lambda},\lambda_{\gamma},\bar{\lambda}_{\gamma}}\int\frac{d^3p_+}{(2\pi)^3}\frac{d^3k_+}{(2\pi)^3}\frac{d^3p_-}{(2\pi)^3}\,(2\pi)^5 f_e^{(\text{det})}(\bp_+ - \bp_-/2,\bar{\lambda})(f_e^{(\text{det})}(\bp_+ + \bp_-/2,\lambda'))^*\cr
& \displaystyle \times |\bk_+ - \bp_-/2,\lambda_{\gamma}\ra \la \bk_+ + \bp_-/2,\bar{\lambda}_{\gamma}|\times\, \delta\Big(\varepsilon - \varepsilon(\bp_+ + \bp_-/2) - \omega(\bk_+ - \bp_-/2)\Big)\times \cr
& \displaystyle \delta\Big(\varepsilon(\bp_+ + \bp_-/2) + \omega(\bk_+ - \bp_-/2) - \varepsilon(\bp_+ - \bp_-/2) - \omega(\bk_+ + \bp_-/2)\Big)\, \delta(\bp - \bp_+ - \bk_+)\times \cr
& \displaystyle \times T_{fi}^{(\lambda'\lambda_{\gamma})}(\bp_+ + \bp_-/2, \bk_+ - \bp_-/2)\left(T_{fi}^{(\bar{\lambda}\bar{\lambda}_{\gamma})}(\bp_+ - \bp_-/2, \bk_+ + \bp_-/2)\right)^*.
\label{supp:rhogpu2}
\eea

This time, we make the same paraxial approximation for the final electron packet and we obtain
\bea
& \displaystyle |\gamma\ra \la\gamma| \approx T \sum\limits_{\lambda',\bar{\lambda},\lambda_{\gamma},\bar{\lambda}_{\gamma}}\int d\Gamma\, n_{\text{det}}^{(\lambda'\bar{\lambda})}({\br} = 0,\bp', t=0) |\bk,\lambda_{\gamma}\ra \la \bk,\bar{\lambda}_{\gamma}|\,T_{fi}^{(\lambda'\lambda_{\gamma})}(\bp', \bk)\left(T_{fi}^{(\bar{\lambda}\bar{\lambda}_{\gamma})}(\bp', \bk)\right)^*,
\label{supp:gg}
\eea
where $d\Gamma = (2\pi)^4\,\delta^{(4)}(p - p' - k)\,\frac{d^3p'}{(2\pi)^3}\frac{d^3k}{(2\pi)^3}$ from Eq.(\ref{supp:Gamma}), while
\bea
n_{\text{det}}^{(\lambda'\bar{\lambda})}({\br} = 0,\bp', t=0) = \int\frac{d^3p_-}{(2\pi)^3}\, f_e^{(\text{det})}(\bp' - \bp_-/2,\bar{\lambda})(f_e^{(\text{det})}(\bp' + \bp_-/2,\lambda'))^*\, e^{-i\bp_-\cdot {\br}}\Big|_{{\br} = 0}
\label{supp:WignerDetector}
\eea
is a Wigner function of the electron detector. As the realistic packets are usually much wider than the Compton wavelengths of the corresponding particles while the momentum uncertainties are much smaller than $m_e$, the paraxial expression (\ref{supp:gg}) should suffice in the majority of practical cases. Clearly, the paraxial photon state has a vanishing TAM, $\la\hat{J}_z\ra=0$. Comparing this result with Eq.(\ref{supp:rhog2}), one can interpret the helicity-diagonal function 
\[
n_{\text{det}}^{(\lambda'=\bar{\lambda})}({\br} = 0,\bp', t=0)
\]
as the above probability\footnote{More precisely, the probability in p-space is given by \cite{PRD}
$|f_e^{(\text{det})}(\bp',\lambda')|^2 = \sqrt{2\varepsilon({\bm p'})}\int d^3x\, n_{\text{det}}^{(\lambda'=\bar{\lambda})}({\br}, \bp', t)$, 
which does not depend on $t$.}. If this were the case, the only formal difference between the paraxial expression (\ref{supp:gg}) for the pure state and Eq.(\ref{supp:rhog2}) for the mixed state within the POVM scheme would be an additional summation over $\bar{\lambda}$ in Eq.(\ref{supp:gg}), which can be removed supposing that $f_e^{(\text{det})}(\bp',\lambda') \propto \delta_{\lambda',\bar{\lambda}'}$. 

In contrast to the above POVM approach, however, the Wigner function is \textit{not} necessarily everywhere positive, although in the paraxial regime its negative regions (if any) are usually marginal. Moreover, this function -- even if everywhere positive -- can still depend on a phase of the detector function $f_e^{(\text{det})}$ (say, on the OAM $\ell$, such that $\text{arg}\,f_e^{(\text{det})}(\bp') = \ell\phi'$, see \cite{PRD}) if this phase or the quantum numbers it depends on are measured. Thus, Eq.(\ref{supp:gg}) is \textit{more general} than Eq.(\ref{supp:rhog2}), and this Wigner function 
describes a ``quantum detector'', which may \textit{not} lose information about the electron, unlike the classical scrambler. Such a quantum detector may not necessarily represent a classical macroscopic system, but it can be an atom or a particle interacting with the electron in a way sensitive to the needed quantum numbers. Recall that the standard POVM approach implies loss of information about the system, whereas the use of the more general formula (\ref{supp:rhogpu2}) implies measuring two conjugate variables jointly -- oftentimes momenta and coordinates -- which is why the photon can stay in a pure state. The generalized measurements preserving information and keeping the system in a pure state are often called \textit{informationally complete} measurements \cite{MIC, Yashin}. \textbf{In other words, to leave the photon in a pure state we need to perform quantum tomography of the electron state and vice versa}.

The paraxial counterpart (\ref{supp:gg}) of the general result (\ref{supp:rhogpu2}) still implies loss of information and it does \textit{not} describe an important phenomenon of a phase transfer (in particular, of the vorticity because $\la \hat{J}_z\ra = 0$) from the measured electron to the photon, even if the electron is detected in the cylindrical basis as a twisted state with a definite TAM. 
Clearly, the final photon can become twisted if starting from the general non-paraxial state Eq.(\ref{supp:1partPhGenPW}) -- see \cite{EPJC} for more detail and specific examples. Thus, the POVM scheme is a limiting case of the more general approach with the informationally complete generalized measurements within the paraxial approximation.

To put it differently, the general formula (\ref{supp:rhogpu2}) does \textit{not} reduce to Eq.(\ref{supp:gg}) when the different momentum components $\{p'_x,p'_y,p'_z\}$ or $\{p'_{\perp},p'_{\phi},p'_z\}$ are measured with different errors $\{\sigma_x,\sigma_y,\sigma_z\}$ or $\{\sigma_{\perp},\sigma_{\phi},\sigma_z\}$, respectively, and some of the latter are \textit{not} small. Say, when the momentum azimuthal angle $\phi'$ is not measured at all (that is, its error $\sigma_{\phi} \to 2\pi$) and
\bea
f_e^{(\text{det})}(\bp',\lambda') \propto \delta(p'_z - p'_{\parallel})\frac{1}{p'_{\perp}}\delta(p'_{\perp} - \bar{p}'_{\perp})\,\delta_{\lambda'\bar{\lambda}'},
\label{supp:feunm}
\eea
we cannot employ the above expansion in series of $\bp_-$. Such a detector function corresponds to the delocalized Bessel beam. That is why transfer of the TAM is not described by the paraxial formula (\ref{supp:gg}). 

Summarizing, realistic description of both the incoming electron and of the outgoing one as the spatially localized wave packets make transfer of the TAM to the final photon feasible, whether the incoming electron is twisted or not and if this TAM obeys the conservation law $\la\hat{J}_z\ra = \la\hat{J}_z\ra^e_{\text{in}} - \la\hat{J}_z\ra^e_{\text{det}}$. Table 1 demonstrates the comparison between the common features of the projective measurements, generalized measurements and the special case of the latter - POVMs. Informational completeness of the measurement here only affects the TAM bandwidth $\la\hat{J}_z^2\ra-\la\hat{J}_z\ra^2$ (which can, of course, be quite broad), but the loss of information within the POVM scheme itself does \textit{not} automatically result in the vanishing mean TAM of the final photon.

\begin{table}
	\begin{center}
		\begin{tabular}{|c|c|c|c|}
			\hline
			Measured quantity, error  & $L_z$, $\Delta L_z=0$ & $L_z$, $\Delta L_z\ne 0$ & No measurement of $L_z$ \\
			\hline
			$\phi_p$, $\Delta \phi_p=0$ & impossible & PM in PW-basis, $\Delta L_z\to \infty$ &  PM in PW-basis\\
			\hline
			$\phi_p$, $\Delta \phi_p\ne 0$ & PM in cyl. basis, $\Delta \phi_p = 2\pi$ & GM with no info loss & POVM with info loss \\
			\hline
			No measurement of $\phi_p$ & PM in cyl. basis & POVM with info loss & POVM with info loss \\
			\hline
		\end{tabular}
		\caption{The measurement schemes of two canonically conjugated variables and their key features. PM -- projective measurement, GM -- generalized measurement, PW - plane-wave.}
	\end{center}
	\label{supp:tab:parameters-1mm}
\end{table}
\renewcommand{\refname}{Supplementary References}
\begingroup
\makeatletter
\let\suppOriginalLabel\label
\def\label#1{%
  \def\suppGivenLabel{#1}%
  \def\suppLastLabel{LastBibItem}%
  \ifx\suppGivenLabel\suppLastLabel
    \suppOriginalLabel{supp:LastBibItem}%
  \else
    \suppOriginalLabel{#1}%
  \fi
}
\makeatother
\putbib  
\endgroup
\end{bibunit}
\endgroup
\end{document}